\documentclass[11pt]{article} 
\usepackage{epsfig}
\begin{document}
\input amssym.def 
\input amssym
\hfuzz=5.0pt
%
%
%
%
\def\vec#1{\mathchoice{\mbox{\boldmath$\displaystyle\bf#1$}}
{\mbox{\boldmath$\textstyle\bf#1$}}
{\mbox{\boldmath$\scriptstyle\bf#1$}}
{\mbox{\boldmath$\scriptscriptstyle\bf#1$}}}
\def\mbf#1{{\mathchoice {\hbox{$\rm\textstyle #1$}}
{\hbox{$\rm\textstyle #1$}} {\hbox{$\rm\scriptstyle #1$}}
{\hbox{$\rm\scriptscriptstyle #1$}}}}
\def\operatorname#1{{\mathchoice{\rm #1}{\rm #1}{\rm #1}{\rm #1}}}
\chardef\ii="10
\def\widehat{\mathaccent"0362 }
\def\widetilde{\mathaccent"0365 }
\def\vphi{\varphi}
\def\vrho{\varrho}
\def\vtheta{\vartheta}
\def\ih{{\i\over\hbar}}
\def\hi{\frac{\hbar}{\i}}
\def\CD{{\cal D}}
\def\CE{{\cal E}}
\def\CH{{\cal H}}
\def\CL{{\cal L}}
\def\CP{{\cal P}}
\def\CV{{\cal V}}
\def\half{{1\over2}}
\def\bhalf{\hbox{$\half$}}
\def\viert{{1\over4}}
\def\bviert{\hbox{$\viert$}}
\def\hhbox#1#2{\hbox{$\frac{#1}{#2}$}}
\def\dfrac#1#2{\frac{\displaystyle #1}{\displaystyle #2}}
\def\intT{\ih\int_0^\infty\d\,T\,e^{\i ET/\hbar}}
\def\pathint#1{\int\limits_{#1(t')=#1'}^{#1(t'')=#1''}\CD #1(t)}
\def\hbarm{{\dfrac{\hbar^2}{2m}}}
\def\hbarmq{{\dfrac{\hbar^2}{2mq}}}
\def\mzwei{\dfrac{m}{2}}
\def\overh{\dfrac1\hbar}
\def\ihbar{\dfrac\i\hbar}
\def\intt{\int_{t'}^{t''}}
\def\tn{\tilde n}
\def\pmb#1{\setbox0=\hbox{#1}
    \kern-.025em\copy0\kern-\wd0
    \kern.05em\copy0\kern-\wd0
    \kern-.025em\raise.0433em\box0}
\def\pathintG#1#2{\int\limits_{#1(t')=#1'}^{#1(t'')=#1''}\CD_{#2}#1(t)}
\def\limN{\lim_{N\to\infty}}
\def\Norm{\bigg({m\over2\pi\i\epsilon\hbar}\bigg)}
\def\hbaram{{\hbar^2\over8m}}
\def\bbbr{{\rm I\!R}}                                
\def\bbbn{{\rm I\!N}}                                
\def\bbbz{{\mathchoice {\hbox{$\sf\textstyle Z\kern-0.4em Z$}}
{\hbox{$\sf\textstyle Z\kern-0.4em Z$}}
{\hbox{$\sf\scriptstyle Z\kern-0.3em Z$}}
{\hbox{$\sf\scriptscriptstyle Z\kern-0.2em Z$}}}}    
\def\bbbc{{\mathchoice {\setbox0=\hbox{\rm C}\hbox{\hbox
to0pt{\kern0.4\wd0\vrule height0.9\ht0\hss}\box0}}
{\setbox0=\hbox{$\textstyle\hbox{\rm C}$}\hbox{\hbox
to0pt{\kern0.4\wd0\vrule height0.9\ht0\hss}\box0}}
{\setbox0=\hbox{$\scriptstyle\hbox{\rm C}$}\hbox{\hbox
to0pt{\kern0.4\wd0\vrule height0.9\ht0\hss}\box0}}
{\setbox0=\hbox{$\scriptscriptstyle\hbox{\rm C}$}\hbox{\hbox
to0pt{\kern0.4\wd0\vrule height0.9\ht0\hss}\box0}}}}
\def\CP{{\cal P}}
\def\CQ{{\cal Q}}
\def\Ai{\operatorname{Ai}} 
\def\Cl{\operatorname{Cl}} 
\def\SU{\operatorname{SU}} 
\def\dt{\d t}
\def\d{\operatorname{d}}
\def\e{\operatorname{e}}
\def\i{\operatorname{i}}
\def\max{\operatorname{max}}
\def\DI{D_{\,\rm I}}
\def\DII{D_{\,\rm II}}
\def\DIII{D_{\,\rm III}}
\def\DIV{D_{\,\rm IV}}
\def\vphi{\varphi}
\def\ps{\operatorname{ps}}
\def\Ps{\operatorname{Ps}}
\def\Si{\operatorname{Si}}
\def\energyldrei{\e^{-\i\hbar T(p^2+1)/2m}}
\def\ints{\int_0^{s''}}
\def\OO{\operatorname{O}}
\def\SO{\operatorname{SO}}
\def\operatorname#1{{\mathchoice{\rm #1}{\rm #1}{\rm #1}{\rm #1}}}
\def\bbbone{{\mathchoice {\rm 1\mskip-4mu l} {\rm 1\mskip-4mu l}
{\rm 1\mskip-4.5mu l} {\rm 1\mskip-5mu l}}}
\def\pathint#1{\int\limits_{#1(t')=#1'}^{#1(t'')=#1''}\CD #1(t)}
\def\pathints#1{\int\limits_{#1(0)=#1'}^{#1(s'')=#1''}\CD #1(s)}
 
\begin{titlepage}
\centerline{\normalsize DESY 04--221 \hfill ISSN 0418 - 9833}
\centerline{\hfill November 2004}
\vskip.3in
\message{TITLE:}
\begin{center}
{\Large Path Integration on Darboux Spaces}
\end{center}
\message{Path Integration on Darboux Spaces}
\vskip.5in
\begin{center}
{\Large Christian Grosche}
\vskip.2in
{\normalsize\em II.\,Institut f\"ur Theoretische Physik}
\vskip.05in
{\normalsize\em Universit\"at Hamburg, Luruper Chaussee 149}
\vskip.05in
{\normalsize\em 22761 Hamburg, Germany}
\end{center}
\normalsize
\vfill
\begin{center}
{\bf Abstract}
\end{center}
In this paper the Feynman path integral technique is applied to 
two-dimensional spaces of non-constant curvature: these spaces are
called Darboux spaces $\DI$--$\DIV$. We start each consideration
in terms of the metric and then analyze the quantum theory in the
separable coordinate systems. The path integral in each case is formulated
and then solved in the majority of cases, the exceptions being
quartic oscillators where no closed solution is known. The required 
ingredients are the path integral solutions of the linear potential, 
the harmonic oscillator, the radial harmonic oscillator, the modified 
P\"oschl--Teller potential, and for spheroidal wave-functions, respectively.
The basic path integral solutions, which appear here in a complicated way, 
have been developed in recent work and are known. The final solutions 
are represented in terms of the corresponding  Green's  functions and the
expansions into the wave-functions, respectively. We also sketch some
limiting cases of the Darboux spaces, where spaces of constant negative
and zero curvature emerge. 
\end{titlepage}
 
 
\tableofcontents

\setcounter{page}{1}
\setcounter{equation}{0}
\section{Introduction}
\message{Introduction}

\subsection{General Overview and Recent Work}
\message{General Overview and Recent Work}
In recent years there have been an enormous success in developing path
integral techniques and in solving Feynman path integrals.
After its invention by Feynman \cite{FEYb}, the solution of the
harmonic oscillator has been for a long time the only accessible path
integral solution. Several textbooks made the path integral more
popular, e.g. Feynman and Hibbs \cite{FH} and Schulman \cite{SCHUHd}. 
However, as a tool in quantum mechanics the path integral remained 
in a dormant state, whereas its main applications were in field
theory, e.g. \cite{FEYe,GY,IZ,LEE}. Only later on, the
technique of radial path integrals, i.e. the Cartesian path integral
formulated in spherical coordinates, were developed in the 1960's 
by Edwards and Gulyaev \cite{EG} and Peak and Inomata \cite{PI}. 

Matters shifted with the first calculation of the path integral for the
Hydrogen atom \cite{DKa,DKb} by Duru and Kleinert. In the textbook \cite{KLEo} 
a lot of applications and a summary of the development of this
technique can be found, including many references on the subject. 
The principal development in their approach was a technique called space-time
transformation, sometimes also called ``Duru--Kleinert transformation'', which
does not only performs a coordinate transformation in the path
integral but also transforms the time-sclicings  $\epsilon=T/N$ into new
time-sclicings $\delta$. This combined coordinate and time-transformation 
usually ends up in manipulating the action in the path integral in
such a way that a given problem is transformed into 
one of the basic path integrals. These basic path integrals are roughly
speaking, the harmonic oscillator, the radial harmonic oscillator
\cite{PI}, the (modified) P\"oschl--Teller potential path integrals
\cite{BJb,DURb,FLM}, and the Spheroidal Path Integral \cite{GROad,GKPSa}. 
In our book ``Handbook of Feynman Path Integrals'' \cite{GRSh}
we have given a thorough overview of all the techniques how to handle
and manipulate path integrals, and, most important, an up-to-date list
of several hundreds solvable path integrals and the corresponding
references. 

The separation of a particular quantum mechanical potential problem into
more than one coordinate system has the consequence that there are
additional integrals of motion and that the spectrum is degenerate. The
Noether theorem \cite{NOETHER} connects the particular symmetries of a
Lagrangian, i.e., the invariances with respect to the dynamical
symmetries, with conservation laws in classical mechanics and with
observables in quantum mechanics, respectively. In the case of the
isotropic harmonic oscillator one has in addition to the conservation
of energy and the conservation of the angular momentum, the conservation
of the quadrupole moment; in the case of the Coulomb problem one has
in addition to the conservation of energy and the angular momentum, the
conservation of the Pauli-Runge-Lenz vector. In total, the additional
conserved quantities in these two examples add up to five functionally
independent integrals of motion in classical mechanics, respectively
observables in quantum mechanics. 

A topic which appeared in the formulation of the radial path
integral and for the (modified) P\"oschl--Teller potential 
\cite{BJb,DURb,FLM,KLEMUS} was path integration over group
spaces. This included  the formulation and evaluation of the path
integral in spaces of constant curvature, the curvature being positive
(spheres), negative (hyperboloids), or zero (Euclidean and
pseudo-Euclidean space).  This opened a new rich field of investigations. 
It required separation of variables in the Schr\"odinger-, respectively the
Laplace--Beltrami equation $\Delta_{LB}-E$,  in various coordinate
systems in these spaces. The case of two- and three-dimensional 
Euclidean space can be found e.g. in \cite{MOON} and in the textbooks
\cite{MILLf,MOFE}: In two-dimensional flat space there are four coordinate
systems and in three-dimensional space there are 11 coordinate systems 
which separate the Helmholtz, respectively the Schr\"odinger equation.%
\footnote{As pointed out in \cite{BKM} there are 17 types of
coordinate systems which $R$-separate the Laplace equation, 
i.e.  $\Delta_{LB}\Phi=0$. However, these additional systems are very
complicated except the case of the toroidal coordinate system.}

A thorough study of separation of variables of the Laplace--Beltrami
equation in spaces of non-zero constant curvature was first performed
by Olevski\u\ii\ \cite{OLE}, who studied the two- and
three-dimensional cases of constant curvature. In particular, he found
that on the two-dimensional hyperboloid there are nine coordinate
systems and on the three-dimensional hyberboloid 34 coordinate
systems which allow separation of variables in the Laplace--Beltrami
equation. However, only in several cases which exhibit symmetry properties
closed solutions in terms of higher transcendental functions are
known. In many coordinate systems very little is known about the 
corresponding solutions of the Eigenvalue equation in terms of special 
functions and no closed solution is known. These coordinate systems 
are usually parameter-dependent, e.g. ellipsoidal or paraboloid systems. 
Usually, only in the non-parameter-dependent coordinate systems, like 
spherical and parabolic coordinates, a well-developed theory of higher
transcendental functions is known. In these cases, the relevant
eigen-functions can be expressed in terms of hypergeometric and
degenerate hypergeometric functions and we find Legendre polynomials
(in cases with a discrete spectrum) and functions (in cases with a
continuous spectrum), Bessel- and Whittaker functions (in cases with a
continuous spectrum) and numerous polynomial solutions (in cases with
a discrete spectrum), and so on.

The only exception of parametric coordinate systems, where a developed
theory of the corresponding higher transcendental functions exists,
are the elliptic and spheroidal coordinate systems in flat space 
\cite{MESCH} and on spheres \cite{GKPSa}. The important point is that
this theory allows one to expand the exponentiated invariant distance
in terms of elliptic and
spheroidal wave-functions. These functions are one-parameter-generalized 
functions of the well-known spherical harmonics and Bessel
functions. It also allows one to formulate the ``Spheroidal Path
Integral'' which can be added to the list of Basic Path Integrals. 

Based on \cite{OLE}, the theory of \cite{MESCH}, and the thorough
study of coordinate systems in spaces of constant curvature \cite{KAL},
we were able to find the solution of the path integral formulations
in these spaces, expressed in its separating coordinate systems
\cite{GROab,GROad}. This study included the two- and three-dimensional
Euclidean and pseudo-Euclidean spaces, the two- and
three-dimensional spheres, the two- and three-dimensional
hyperboloids, imaginary Lobachevsky  space, respectively the
single-sheeted hyperbolic \cite{GROaa}, $\SU(u,v)$-path integration
\cite{GROl}, hyperbolic spaces of rank one \cite{GROn}, and 
Hermitian hyperbolic spaces \cite{GROn,GROar}.

Let us finally note that due to Kalnins, Miller, Winternitz, and 
coworkers there is in a series of papers an extensive study on
separation of variables of the Schr\"odinger-, respectively the
Helmholtz equation in spaces of constant curvature, 
called ''Lie Theory and Separation of Variables''. We just would
like to refer to Refs.~\cite{BKM,BKW,DORW,KAMI,MILLf} and the textbooks
\cite{GROad,KAL} with their extensive reference list on this subject.

\subsection{Introducing Darboux Spaces}
\message{Introducing Darboux Spaces}
An extension of the study of path integration on spaces of constant 
curvature is the investigation of path integral formulations in spaces 
of non-constant curvature. Kalnins et al. 
\cite{KalninsKMWinter,KalninsKWinter} denoted four types of 
two-dimensional spaces of non-constant curvature, labeled by
$\DI$--$\DIV$, which are called Darboux spaces \cite{KOENIGS}. 
In terms of the infinitesimal distance they are described~by:
\begin{eqnarray}
({\rm I})  \qquad  \d s^2&=& (x+y)\d x\d y\enspace,
\label{DarbouxI}
\\[2mm]
({\rm II}) \qquad \d s^2&=& \bigg(\frac{a}{(x-y)^2}+b\bigg) \d x\d y\enspace,
\label{DarbouxII}
\\[2mm]
({\rm III})\qquad \d s^2&=& \big(a\,\e^{-(x+y)/2}+b\,\e^{-x-y}\big)
\d x\d y\enspace,
\label{DarbouxIII}
\\[2mm]
({\rm IV}) \qquad \d s^2&=& -\frac{a\big(\e^{(x-y)/2}+\e^{(y-x)/2}\big)+b}
{\big(\e^{(x-y)/2}-\e^{(y-x)/2}\big)^2}\d x\d y\enspace.
\label{DarbouxIV}
\end{eqnarray}
$a$ and $b$ are additional (real) parameters.
Kalnins et al. \cite{KalninsKMWinter,KalninsKWinter} studied not only the
solution of the free motion, but also emphasized on the super-integrable
systems in theses spaces. Super-integrable means that in two dimensions
at least three constants of motion must exist, which is by construction
already fulfilled for the free motion. They found appropriate
coordinate systems, and we will consider all of them. In the majority
of the cases we will be able to find a solution, however in some cases
this will not be possible due to the quartic anharmonicity of the
problems in question.

\goodbreak
The Gaussian curvature in a space with metric
$\d s^2=g(u,v)(\d u^2+\d v^2)$ is given by ($g=\det g(u,v)$)
\begin{equation}
K=-\frac{1}{2g}\bigg(\frac{\partial^2}{\partial u^2}
+\frac{\partial^2}{\partial v^2}\bigg)\ln g\enspace.
 \label{Gaussian-curvature}
\end{equation}
Equation (\ref{Gaussian-curvature}) will be used to discuss shortly the
curvature properties of the Darboux spaces, including their limiting cases of
constant curvature.

In the following sections we discuss each of the four Darboux spaces,
we set up the Lagrangian, the Hamiltonian, the quantum operator, and 
formulate and solve (if this is possible) the corresponding path integral.
We also discuss some of the limiting cases of the Darboux-spaces,
i.e. where we obtain a space of constant (zero or negative) curvature.
In particular, for $\DII$ we consider the limiting case $b=0$ a little
bit more explicitly: it gives the two-dimensional hyperboloid (with
constant negative curvature). The other limiting case on $\DII$
($a=0$, i.e. $\bbbr^2$) is only sketched. In the case of $\DI$ there
is no limiting case, because we have no free parameter in the metric
to choose from.  In the two remaining Darboux-spaces, $\DIII$ and
$\DIV$, the limiting cases of $\DIII$ is not very difficult since it is
the zero-curvature case $\bbbr^2$ which emerges. In $\DIV$ we 
sketch the matter with some notes. 

In order to make the paper self-contained we provide in  the
Appendices some material about the Basic Path Integral techniques and
solutions. We set up the path integral formulation for general coordinates, 
including our lattice definition of the time-sliced path integral and 
shortly describe several transformation techniques, including coordinate 
transformation  and the space-time transformation. Also, we summarize 
some important path integral solutions, like the (radial) harmonic 
oscillator, the linear potential, and the modified P\"oschl--Teller
potential, including the corresponding Green's functions. These
solutions including their generalization to related potentials are
indispensable tools in the path integral investigation of Darboux spaces.


\setcounter{equation}{0}
\section{Darboux Space $\DI$}
\message{Darboux Space D_I}
We start with the consideration of the Darboux Space $\DI$ and
consider the following coordinate systems
\begin{eqnarray}
\hbox{($(u,v)$-Coordinates:)}&&
x=u+\i v\enspace,\quad y=u-\i v\enspace,\qquad\,\,
(u\geq a)\,,
\\
\hbox{(Rotated $(r,q)$-Coordinates:)}&&
u=r\cos\vtheta+q\sin\vtheta\enspace,\\  &&
v=-r\sin\vtheta +q\cos\vtheta,\qquad\qquad\,\,\,\, (\vtheta\in[0,\pi]),
\\
\hbox{(Displaced parabolic:)}&&
u=\half(\xi^2-\eta^2)+a\enspace,\,\,
v=\xi\eta\enspace,\quad (\xi\in\bbbr,\eta>0,a>0)\,.\qquad
\end{eqnarray}
The infinitesimal distance, i.e., the metric is given by
\begin{eqnarray}
\d s^2&=& (x+y)\d x\d y\enspace,\\
\hbox{($(u,v)$-Coordinates:)}
&=&2u(\d u^2+ \d v^2)\enspace,\\
\hbox{(Rotated $(r,q)$-Coordinates:)}
&=&2(r\cos\vtheta+q\sin\vtheta)(\d r^2+ \d q^2)\enspace,\\
\hbox{(Displaced parabolic:)}
&=&(\xi^2-\eta^2+2a)(\xi^2+\eta^2)(\d\xi^2+ \d\eta^2)\enspace.
\qquad\qquad\qquad\qquad
\end{eqnarray}
We find e.g. in the $(u,v)$-system for the Gaussian curvature
\begin{equation}
K=\frac{1}{u^4}\enspace.
\end{equation}
There is no further parameter in the metric, therefore this space is
  of non-constant curvature throughout for all $u>a$ with $a$ some
  real constant $a>0$\footnote{In \cite{KalninsKWinter} the condition
  $a\geq\half$ is 
  imposed in order to embed $\DI$ into a three-dimensional space with
  coordinates $X,Y,Z$ such that $\d X^2+\d Y^2+\d Z^2=2u(\d u^2+\d v^2)$.
  For $\d X^2+\d Y^2+\d Z^2$ we have $v\in[0,2\pi)$.
  For $\d X^2+\d Y^2-\d Z^2$ we have $v\in\bbbr$.}.

\subsection{The Path Integral in $(u,v)$-Coordinates on $\DI$}
\message{The Path Integral in (u,v)-Coordinates on D_I}
In order to set up the path integral formulation we follow our
canonical procedure as presented in \cite{GRSh}. The Lagrangian and
Hamiltonian are given by, respectively:
\begin{equation}
\CL(u,\dot u, v,\dot v)=mu(\dot u^2+\dot v^2), \quad
\CH(u,p_u,v,p_v)=\frac{1}{4mu}(p_u^2+p_v^2)\enspace,
\end{equation}
and we must require $u>a$ for some $a>0$, and $v\in[0,2\pi]$ can be
considered as a cyclic variable \cite{KalninsKWinter}.
The canonical momenta are
\begin{equation}
p_u=\hi\bigg(\frac{\partial}{\partial u}+\frac{1}{2u}\bigg),
\qquad 
p_v=\hi\frac{\partial}{\partial v}\enspace,
\end{equation}
and for the quantum Hamiltonian we find
\begin{eqnarray}
H=-\frac{\hbar^2}{2m}\frac{1}{2u}
\bigg(\frac{\partial^2}{\partial u^2}+\frac{\partial^2}{\partial
  v^2}\bigg)
=\frac{1}{2m}\frac{1}{\sqrt{2u}}(p_u^2+p_v^2)\frac{1}{\sqrt{2u}}\enspace.
\end{eqnarray}
We formulate the path integral (first ignoring the half-space
constraint):
\begin{eqnarray}
K(u'',u',v'',v';T)
&=&
\lim_{N\to\infty}\bigg(\frac{m}{2\pi\i\epsilon\hbar}\bigg)^N
\prod_{j=1}^{N-1}\int 2u_j\d u_j\d v_j
\exp\left[\frac{\i m}{\hbar}\sum_{j=1}^N
     \widehat{u_j}(\Delta^2u_j+\Delta^2v_j)\right]\quad
\\
&=&\pathint{u}\pathint{v}2u\exp\left[\frac{\i m}{\hbar}
    \int_0^Tu(\dot u^2+\dot v^2)\dt\right]
\end{eqnarray}
($\widehat{u_j}=\sqrt{u_ju_{j-1}}\,$). 
I have displayed the path integral in our lattice definition,
which will be used throughout this paper. Due to this lattice definition of
the path integral, we have no additional $\hbar^2$-potential because
the dimension of the space of non-constant curvature equals $2$,
c.f. (\ref{DeltaVPF}). In this path integral we perform a time
transformation according to $\Delta t_{(j)}=2\widehat{u_j}\Delta
s_{(j)}$, i.e. with time-transformation function $f(u)=2u=\sqrt{g}$
($g$ the determinant of the metric tensor), and we obtain: 
\begin{eqnarray}
&&K(u'',u',v'',v';T)
=\int_{-\infty}^\infty\frac{\d E}{2\pi\hbar}\,\e^{-\i ET/\hbar}
   \int_0^\infty \d s'' K(u'',u',v'',v';s'')
\\
&&\hbox{with $K(s'')$ given by:}\nonumber\\
&&K(u'',u',v'',v';s'')=
\pathints{u}\pathints{v}\exp\left\{\ihbar\int_0^{s''}\bigg[
\frac{m}{2}(\dot u^2+\dot v^2)+2uE\bigg]\d s\right\}
\nonumber\\
&&\qquad\qquad
=\sum_{l=-\infty}^\infty\frac{\e^{\i l(v''-v')}}{2\pi}
\exp\bigg(-\ihbar\frac{l^2\hbar^2}{2m}s''\bigg)
\pathints{u}\exp\left[\ihbar\int_0^{s''}\bigg(
\frac{m}{2}\dot u^2+2uE\bigg)\d s\right]\enspace.
\label{Kuvs}
\nonumber\\  &&
\end{eqnarray}
The feature that the time-transformation function equals $f=\sqrt{g}$
is a general feature of the Darboux-space path integration.
I have separated the $v$-dependent part of the path integral in
circular waves. The remaining path integral in the variable $u$ 
is a path integral for the linear potential. Let us denote the path
integrals in the variables $u$ and $v$ by $K_u(s'')$ and $K_v(s'')$,
respectively. We obtain for the product of the two kernels with
corresponding energy-dependent Green-functions $G_u(E;\CE_u)$ and 
$G_v(E;\CE_v)$: 
\begin{eqnarray}
&&K(u'',u',v'',v';T)
\nonumber\\  &&=\int_{-\infty}^\infty\frac{\d E}{2\pi\hbar}\,\e^{-\i ET/\hbar}
   \int_0^\infty \d s'' K_v(v'',v';s'')\cdot K_u(u'',u';s'')
\nonumber\\  &&
=\int_{-\infty}^\infty\frac{\d E}{2\pi\hbar}\,\e^{-\i ET/\hbar}\int_0^\infty \d s''
\nonumber\\  &&\qquad\times
\frac{1}{2\pi\i}\int\d\CE_v\,\e^{-\i \CE_vs''/\hbar}G_v(E;v'',v';\CE_v)\cdot
\frac{1}{2\pi\i}\int\d\CE_u\,\e^{-\i \CE_us''/\hbar}G_u(E;u'',u';\CE_u)\qquad
\nonumber\\  &&
=\int_{-\infty}^\infty\frac{\d E}{2\pi\hbar}\,\e^{-\i ET/\hbar}
\frac{\hbar}{2\pi\i}\int\d\CE G_v(E;v'',v';-\CE) G_u(E;u'',u';\CE)\enspace.
\end{eqnarray}
Alternatively, in terms of the Green's  function we get
\begin{equation}
G(u'',u',v'',v';E)
=\frac{\hbar}{2\pi\i}\int\d\CE G_v(E;v'',v';\CE) G_u(E;u'',u';-\CE)\enspace.
\end{equation}
Inserting now the explicit form of the $v$-dependent kernel we obtain:
\begin{equation}
K(u'',u',v'',v';T)
=\int_{-\infty}^\infty\frac{\d E}{2\pi\hbar}\,\e^{-\i ET/\hbar}
\sum_{l=-\infty}^\infty\frac{\e^{\i l(v''-v')}}{2\pi}
G_u\bigg(E;u'',u';-\frac{l^2\hbar^2}{2m}\bigg)\enspace.
\end{equation}
For the complete solution we must know the kernel $G_u(u'',u';\CE)$
which we obtain in the following way. The Green's  function for the
linear potential $V(x)=kx$ is given by \cite{GRSh} 
\begin{eqnarray}
G^{(k)}(x'',x';\CE)&=&\frac{4m}{3\hbar}
\bigg[\bigg(x'-\frac{\CE}{k}\bigg)\bigg(x''-\frac{\CE}{k}\bigg)\bigg]^{1/2}
\nonumber\\ &&\qquad\times
I_{1/3}\left[\frac{\sqrt{8mk}}{3\hbar}\bigg(x_<-\frac{\CE}{k}\bigg)^{3/2}\right]
K_{1/3}\left[\frac{\sqrt{8mk}}{3\hbar}\bigg(x_>-\frac{\CE}{k}\bigg)^{3/2}\right]
\enspace.\quad
\end{eqnarray}
$I_\nu$ and $K_\nu$ are modified Bessel-functions \cite{GRA}, and $x_<$ and 
$x_>$ denote the smaller and larger of $x'$ and $x''$, respectively. 
We have to identify $\CE=-L^2\hbar^2/2m$, $k=-2E$, and $x=u$.
In addition, we have to recall that the motion in $u$ takes place only
in the half-space $u>a$. In order to construct
the Green's  function in the half-space $x>a$ we have to put Dirichlet
boundary-conditions at $x=a$ \cite{GROr,GROs}. Therefore the Green
function for the linear potential in the half-space $x>a$ is given by:
\begin{equation}
G_{(x=a)}(u'',u';\CE)=G^{(k)}(u'',u';\CE)
-\frac{G^{(k)}(u'',a;\CE)G^{(k)}(a,u';\CE)}{G^{(k)}(a,a;\CE)}\enspace.
\label{Ghalfspace}
\end{equation}
Therefore we obtain finally:
\begin{eqnarray}
&&G(u'',u',v'',v';E)=
\sum_{l=-\infty}^\infty\frac{\e^{\i l(v''-v')}}{2\pi}
\frac{4m}{3\hbar}
\bigg[\bigg(u'-\frac{l^2\hbar^2}{4mE}\bigg)
\bigg(u''-\frac{l^2\hbar^2}{4mE}\bigg)\bigg]^{1/2}
\nonumber\\  &&\qquad\times
\left[\tilde I_{1/3}\bigg(u_<-\frac{l^2\hbar^2}{4mE}\bigg)
      \tilde K_{1/3}\bigg(u_>-\frac{l^2\hbar^2}{4mE}\bigg)
\vphantom{\dfrac{\tilde I_{1/3}\bigg(a-\frac{l^2\hbar^2}{4mE}\bigg)}
      {\tilde K_{1/3}\bigg(a-\frac{l^2\hbar^2}{4mE}\bigg) }}\right. 
\nonumber\\  &&\qquad\qquad\qquad\qquad\left.
-\dfrac{\tilde I_{1/3}\bigg(a-\frac{l^2\hbar^2}{4mE}\bigg)}
      {\tilde K_{1/3}\bigg(a-\frac{l^2\hbar^2}{4mE}\bigg) }
\tilde K_{1/3}\bigg(u'-\frac{l^2\hbar^2}{4mE}\bigg)
\tilde K_{1/3}\bigg(u''-\frac{l^2\hbar^2}{4mE}\bigg)\right]
\enspace.\qquad\qquad
\label{G-DarbouxI}
\end{eqnarray}
And $\tilde I_\nu(z)$ denotes
$$
\tilde I_\nu(z)=I_\nu\bigg(\frac{4\sqrt{-mE}}{3\hbar}z^{3/2}\bigg)\enspace,
$$
with $\tilde K_\nu(z)$ similarly. The Green's  function (\ref{G-DarbouxI})
is rather complicated due to the boundary condition at $u=a$, 
and we will not evaluate the free particle
wave-functions. Let us only consider the following point.
According to Langguth and Inomata \cite{LAI} the modified Bessel function 
$I_\nu(z)$ has the asymptotic expansion
\begin{equation}
I_\nu(z)\simeq \frac{1}{\sqrt{2\pi}} \e^{z-\nu^2/z}\qquad
\hbox{(for $|z|\to\infty$),}
\end{equation}
provided $\Re(z)>0$. However, if $-3\pi/2<\arg(z)<\pi/2$ one has
\begin{equation}
I_\nu(z)\simeq \frac{1}{\sqrt{2\pi}} \e^{z-\nu^2/z}
+\frac{1}{\sqrt{2\pi}} \e^{-z+\nu^2/z+\i\pi(\nu+1/2)}\qquad
\hbox{(for $|z|\to\infty$).}
\end{equation}
(A similar consideration is valid for $K_\nu$.)
In our Green's  function (\ref{G-DarbouxI}) we now see that due to the
$-$-sign in the square-root expression in the argument of the modified
Bessel functions, its argument becomes purely imaginary, from which
follows that for large $u$ we get ``plane'' waves $\propto\e^{\pm\i\kappa
  u^{3/2}}$ with some wave-number $\kappa$, including in- and out-coming waves.
This is the well-known feature of ``plane-waves'' in the free particle
motion, in the present case modified by the curvature of the space.

Let us note another particularity of the quantum motion in
$\DI$. According to \cite{KalninsKWinter} the coordinates $(u,v)$ can
only be uniquely determined if we embed the space $\DI$ into a
three-dimensional Euclidean space with definite or indefinite metric
respectively. In the present calculation we have chosen the definite
case. However, if we chose a metric according $d^2X+d^2Y-d^2T=2u(d^2u+d^2v)$, 
it is found that the variable $v$ can vary 
in its range over the entire real line, i.e.~$v\in\bbbr$. 
In the separation in the path integral (\ref{Kuvs}) the only
difference would be that the summation over the discrete quantum
number $l$ is replaced by an integration over the continuous quantum
number $k$, say, including the replacement $l\to k$ in all following
formulas from (\ref{Kuvs}) on. 

The problem of the exact range of the variables $(u,v)$ we will encounter 
several times, and for this reason we will leave this range unspecified. 
Implicitly we assume that when terms $\propto u^{-2}$ appear 
that $u$ is in the range $u>0$, i.e, a radial variable.
In the other coordinate systems like parabolic, spherical, elliptic,
etc, the usual range of variables is assumed. If however, a coordinate
is treated within the range of $\bbbr$, but is in fact restricted to
be positive or larger than a definite number, then (\ref{Ghalfspace}) 
must be applied to find the proper quantum solution.

\subsection{The Path Integral in Rotated $(r,q)$-Coordinates on $\DI$}
\message{The Path Integral in Rotated (u,v)-Coordinates on D_I}
In order to set up the path integral formulation we follow again our
canonical procedure. The Lagrangian and Hamiltonian are given by, respectively:
\begin{eqnarray}
\CL(r,\dot r, q,\dot q)
=m(r\cos\vtheta+q\sin\vtheta)(\dot r^2+\dot q^2)\enspace,
\\
\CH(r,p_r,q,p_q)
=\frac{1}{4m(r\cos\vtheta+q\sin\vtheta)}(p_r^2+p_q^2)\enspace.
\end{eqnarray}
The canonical momenta are
\begin{eqnarray}
p_r&=&\hi\bigg(\frac{\partial}{\partial r}
+\frac{\cos\vtheta}{2(r\cos\vtheta+q\sin\vtheta)}\bigg)\enspace,
\\
p_q&=&\hi\bigg(\frac{\partial}{\partial q}
+\frac{\sin\vtheta}{2(r\cos\vtheta+q\sin\vtheta)}\bigg)\enspace,
\end{eqnarray}
The quantum Hamiltonian has the form
\begin{eqnarray}
H&=&-\frac{\hbar^2}{2m}\frac{1}{2(r\cos\vtheta+q\sin\vtheta)}
\bigg(\frac{\partial^2}{\partial r^2}
 +\frac{\partial^2}{\partial q^2}\bigg)\enspace, 
\\
&=&\frac{1}{2m}\frac{1}{\sqrt{2(r\cos\vtheta+q\sin\vtheta)}}(p_r^2+p_q^2)
               \frac{1}{\sqrt{2(r\cos\vtheta+q\sin\vtheta)}}\enspace.
\end{eqnarray}
The path integral is formulated in the usual form:\footnote{We will
  assume $r,q\in\bbbr$, otherwise (\ref{Ghalfspace}) must be applied.}
\begin{eqnarray}
K(r'',r',q'',q';T)
&=&\pathint{r}\pathint{q} 2(r\cos\vtheta+q\sin\vtheta)
\nonumber\\  &&\qquad\times
\exp\left[\frac{im}{\hbar}\int_0^T
(r\cos\vtheta+q\sin\vtheta)(\dot r^2+\dot q^2)\dt\right]
\nonumber\\  
&=&\int\frac{\d E}{2\pi\hbar}\,\e^{-\i ET/\hbar}\int_0^\infty \d s''
   K(r'',r',q'',q';s'')
\\
\nonumber\\  
\nonumber\\  
K(r'',r',q'',q';s'')
&=&\pathints{r}
\exp\left[\ih\int_0^{s''}\bigg(\frac{m}{2}\dot r^2
+2Er\cos\vtheta\bigg)\d s\right]
\nonumber\\  &&\qquad\times
\pathints{q}
\exp\left[\ih\int_0^{s''}\bigg(\frac{m}{2}\dot q^2
+2Eq\sin\vtheta\bigg)\d s\right]\enspace.\qquad
\end{eqnarray}
I have time-transformed the path integral with the time-transformation 
function $f(r,q)=$ \linebreak $2(r\cos\vtheta+q\sin\vtheta)=\sqrt{g}$. 
Both path integrals are path integrals in $r$ and $q$, respectively,
for the linear potential. In order to find the Green's  function we set
the Green's  function in the variable $r$ as $G_r(\CE)$ and in the
variable $q$ as $G_q(\CE)$, respectively. We get similarly as before:
\begin{equation}
K(r'',r',q'',q';T)
=\int_{-\infty}^\infty\frac{\d E}{2\pi\hbar}\,\e^{-\i ET/\hbar}
\frac{\hbar}{2\pi\i}\int\d\CE G_r(r'',r';-\CE) G_q(q'',q';\CE)\enspace.
\end{equation}
Together with the solution of the linear potential, this gives the
solution
\begin{eqnarray}
&&K(r'',r',q'',q';T)=\int\d E\,\e^{-\i ET/\hbar}\int\d\CE
\nonumber\\  &&\qquad\times
\bigg(\frac{m}{\hbar^2}\sqrt{\frac{2}{E\cos\vtheta}}\,\bigg)^{2/3}
\nonumber\\  &&\qquad\times
\Ai\left[-\bigg(r''-\frac{\CE}{2E\cos\vtheta}\bigg)
\bigg(\frac{4mE\cos\vtheta}{\hbar^2}\bigg)^{1/3}\right]
\Ai\left[-\bigg(r'-\frac{\CE}{2E\cos\vtheta}\bigg)
\bigg(\frac{4mE\cos\vtheta}{\hbar^2}\bigg)^{1/3}\right]
\nonumber\\  &&\qquad\times
\bigg(\frac{m}{\hbar^2}\sqrt{\frac{2}{E\sin\vtheta}}\,\bigg)^{2/3}
\nonumber\\  &&\qquad\times
\Ai\left[-\bigg(q''+\frac{\CE}{2E\sin\vtheta}\bigg)
\bigg(\frac{4mE\sin\vtheta}{\hbar^2}\bigg)^{1/3}\right]
\Ai\left[-\bigg(q'+\frac{\CE}{2E\sin\vtheta}\bigg)
\bigg(\frac{4mE\sin\vtheta}{\hbar^2}\bigg)^{1/3}\right]\enspace.
\nonumber\\ 
\end{eqnarray}
and $\Ai(z)$ denotes the Airy function ($\xi=\frac{2}{3}z^{3/2}$):
\begin{eqnarray}
 \Ai(z)&=&\frac{1}{3}\sqrt{z}\,
\Big[I_{-1/3}(\xi)-I_{1/3}(\xi)\Big]=\frac{1}{\pi}\sqrt{\frac{z}{3}}\,
K_{1/3}(\xi)\enspace,
\\
\Ai(-z)&=&\frac{1}{3}\sqrt{z}\,\Big[J_{-1/3}(\xi)-J_{1/3}(\xi)\Big]\enspace.
\end{eqnarray}

\subsection{The Path Integral in Displaced Parabolic Coordinates on $\DI$}
\message{The Path Integral in Parabolic Coordinates on D_I}
In parabolic coordinates the Lagrangian and the Hamiltonian are given by:
\begin{eqnarray}
\CL(\xi,\dot\xi,\eta,\dot\eta)&=&
\frac{m}{2}(\xi^2-\eta^2+2a)(\xi^2+\eta^2)
(\dot\xi^2+\dot\eta^2)\enspace,
\\
\CH(\xi,p_\xi,\eta,p_\eta)&=&
\frac{1}{2m}\frac{p_\xi^2+p_\eta^2}{(\xi^2-\eta^2+2a)(\xi^2+\eta^2)}\enspace.
\end{eqnarray}
The canonical momenta are
\begin{eqnarray}
p_\xi&=&\hi\bigg(\frac{\partial}{\partial\xi}
+\frac{\xi}{\xi^2-\eta^2+2a}+\frac{\xi}{\xi^2+\eta^2}\bigg)
\\
p_\eta&=&\hi\bigg(\frac{\partial}{\partial\eta}
-\frac{\eta}{\xi^2-\eta^2+2a}+\frac{\eta}{\xi^2+\eta^2}\bigg)
\end{eqnarray}
The quantum Hamiltonian has the form
\begin{eqnarray}
H&=&-\frac{\hbar^2}{2m}\frac{1}{(\xi^2-\eta^2+2a)(\xi^2+\eta^2)}
\bigg(\frac{\partial^2}{\partial\xi^2}
 +\frac{\partial^2}{\partial\eta^2}\bigg)\enspace,
\\
&=&\frac{1}{2m} \frac{1}{\sqrt{(\xi^2-\eta^2+2a)(\xi^2+\eta^2)}}
(p_\xi^2+p_\eta^2)\frac{1}{\sqrt{(\xi^2-\eta^2+2a)(\xi^2+\eta^2)}}\enspace.
\end{eqnarray}
The path integral formulation is as follows (with the
implemented time-transformation function $f(\xi,\eta)=
(\xi^2-\eta^2+2a)(\xi^2+\eta^2)=\sqrt{g}\,$):
\begin{eqnarray}
&&K(\xi'',\xi',\eta'',\eta';T)=
\pathint{\xi}\pathint{\eta}(\xi^2-\eta^2+2a)(\xi^2+\eta^2)
\nonumber\\  &&\qquad\qquad\qquad\qquad\times
\exp\left[\frac{\i m}{2\hbar}\int_0^T(\xi^2-\eta^2+2a)(\xi^2+\eta^2)
(\dot\xi^2+\dot\eta^2)\dt\right]
\nonumber\\
&&=\int\frac{\d E}{2\pi\hbar}\,\e^{-\i ET/\hbar}\int_0^{s''}\d s''
\pathints{\xi}\pathints{\eta}
\nonumber\\  &&\qquad\quad\qquad\qquad\times
\exp\left\{\ih\int_0^{s''}\bigg[\frac{m}{2}(\dot\xi^2+\dot\eta^2)
+E\Big(\xi^4-\eta^4+2a(\xi^2+\eta^2)\Big)\bigg]\d s\right\}\enspace.\qquad
\end{eqnarray}
This is a path integral of a quartic anharmonic oscillator which
cannot be solved.


\setcounter{equation}{0}
\section{Darboux Space $\DII$}
\message{Darboux Space D_II}
In this section we consider the Darboux Space $\DII$ (\ref{DarbouxII}).
We have the following four coordinate systems
\begin{eqnarray}
\hbox{($(u,v)$-Coordinates:)}&& x=\half(v+\i u),\quad y=\half(v-\i u)\,,
\\
\hbox{(Polar:)}&& u=\vrho\cos\vtheta,\quad v=\vrho\sin\vtheta\enspace,\quad
\qquad\qquad\quad\,\,\,\,(\vrho>0,
\vtheta\in(-\hbox{$\frac{\pi}{2}$},\hbox{$\frac{\pi}{2}$}))\enspace,
\\
\hbox{(Parabolic:)}&& u=\xi\eta,\quad v=\half(\xi^2-\eta^2)\enspace,\quad 
\qquad\qquad\quad\,\,\,(\xi>0,\eta>0)\,,
\\
\hbox{(Elliptic:)}&& u=d\cosh\omega\cos\vphi,\quad 
                     v=d\sinh\omega\sin\vphi\enspace,\quad
(\omega>0,\vphi\in(-\hbox{$\frac{\pi}{2}$},\hbox{$\frac{\pi}{2}$}))\,.
\qquad
\end{eqnarray}
$2d$ is the interfocal distance in the elliptic system. Separation of 
variables is possible in all four coordinate systems. For convenience 
we also display in the following the special case of the parameters
$a=-1$ and $b=1$ \cite{KalninsKMWinter}. The infinitesimal distance
is given in these four cases:
\begin{eqnarray}
\d s^2&=& \bigg(\frac{a}{(x-y)^2}+b\bigg) \d x\d y
\nonumber\\
\hbox{($(u,v)$-Coordinates:)}     
 &=&\frac{bu^2-a}{u^2}(\d u^2+\d v^2)
      = \frac{u^2+1}{u^2}(\d u^2+\d v^2)\enspace,
\\
\hbox{(Polar:)}
&=&\frac{b\vrho^2\cos^2\vtheta-a}{\vrho^2\cos^2\vtheta}
(\d\vrho^2+\vrho^2\d\vtheta^2)
=\frac{\vrho^2\cos^2\vtheta+1}{\vrho^2\cos^2\vtheta}
(\d\vrho^2+\vrho^2\d\vtheta^2)\enspace,
\\
\hbox{(Parabolic:)}
&=&\frac{b\xi^2\eta^2-a}{\xi^2\eta^2}(\xi^2+\eta^2)(\d\xi^2+\d\eta^2)
\nonumber\\
&=&\bigg[\bigg(b\xi^2-\frac{a}{\xi^2}\bigg)
+\bigg(b\eta^2-\frac{a}{\eta^2}\bigg)\bigg](\d\xi^2+\d\eta^2)\enspace, 
\\
&=&\frac{1+\xi^2\eta^2}{\xi^2\eta^2}(\xi^2+\eta^2)(\d\xi^2+\d\eta^2)\enspace, 
\\
\hbox{(Elliptic:)}
&=&\frac{bd^2\cosh^2\omega\cos^2\vphi-a}{\cosh^2\omega\cos^2\vphi}
        (\cosh^2\omega-\cos^2\vphi)(\d\omega^2+\d\vphi^2)\enspace,
\nonumber\\
&=&\bigg[\bigg(bd^2\cosh^2\omega+\frac{a}{\cosh^2\omega}\bigg)\!-\!
\bigg(bd^2\cos^2\vphi+\frac{a}{\cos^2\vphi}\bigg)\bigg]
(\d\omega^2+\d\vphi^2)
\,.\qquad
\end{eqnarray}
We can see that the case $a=-1$, $b=0$ leads to the case
of the Poincar\'e upper half-plane \cite{GROf,GROab,GROad}, i.e. the
two-dimensional hyperboloid $\Lambda^{(2)}$. 
In this case separation of variables is possible in nine coordinate systems
\cite{OLE}; this has been extensively discussed in \cite{GROad,GROPOc}.
The parabolic case corresponds to the semi-circular-parabolic system
and the elliptic case to the elliptic-parabolic system on the two-dimensional
hyperboloid. On the other hand, the case $a=0$, $b=1$ just gives the usual
two-dimensional Euclidean plane with its four coordinate system which allow
separation of variables of the Laplace-Beltrami equation, i.e., the
Cartesian, polar, parabolic, and elliptic system. Hence, the Darboux space II
contains as special cases a space of constant zero curvature (Euclidean
plane) and a space of constant negative curvature (the hyperbolic plane).
A discussion of a more general case on the question of contracting Lie 
algebras corresponding to coordinate systems in spaces of constant curvature
can be found in \cite{IPSWa}. This includes the emerging of
coordinate systems in flat space from curved spaces.

\begin{table}[h]
\caption{\label{cosytab1} Limiting Cases of Coordinate Systems on $\DII$}
\begin{eqnarray}\begin{array}{l}\vbox{\small\offinterlineskip
\halign{&\vrule#&$\strut\ \hfil\hbox{#}\hfill\ $\cr
\noalign{\hrule}
height2pt&\omit&&\omit&&\omit&&\omit&\cr
&Metric:&&$\DII$      &&$\Lambda^{(2)} ($a=-1,b=0$)$ && $E_2$ ($a=0,b=1$)&\cr
height2pt&\omit&&\omit&&\omit&&\omit&\cr
\noalign{\hrule}\noalign{\hrule}
height2pt&\omit&&\omit&&\omit&&\omit&\cr
&$\dfrac{bu^2-a}{u^2}(\d u^2+\d v^2)$  
        &&$(u,v)$-System     &&Horicyclic    &&Cartesian     &\cr
height2pt&\omit&&\omit&&\omit&&\omit&\cr
\noalign{\hrule}
height2pt&\omit&&\omit&&\omit&&\omit&\cr
&$\dfrac{b\vrho^2\cos^2\vtheta-a}{\vrho^2\cos^2\vtheta}(\d\vrho^2+\d\vtheta^2)$
        &&Polar       &&Equidistant   &&Polar         &\cr
height2pt&\omit&&\omit&&\omit&&\omit&\cr
\noalign{\hrule}
height2pt&\omit&&\omit&&\omit&&\omit&\cr
&$\dfrac{b\xi^2\eta^2-a}{\xi^2\eta^2}(\xi^2+\eta^2)(\d\xi^2+\d\eta^2)$
        &&Parabolic   &&Semi-circular parabolic    
                                      &&Parabolic     &\cr
height2pt&\omit&&\omit&&\omit&&\omit&\cr
\noalign{\hrule}
height2pt&\omit&&\omit&&\omit&&\omit&\cr
&$\dfrac{bd^2\cosh^2\omega\cos^2\vphi-a}{\cosh^2\omega\cos^2\vphi}$
                                             && && && &\cr
&$\times (\cosh^2\omega-\cos^2\vphi)(\d\omega^2+\d^2\vphi^2)$
        &&Elliptic   &&Elliptic-parabolic    
                                      &&Elliptic     &\cr
height2pt&\omit&&\omit&&\omit&&\omit&\cr
\noalign{\hrule}}}\end{array}\nonumber\end{eqnarray}
\end{table}
\noindent
We find for the Gaussian curvature in the $(u,v)$-system
\begin{equation}
K=\frac{a(a-3bu^2)}{(a-2bu^2)^3}\enspace.
\end{equation}
For $b=0$ we find $K=1/a$ which is indeed a space of constant curvature, 
and the quantity $a$ measures the curvature. In particular, for the 
unit-two-dimensional hyperboloid we have $K=1/a$, with $a=-1$ as the
special case of $\Lambda^{(2)}$. In the following we will assume that
$a<0$ in order to assure the positive definiteness of the metric (1.2).

\subsection{The Path Integral in $(u,v)$-Coordinates on $\DII$}
\message{The Path Integral in (u,v)-Coordinates on D_II}
We start with the  $(u,v)$-coordinate system. We formulate the
classical Lagrangian and Hamiltonian, respectively:
\begin{eqnarray}
\CL(u,\dot u,v,\dot v)&=&\frac{m}{2}\frac{bu^2-a}{u^2}
                         (\dot u^2+\dot v^2)\enspace,
\\
\CH(u,p_u,v,p_v)&=& \frac{1}{2m}\frac{u^2}{bu^2-a}(p_u^2+p_v^2)\enspace.
\end{eqnarray}
The canonical momenta are
\begin{equation}
p_u=\hi\bigg(\frac{\partial}{\partial u}
+\frac{bu}{bu^2-a}-\frac{1}{u}\bigg),\quad
p_v=\hi\frac{\partial}{\partial v}\enspace.
\end{equation}
The quantum Hamiltonian has the form
\begin{eqnarray}
H&=&-\frac{\hbar^2}{2m}\frac{u^2}{bu^2-a}
\bigg(\frac{\partial^2}{\partial u^2}+\frac{\partial^2}{\partial v^2}\bigg)
\\
&=&\frac{1}{2m}\frac{u}{\sqrt{bu^2-a}}(p_u^2+p_v^2)\frac{u}{\sqrt{bu^2-a}}
\enspace.
\end{eqnarray}
We write down path integral, perform a time-transformation with
$f(u)=(bu^2-a)/u^2=\sqrt{g}$ and insert the path integral solution for the
free particle in the variable $v$ and the radial harmonic oscillator 
\cite{GRSh,PI} in the variable $u$ ($\lambda^2-1/4=2maE/\hbar^2$):
\begin{eqnarray}
&&K(u'',u',v'',v';T)=\pathint{u}\pathint{v}\frac{bu^2-a}{u^2}
\exp\left[\frac{\i m}{2\hbar}
     \int_0^T\frac{bu^2-a}{u^2}(\dot u^2+\dot v^2)\dt\right]
\nonumber\\  &&
=\int_{-\infty}^\infty\frac{\d E}{2\pi\hbar}\,\e^{-\i ET/\hbar}
   \int_0^\infty \d s''
\nonumber\\  &&\qquad\times
   \pathints{u}\pathints{v}
   \exp\left\{\ih\int_0^{s''}\bigg[
   \frac{m}{2}(\dot u^2+\dot v^2)-\frac{aE}{u^2}\bigg]\d s
    +\ih bEs''\right\}
\nonumber\\  &&
=\int_{-\infty}^\infty\frac{\d E}{2\pi\hbar}\,\e^{-\i ET/\hbar}
\int_0^\infty\frac{ds''}{2\pi}\int_{-\infty}^\infty\d k\,\e^{\i k(v''-v')}
\exp\bigg(\ih bEs''-\ih\frac{\hbar^2k^2}{2m}s''\bigg)
\nonumber\\ &&\qquad\times
\pathints{u}\exp\left[\ih\int_0^{s''}\bigg(\frac{m}{2}\dot u^2
 -\hbar^2\frac{\lambda^2-\viert}{2mu^2}\bigg)\d s\right]
\nonumber\\  &&
=\int_{-\infty}^\infty\frac{\d E}{2\pi\hbar}\,\e^{-\i ET/\hbar}
\int_0^\infty ds''\int_{-\infty}^\infty\d k\,\e^{\i k(v''-v')}
\frac{m\sqrt{u'u''}}{\i\hbar s''}
\nonumber\\ &&\qquad\times
\exp\left[\ih\bigg(bE-\frac{\hbar^2k^2}{2m}\bigg)s''
  +\ih\frac{m}{2s''}({u'}^2+{u''}^2)\right]
  I_\lambda\bigg(\frac{mu'u''}{\i\hbar s''}\bigg)\enspace.
\label{KLambda}
\end{eqnarray}
We can see that the case $a=0$ yields the solution of the free
particle in $\bbbr^2$ since $\lambda=\pm\half$, and a proper
combination of $I_{\pm\half}$ give exponentials.
Together with the integral \cite[p.719]{GRA}
\begin{equation}
\int_0^\infty\e^{-a/x-bx}J_\nu(cx)\frac{\d x}{x}
=2J_\nu\bigg[\sqrt{2a(\sqrt{b^2+d^2}-b)}\,\bigg]
  K_\nu\bigg[\sqrt{2a(\sqrt{b^2+d^2}+b)}\,\bigg]
\end{equation}
we obtain for the Green's  function ($\lambda$ resolved and taken on the
cut, $a<0$)
\begin{eqnarray}
&&G(u'',u',v'',v';E)
\nonumber\\  &&
=\frac{2m\sqrt{u'u''}}{\i\hbar}\int_{-\infty}^\infty\d k\,\e^{\i k(v''-v')}
I_\lambda\left(\sqrt{k^2-\frac{2mbE}{\hbar^2}}\,u_<\right)
K_\lambda\left(\sqrt{k^2-\frac{2mbE}{\hbar^2}}\,u_>\right)
\\  &&=
\frac{\hbar}{\pi^3}\int_{-\infty}^\infty\d k\,\e^{\i k(v''-v')}
\nonumber\\  &&\qquad\times
\int_0^\infty\frac{p\sinh\pi p\d p}{\frac{\hbar^2}{2m|a|}(p^2+\viert)-E}
K_{\i p}\left(\sqrt{k^2-\frac{2mbE}{\hbar^2}}\,u'\right)
K_{\i p}\left(\sqrt{k^2-\frac{2mbE}{\hbar^2}}\,u''\right)\enspace,\qquad
\label{GDarbouxIIuv}
\end{eqnarray}
with 
\begin{equation}
\lambda=\sqrt{\viert-\frac{2m|a|E}{\hbar^2}}\equiv \i p\enspace.
\end{equation}
The wave functions and the energy spectrum are read off:
\begin{eqnarray}
\Psi(u,v)&=&\frac{\e^{\i k v}}{\sqrt{2\pi}}
\cdot\frac{\sqrt{2p\sinh\pi p}}{\pi}
K_{\i p}\left(\sqrt{k^2-\frac{2mbE}{\hbar^2}}\,u\right)\enspace,
\\
E&=&\frac{\hbar^2}{2m|a|}\bigg(p^2+\frac14\bigg)\enspace.
\end{eqnarray}
Here I have used the following identity as used in \cite{GRSa}
utilizing \cite[p.194]{MOS} and \cite[p.819, p.732]{GRA}, respectively
\begin{eqnarray}
I_\lambda(ax)K_\lambda(bx)&=&
\frac{1}{\pi\sqrt{ab}}\int_0^\infty\dt\,
 \CQ_{\lambda-1/2}\bigg(\frac{a^2+b^2+t^2}{2ab}\bigg)\cos xt
\nonumber\\  &=&
\frac{1}{\pi\sqrt{ab}}\int_0^\infty\dt\cos xt
\int_0^\infty \frac{p\tanh\pi p\,\d p}{\lambda^2+p^2}
\CP_{\i p-1/2}\bigg(\frac{a^2+b^2+t^2}{2ab}\bigg)
\nonumber\\  &=&
\frac{2}{\pi^2}\int_0^\infty \frac{p\tanh\pi p\,\d p}{\lambda^2+p^2}
K_{\i p}(ax)K_{\i p}(bx)\enspace.
\end{eqnarray}
$\CP_\mu,\CQ_\mu$ are Legendre-functions of the first and second kind,
respectively \cite[p.999]{GRA}. 
The case $a=-1$, $b=1$ gives the case of \cite{KalninsKMWinter}, i.e. the
Green's  function:
\begin{eqnarray}
&&G(u'',u',v'',v';E)=
\int_{-\infty}^\infty\d k\,\e^{\i k(v''-v')}
\nonumber\\  &&\qquad\times
\int_0^\infty\frac{\d p}{\pi^2}
\frac{2p\sinh\pi p}{\frac{\hbar^2}{2m}(p^2+\viert)-E}
K_{\i p}\left(\sqrt{k^2-(p^2+\bviert)}\,u''\right)
K_{\i p}\left(\sqrt{k^2-(p^2+\bviert)}\,u'\right)\enspace.\qquad
\end{eqnarray}
And the wave functions read
\begin{equation}
\Psi(u,v)=\frac{\e^{\i k v}}{\sqrt{2\pi}}
\cdot\frac{\sqrt{2p\sinh\pi p}}{\pi}
K_{\i p}\left(\sqrt{k^2-(p^2+\bviert)}\,u\right)\enspace.
\end{equation}
If we take in (\ref{GDarbouxIIuv}) $a=-1$, $b=0$ we obtain the
solution for the Poincar\'e upper half-plane \cite{GRSa}. 
We can evaluate (\ref{KLambda}) by means of \cite[p.719]{GRA} yielding
\begin{equation}
G(u'',u',v'',v';E)={m\over\sqrt{2\pi}}
  \int_0^\infty e^{-z\cosh d}I_{\i p}(z){dz\over\sqrt{z}}
  ={m\over\pi}\CQ_{-{1\over2}-\i p}(\cosh d)\enspace,
\end{equation}
where 
\begin{equation}
\cosh d=\frac{(v''-v')^2+{u'}^2+{u''}^2}{2u'u''}\enspace,
\end{equation}
which is the Poincar\'e distance on the hyperboloid.
For $b\not=0$ such an expression cannot be found.

\subsection{The Path Integral in Polar Coordinates on $\DII$}
\message{The Path Integral in Polar Coordinates on D_II}
In polar coordinates the classical Lagrangian and Hamiltonian are given by
\begin{eqnarray}
\CL(r,\dot r,\vtheta,\dot\vtheta)&=&\frac{m}{2}
  \bigg(b-\frac{a}{\vrho^2\cos^2\vtheta}\bigg)
(\dot \vrho^2+\vrho^2\dot\vtheta^2)\enspace,
\\
\CH(\vrho,p_\vrho,\vtheta,p_\vtheta)&=&\frac{1}{2m}
\bigg(b-\frac{a}{\vrho^2\cos^2\vtheta}\bigg)^{-1}
\bigg(p_\vrho^2+\frac{1}{\vrho^2}p_\vtheta^2\bigg)\enspace.
\end{eqnarray}
The momentum operators are
\begin{eqnarray}
p_\vrho&=&\hi\bigg[\frac{\partial}{\partial\vrho}
+\bigg(\frac{b\vrho\cos^2\vtheta}{b\cos^2\vtheta\vrho^2-a}-\frac{1}{2\vrho}
\bigg)\bigg]\enspace,
\\
p_\vtheta&=&\hi\bigg[\frac{\partial}{\partial\vtheta}+\bigg(\tan\vtheta
-\frac{b\vrho^2\sin\vtheta\cos\vtheta}{b\vrho^2\cos^2\vtheta-a}\bigg)\bigg]
\enspace,
\end{eqnarray}
and the quantum Hamiltonian is given by:
\begin{eqnarray}
H&=&-\frac{\hbar^2}{2m}\bigg(b-\frac{a}{\vrho^2\cos^2\vtheta}\bigg)^{-1}
 \bigg(\frac{\partial^2}{\partial \vrho^2}
  +\frac{1}{\vrho}\frac{\partial}{\partial\vrho}
  +\frac{1}{\vrho^2}\frac{\partial^2}{\partial\vtheta^2}\bigg)
\\
&=&\frac{1}{2m}\bigg(b-\frac{a}{\vrho^2\cos^2\vtheta}\bigg)^{-1/2}
\bigg(p_\vrho^2+\frac{1}{\vrho^2}p_\vtheta^2\bigg)
\bigg(b-\frac{a}{\vrho^2\cos^2\vtheta}\bigg)^{-1/2}
-\bigg(b-\frac{a}{\vrho^2\cos^2\vtheta}\bigg)^{-1}
\frac{\hbar^2}{8m\vrho^2}\enspace.
\nonumber\\
\end{eqnarray}
Hence, we get for the path integral 
\begin{eqnarray}
&&\!\!\!\!\!\!\!\!\!\!\!\!
K(\vrho'',\vrho',\vtheta'',\vtheta';T)=
\pathint{\vrho}\pathint{\vtheta}\vrho\bigg(b-\frac{a}{\vrho^2\cos^2\vtheta}\bigg)
\nonumber\\  &&\!\!\!\!\!\!\!\!\!\!\!\!\qquad\times
\exp\left\{\ih\int_0^T\left[
\frac{m}{2}\bigg(b-\frac{a}{\vrho^2\cos^2\vtheta}\bigg)
(\dot\vrho^2+\vrho^2\dot\vtheta^2) 
+\bigg(b-\frac{a}{\vrho^2\cos^2\vtheta}\bigg)^{-1}\frac{\hbar^2}{8m\vrho^2}
\right]\dt\right\}\,.
\end{eqnarray}
However, this coordinate representation is not very well suited for
our purposes, except that we recover for $a=0$ polar coordinate in $\bbbr^2$. 
We introduce $\vrho=\e^{\tau_2}$ and $\cos\vtheta=1/\cosh\tau_1$.
This gives the transformed path integral
\begin{eqnarray}
&&\!\!\!\!\!\!\!\!\!\!\!\!
\pathint{\tau_1}\pathint{\tau_2}\cosh\tau_1
\bigg(\frac{b\,\e^{2\tau_2}}{\cosh^2\tau_1}-a\bigg)
\nonumber\\  &&\!\!\!\!\!\!\!\!\!\!\!\!\quad \times
\exp\left\{\ih\int_0^T\left[
\frac{m}{2}\bigg(\frac{b\,\e^{2\tau_2}}{\cosh^2\tau_1}-a\bigg)
(\dot\tau_1^2+\cosh^2\tau_1\dot\tau_2^2)
-\bigg(\frac{b\,\e^{2\tau_2}}{\cosh^2\tau_1}-a\bigg)^{-1}\frac{\hbar^2}{8m}
\bigg(1+\frac{1}{\cosh^2\tau_1}\bigg)\right]\dt\right\}
\nonumber\\  &&\!\!\!\!\!\!\!\!\!\!\!\!=
\int_{-\infty}^\infty\frac{\d E}{2\pi\hbar}\,\e^{-\i ET/\hbar}
\int_0^\infty \d s'' K(\tau_1'',\tau_1',\tau_2'',\tau_2';s'')
\\
&&\!\!\!\!\!\!\!\!\!\!\!\!\hbox{with the time-transformed path integral 
$K(s'')$ given by  $\Big(\lambda=\sqrt{\viert-2m|a|E/\hbar^2}$, $a<0\Big)$:}
\nonumber\\
&&\!\!\!\!\!\!\!\! \!\!\!\!
K(\tau_1'',\tau_1',\tau_2'',\tau_2';s'')
=\pathints{\tau_1}\pathints{\tau_2}\cosh\tau_1
\nonumber\\  &&\!\!\!\!\!\!\!\!\!\!\!\!\qquad\times
\exp\left\{\ih\int_0^{s''}\left[
\frac{m}{2}(\dot\tau_1^2+\cosh^2\tau_1\dot\tau_2^2)
+Eb\frac{\e^{2\tau_2}}{\cosh^2\tau_1}-aE
-\frac{\hbar^2}{8m}\bigg(1+\frac{1}{\cosh^2\tau_1}\bigg)\right]
\dt\right\}\enspace.
\end{eqnarray}
The path integral in $\tau_2$ has now the form of the path integral for
Liouville quantum mechanics \cite{GRSa,GRSh} yielding 
\begin{eqnarray}
&&K(\tau_1'',\tau_1',\tau_2'',\tau_2';s'')
=\sqrt{\cosh\tau_1'\cosh\tau_1'}
\exp\left[\ih \bigg(|a|E-\frac{\hbar^2}{8m}\bigg)s''\right]
\nonumber\\  &&\qquad\times
\frac{2}{\pi^2}\int_0^\infty\d k\,k\sinh\pi k 
K_{\i k}\Bigg(\frac{\sqrt{-2mbE}}{\hbar}\,\e^{\tau_2'}\Bigg)
K_{\i k}\Bigg(\frac{\sqrt{-2mbE}}{\hbar}\,\e^{\tau_2''}\Bigg)
\nonumber\\  &&\qquad\times
\pathints{\tau_1}\exp\left[\ih\int_0^{s''}\bigg(
\frac{m}{2}\dot\tau_1^2-\frac{\hbar^2}{2m}\frac{k^2+\viert}{\cosh^2\tau_1}
\bigg)\d s\right]
\nonumber\\  &&
=\sqrt{\cosh\tau_1'\cosh\tau_1'}
\exp\left[\ih \bigg(|a|E-\frac{\hbar^2}{8m}\bigg)s''\right]
\nonumber\\  &&\qquad\times
\frac{2}{\pi^2}\int_0^\infty\d k\,k\sinh\pi k 
K_{\i k}\Bigg(\frac{\sqrt{-2mbE}}{\hbar}\,\e^{\tau_2'}\Bigg)
K_{\i k}\Bigg(\frac{\sqrt{-2mbE}}{\hbar}\,\e^{\tau_2''}\Bigg)
\nonumber\\  &&\qquad\times
\half\sum_{\pm}\int_0^\infty\frac{p\sinh\pi p\d p}{\cosh^2k+\sinh^2\pi p}
\e^{-\i p^2\hbar s''/2m}
P_{\i k-1/2}^{\i p}(\pm\tanh\tau_1')
P_{\i k-1/2}^{\i p}(\pm\tanh\tau_1'')\enspace,\qquad\qquad
\end{eqnarray}
where we have inserted the path integral solution for the special case of
the modified P\"oschl--Teller potential~\cite{GROad}.
Performing the $s''$-integration gives the energy-spectrum:
$$E=\frac{\hbar^2}{2m|a|}\bigg(p^2+\viert\bigg)$$
with the Green's function
\begin{equation}
G(\tau_1'',\tau_1',\tau_2'',\tau_2';E)
=\sum_{\pm}\int_0^\infty \d p\int_0^\infty \d k
\frac{\Psi_{p,k,\pm}(\tau_1'',\tau_2'')\Psi_{p,k,\pm}^*(\tau_1',\tau_2')}
    {\frac{\hbar^2}{2m|a|}\big(p^2+\viert\big)-E}\enspace,
\end{equation}
and the wave-functions are given by
\begin{eqnarray}
&&\!\!\!\!\!\!\!\!\!\!\!\!\!\!\!\!\!\!\!\!\!\!\!\!\!\!
\Psi_{p,k,\pm}(\tau_1,\tau_2)
\nonumber\\  &&\!\!\!\!\!\!\!\!\!\!\!\!\!\!\!\!\!\!\!\!\!\!\!\!\!\!=
\frac{\sqrt{2\cosh\tau_1}}{\pi}
\sqrt{k\sinh\pi k} \,
K_{\i k}\Bigg(\i\sqrt{\frac{b}{|a|}\bigg(p^2+\viert\bigg)}\,\e^{\tau_2}\Bigg)
\sqrt{\frac{p\sinh\pi p\d p}{\cosh^2k+\sinh^2\pi p}}
P_{\i k-1/2}^{\i p}(\pm\tanh\tau_1)
         \\  &&\!\!\!\!\!\!\!\!\!\!\!\!\!\!\!\!\!\!\!\!\!\!\!\!\!\!
\Psi_{p,k,\pm}(\vrho,\vtheta)
\nonumber\\  &&\!\!\!\!\!\!\!\!\!\!\!\!\!\!\!\!\!\!\!\!\!\!\!\!\!\!=
\frac{\sqrt{2\sin\vtheta}}{\pi}
\sqrt{k\sinh\pi k} \,
K_{\i k}\Bigg(\i\sqrt{\frac{b}{|a|}\bigg(p^2+\viert\bigg)}\,\vrho\Bigg)
\sqrt{\frac{p\sinh\pi p\d p}{\cosh^2k+\sinh^2\pi p}}
P_{\i k-1/2}^{\i p}(\pm\sin\vtheta).
\end{eqnarray}
Here, we have inserted the original ($\vrho,\vtheta$)-coordinate system.
For $b=0$ we obtain the equidistant coordinate system on the
two-dimensional hyperboloid. For the coordinate $\tau_1$, this is obvious.
For the coordinate $\tau_2$ we observe that the $K_\nu$-Bessel
function can be represented in this limit as \cite[p.1063]{GRA}
\begin{equation}
K_\nu(x)=\sqrt{\pi}\,\e^{-x}(2x)^\nu\psi(\bhalf+\nu,1+2\nu,x)
\propto
\frac{\Gamma(-2\i k)}{\Gamma(\half-\i k)}\sqrt{\pi}\,\e^{\i k\tau_2}
\to
\sqrt{\frac{\pi}{2k\sinh\pi k}}\,\e^{\i k\tau_2}\enspace,
\end{equation}
which gives together with the normalization factors the final result
of \cite{GROad}. 

\subsection{The Path Integral in Parabolic Coordinates on $\DII$}
\message{The Path Integral in Parabolic Coordinates on D_II}
The classical Lagrangian and Hamiltonian are given by
\begin{eqnarray}
\CL(\xi,\dot\xi,\eta,\dot\eta)&=&\frac{m}{2}\frac{b\xi^2\eta^2-a}{\xi^2\eta^2}
(\xi^2+\eta^2)(\dot\xi^2+\dot\eta^2)\enspace,
\\
\CH(\xi,p_\xi,\eta,p_\eta)&=&\frac{1}{2m}\frac{\xi^2\eta^2}{b\xi^2\eta^2-a}
\frac{p_\xi^2+p_\eta^2}{\xi^2+\eta^2}\enspace.
\end{eqnarray}
The canonical momenta are given by
\begin{eqnarray}
p_\xi&=&\hi\bigg(\frac{\partial}{\partial\xi}
+\frac{b\xi+a/\xi^3}{\sqrt{g}}\bigg)\enspace,
\\
p_\eta&=&\hi\bigg(\frac{\partial}{\partial\eta}
+\frac{b\eta+a/\eta^3}{\sqrt{g}}\bigg)\enspace.
\end{eqnarray}
The quantum Hamiltonian has the form:
\begin{eqnarray}
H&=&-\frac{\hbar^2}{2m}
\bigg(b\xi^2+b\eta^2-\frac{a}{\xi^2}-\frac{a}{\eta^2}\bigg)^{-1}
\bigg(\frac{\partial^2}{\partial\xi^2}+\frac{\partial^2}{\partial\eta^2}\bigg)
\\
&=&\frac{1}{2m}
\bigg(b\xi^2+b\eta^2-\frac{a}{\xi^2}-\frac{a}{\eta^2}\bigg)^{-1/2}
(p_\xi^2+p_\eta^2)
\bigg(b\xi^2+b\eta^2-\frac{a}{\xi^2}-\frac{a}{\eta^2}\bigg)^{-1/2}\enspace.
\end{eqnarray}
We obtain for the path integral in parabolic coordinates and a time
transformation (the time-transformation function reads
$f(\xi,\eta)=(b\xi^2\eta^2-a)(\xi^2+\eta^2)/\xi^2\eta^2=\sqrt{g}\,$):
\begin{eqnarray}
&&K(\xi'',\xi',\eta'',\eta';T)
\nonumber\\  &&
=\pathint{\xi}\pathint{\eta}
\frac{b\xi^2\eta^2-a}{\xi^2\eta^2}(\xi^2+\eta^2)
\exp\left[\frac{\i m}{2\hbar}\int_0^T
\frac{b\xi^2\eta^2-a}{\xi^2\eta^2}(\xi^2+\eta^2)
                 (\dot\xi^2+\dot\eta^2)\dt\right]
\nonumber\\  &&
=\int_{-\infty}^\infty\frac{\d E}{2\pi\hbar}\,\e^{-\i ET/\hbar}
\int_0^\infty\d s''K(\xi'',\xi',\eta'',\eta';s'')\enspace,
\end{eqnarray}
with the path integral $K(s'')$ given by 
\begin{eqnarray}
&&K(\xi'',\xi',\eta'',\eta';s'')=K_\xi(\xi'',\xi';s'')K_\eta(\eta'',\eta';s'')
\nonumber\\  &&
=\pathints{\xi}
\exp\left\{\ih\int_0^{s''}\left[\frac{m}{2}\dot\xi^2
+E\bigg(b\xi^2-\frac{a}{\xi^2}\bigg)\right]\d s\right\}
\nonumber\\  &&\qquad\qquad\times
\pathints{\eta}\exp\left\{\ih\int_0^{s''}
\left[\frac{m}{2}\dot\eta^2
+E\bigg(b\eta^2-\frac{a}{\eta^2}\bigg)\right]\d s\right\}\enspace.\qquad\qquad
\label{Parabolic-s-transformed}
\end{eqnarray}
The Green's  function of the kernel $G_\eta(\CE)$ is given by 
($\lambda^2=\viert-2m|a|E/\hbar^2$, $\omega^2=-2bE/m$, $a<0$):
\begin{equation}
G_\xi(\xi'',\xi';\CE)
=\frac{\Gamma[\half(1+\lambda-\CE/\hbar\omega)]}
      {\hbar\omega\sqrt{\xi'\xi''}\Gamma(1+\lambda)}
W_{\CE/2\hbar\omega,\lambda/2}\bigg(\frac{m\omega}{\hbar}\eta_>^2\bigg)
M_{\CE/2\hbar\omega,\lambda/2}\bigg(\frac{m\omega}{\hbar}\eta_<^2\bigg)
\enspace.
\end{equation}
$M_{\mu,\nu}(z),W_{\mu,\nu}(z)$ are Whittaker functions \cite[p.1059]{GRA}.
Inserting the solution for the radial kernel in $\xi$ we get for
$G(E)$ (we can also interchange $\xi$ and $\eta$, however, the final
result will be symmetric in $\xi$ and $\eta$, so this is not necessary) 
\begin{eqnarray}
&&G(\xi'',\xi',\eta'',\eta';E)
=\frac{m\sqrt{\xi'\xi''}}{\i\hbar}
\int_0^\infty\frac{\omega\d s''}{\sin\omega s''}
\exp\bigg[-\frac{m\omega}{2\i\hbar}({\xi'}^2+{\xi'}^2)\cot\omega s''\bigg]
I_\lambda\left(\frac{m\omega\xi'\xi''}{\i\hbar\sin\omega s''}\right)
\nonumber\\  &&\qquad\times
\int\frac{\d\CE}{2\pi\i}\,\e^{-\i\CE s''/\hbar}
\frac{\Gamma[\half(1+\lambda-\CE/\hbar\omega)]}
      {\hbar\omega\sqrt{\eta'\eta''}\,\Gamma(1+\lambda)}
W_{\CE/2\hbar\omega,\lambda/2}\bigg(\frac{m\omega}{\hbar}\eta_>^2\bigg)
M_{\CE/2\hbar\omega,\lambda/2}\bigg(\frac{m\omega}{\hbar}\eta_<^2\bigg)
\nonumber\\  &&
=\frac{\hbar}{\i\pi^2}\int\frac{\d\CE}{2\pi\i}
\sqrt{\frac{\xi'\xi''}{\eta'\eta''}}
\int_0^\infty\frac{\d p\,p\sinh\pi p}{\frac{\hbar^2}{2m|a|}(p^2+\viert)-E}
\nonumber\\  &&\qquad\times
\frac{\Gamma[\half(1+\lambda-\CE)]}{\Gamma(1+\lambda)}
W_{\CE/2,\lambda/2}\bigg(\frac{m\omega}{\hbar}\eta_>^2\bigg)
M_{\CE/2,\lambda/2}\bigg(\frac{m\omega}{\hbar}\eta_<^2\bigg)
\nonumber\\  &&\qquad\times
\int_0^\infty\frac{\omega\d s''}{\sin\omega s''}
\exp\left[-\i\CE\omega s''
-\frac{m\omega}{2\i\hbar}({\xi'}^2+{\xi'}^2)\cot\omega s''\right]
K_{\i p}\left(\frac{m\omega\xi'\xi''}{\i\hbar\sin\omega s''}\right)\enspace.
\end{eqnarray}
where we have used the dispersion-relation \cite{GROb}
\begin{equation}
  I_\lambda(z)={\hbar^2\over\pi^2 m}\int_0^\infty
  {dp\,p\sinh\pi p\over\hbarm(p^2+\viert)-E}K_{\i p}(z)\enspace,
\label{numFad}
\label{DispersionIlambda}
\end{equation}
and have redefined $\CE\to\CE\hbar\omega$.
The $s''$-integral in (3.51) is evaluated in the following way:
We set $u=\omega s''$, followed by a Wick rotation, yielding
\begin{eqnarray}
&&
\int_0^\infty\frac{\d u}{\sinh u}
\exp\bigg[-u\CE-\frac{m\omega}{2\hbar}({\xi'}^2+{\xi'}^2)\coth u\bigg]
K_{\i p}\left(\frac{m\omega\xi'\xi''}{\hbar\sinh u}\right)
\nonumber\\  &&
\hbox{(subsitution $\sinh u=1/\sinh v$:)}
\nonumber\\  &&
=\int_0^\infty\d v\bigg(\coth\frac{v}{2}\bigg)^\CE
\exp\bigg[-\frac{m\omega}{2\hbar}({\xi'}^2+{\xi'}^2)\cosh v\bigg]
K_{\i p}\left(\frac{m\omega}{\hbar}\xi'\xi''\sinh v\right)\qquad\qquad
\nonumber\\  &&
=\frac{\hbar}{2m\omega\xi'\xi''}\Big|\Gamma[\bhalf(1+\i p-\CE)]\Big|^2
W_{\CE/2,\i p/2}\bigg(\frac{m\omega}{\hbar}{\xi''}^2\bigg)
W_{\CE/2,\i p/2}\bigg(\frac{m\omega}{\hbar}{\xi'}^2\bigg)\enspace,
\label{Wickrotated-parabolic}
\end{eqnarray}
where we have applied the integral representation \cite[p.729]{GRA}:
\begin{eqnarray}
&&\int_0^\infty \bigg(\coth\frac{x}{2}\bigg)^{2\nu}
\exp\bigg(-\frac{a_1+a_2}{2}t\cosh x\bigg)
K_{2\mu}(t\sqrt{a_1a_2}\,\sinh x)dx
\nonumber\\  &&\qquad\qquad\qquad\qquad\qquad\qquad\qquad\qquad
=\frac{\Gamma(\half+\mu-\nu)}{t\sqrt{a_1a_2}\Gamma(1+2\mu)}
W_{\nu,\mu}(a_1t)W_{\nu,\mu}(a_2t)\enspace.
\end{eqnarray}
Collecting terms, this gives for $G(E)$
\begin{eqnarray}
&&G(\xi'',\xi',\eta'',\eta';E)
=\frac{\i\hbar}{4\pi^2}(\xi'\xi''\eta'\eta'')^{-1/2}\int\d\CE
\nonumber\\  &&\qquad\times
\int_0^\infty\frac{\d p\,p\sinh\pi p}{\frac{\hbar^2}{2m|a|}(p^2+\viert)-E}
\frac{|\Gamma[\half(1+\i p-\CE)]|^2}{\tilde p^2}
W_{\CE/2,\i p/2}\Big(\i\tilde p{\xi''}^2\Big)
W_{\CE/2,\i p/2}^*\Big(\i\tilde p{\xi'}^2\Big)
\nonumber\\  &&\qquad\times
|\Gamma[\bhalf(1+\i p-\CE)]|^2W_{\CE/2,\i p/2}\big(\i\tilde p\eta_>^2\big)
\frac{W_{\CE/2,\i p/2}\big(\i\tilde p\eta_<^2\big)}
     {\Gamma[\half(1-\i p-\CE)]\Gamma(1+\i p)}\enspace.
\end{eqnarray}
We have abbreviated $\omega=\i\frac{\hbar}{m}
\sqrt{b(p^2+\viert)/|a|}\equiv \i\frac{\hbar}{m}\tilde p$.
In order to evaluate this expression on the cut, we use the following
representation as given in Ref. \cite[p.298]{MOS}
\begin{equation}
W_{\chi,\mu(z)}=\frac{\pi}{\sin(2\pi\mu)}
\left[\frac{M_{\chi,-\mu}(z)}{\Gamma(\half+\mu-\chi)\Gamma(1-2\mu)}
-\frac{M_{\chi,\mu}(z)}{\Gamma(\half-\mu-\chi)\Gamma(1+2\mu)}\right]
\enspace.
\end{equation}
This gives the final expression for the Green's  function:
\begin{eqnarray}
&&G(\xi'',\xi',\eta'',\eta';E)
=\hbar(\xi'\xi''\eta'\eta'')^{-1/2}
\nonumber\\  &&\qquad\times
\int\d\CE\int_0^\infty
\frac{\d p\,p\sinh\pi p}{\frac{\hbar^2}{2m|a|}(p^2+\viert)-E}
\frac{|\Gamma[\half(1+\i p-\CE)]|^4}{2\pi\tilde p^2}
\nonumber\\  &&\qquad\times
W_{\CE/2,\i p/2}\Big(\i\tilde p{\xi''}^2\Big)
W_{\CE/2,\i p/2}^*\Big(\i\tilde p{\xi'}^2\Big)
W_{\CE/2,\i p/2}\Big(\i\tilde p{\xi''}^2\Big)
W_{\CE/2,\i p/2}^*\Big(\i\tilde p{\eta'}^2\Big)\enspace.\qquad\qquad
\end{eqnarray}
The wave-functions can be easily read off from this expression:
\begin{equation}
\Psi_{\CE,p}(\xi,\eta)
=\sqrt{\frac{p\sinh\pi p}{2\pi\xi\eta}}\,
\frac{|\Gamma[\half(1+\i p-\CE)]|^2}{\tilde p}
W_{\CE/2,\i p/2}\Big(\i\tilde p\xi^2\Big)
W_{\CE/2,\i p/2}\Big(\i\tilde p\xi^2\Big)\enspace.
\end{equation}
Note the simplifications in the case $a=-1, b=1$. 

Recall that we have redefined $\CE\to\hbar\omega\CE$ in 
(\ref{Wickrotated-parabolic}). In the limiting case $b=0$ and
$\CE/\hbar\omega$ reinserted, this means that
$\CE/2\hbar\omega\to\infty$ for $\omega\to0$; however the product of
the index and the argument of the Whittaker-functions 
$(\CE/2\hbar\omega)\cdot(m\omega/\hbar)$ is constant. In this limit
the Whittaker-functions yield $K_\nu$- and $H^{(1)}_\nu$-Bessel functions,
as it should be for the semi-circular-parabolic coordinate system on
the two-dimensional hyperboloid. Of course, the parabolic system in
$\bbbr^2$ can be recovered, e.g. staring from (\ref{Parabolic-s-transformed}).
This concludes the discussion. 

\subsection{The Path Integral in Elliptic Coordinates on $\DII$}
\message{The Path Integral in Elliptic Coordinates on D_II}
The classical Lagrangian and Hamiltonian are given by
\begin{eqnarray}
\CL(\omega,\dot\omega,\vphi,\dot\vphi)&=&
\frac{m}{2}\frac{bd^2\cosh^2\omega\cos^2\vphi-a}{\cosh^2\omega\cos^2\vphi}
        (\cosh^2\omega-\cos^2\vphi)(\dot\omega^2+\dot\vphi^2)
\nonumber\\
&=&\frac{m}{2}\left[
\bigg(bd^2\cosh^2\omega+\frac{a}{\cosh^2\omega}\bigg)
-\bigg(bd^2\cos^2\vphi+\frac{a}{\cos^2\vphi}\bigg)\right]
(\dot\omega^2+\dot\vphi^2)\enspace,\qquad
\\
\CH(\omega,p_\omega,\vphi,p_\vphi)&=&
\frac{1}{2m}\frac{\cosh^2\omega\cos^2\vphi}
{(bd^2\cosh^2\omega\cos^2\vphi-a)(\cosh^2\omega-\cos^2\vphi)}
(p_\omega^2+p_\vphi^2)\enspace.
\end{eqnarray}
In the following we use 
$$\sqrt{g}=\frac{bd^2\cosh^2\omega\cos^2\vphi-a}{\cosh^2\omega\cos^2\vphi}
        (\cosh^2\omega-\cos^2\vphi)\enspace.$$
For the momentum operators we obtain
\begin{eqnarray}
p_\omega&=&\hi\left[\frac{\partial}{\partial\omega}
   +\frac{\tanh\omega}{\sqrt{g}}\bigg(bd^2\cosh^2\omega-\frac{a}{\cosh^2\omega}
\bigg)\right]\enspace,
\\
p_\vphi&=&\hi\left[\frac{\partial}{\partial\vphi}
   +\frac{\tan\vphi}{\sqrt{g}}\bigg(bd^2\cos^2\vphi-\frac{a}{\cos^2\vphi}
\bigg)\right]\enspace.
\end{eqnarray}
This gives for the quantum Hamiltonian
\begin{eqnarray}
H&=&-\frac{\hbar^2}{2m}
\frac{\cosh^2\omega\cos^2\vphi}
{(bd^2\cosh^2\omega\cos^2\vphi-a)(\cosh^2\omega-\cos^2\vphi)}
     \bigg(\frac{\partial^2}{\partial\omega^2}
           +\frac{\partial^2}{\partial\vphi^2}\bigg)
\nonumber\\
&=&\frac{1}{2m}\frac{1}{\sqrt[4]{g}}(p_\omega^2+p_\phi^2)\frac{1}{\sqrt[4]{g}}
\enspace.
\end{eqnarray}
Therefore we obtain for the path integral
(the time-transformation function reads $f(\omega,\vphi)=\sqrt{g}\,$)
\begin{eqnarray}
&&K(\omega'',\omega',\vphi'',\vphi';T)
\nonumber\\  &&
=\pathint{\omega}\pathint{\vphi}
\frac{bd^2\cosh^2\omega\cos^2\vphi-a}{\cosh^2\omega\cos^2\vphi}
        (\cosh^2\omega-\cos^2\vphi)\qquad\qquad\qquad\qquad
\nonumber\\ &&\qquad\times
\exp\left[\frac{\i m}{2\hbar}\int_0^T
\frac{bd^2\cosh^2\omega\cos^2\vphi-a}{\cosh^2\omega\cos^2\vphi}
        (\cosh^2\omega-\cos^2\vphi)(\dot\omega^2+\dot\vphi^2)\dt\right]
\nonumber\\   &&
=\int_{-\infty}^\infty\frac{\d E}{2\pi\hbar}\,\e^{-\i ET/\hbar}
\int_0^\infty \d s'' K(\omega'',\omega',\vphi'',\vphi';s'')\enspace,
\end{eqnarray}
with the path integral $K(\omega'',\omega',\vphi'',\vphi';s'')$ given by
($a<0$)
\begin{eqnarray}
&&K(\omega'',\omega',\vphi'',\vphi';s'')
=\pathints{\omega}\pathints{\vphi}
\nonumber\\ &&\qquad\times
\exp\left\{\ih\int_0^{s''}\left[\frac{m}{2}(\dot\omega^2+\dot\vphi^2)
+Ebd^2(\cosh^2\omega-\cos^2\vphi)+|a|E
\bigg(\frac{1}{\cosh^2\omega}+\frac{1}{\cos^2\vphi}\bigg)\right]\d s\right\}.
\nonumber\\   &&
\label{path-elliptic}
\end{eqnarray}
For $a=0$ we recover elliptic coordinates in $\bbbr^2$.
This path integral has the form of the spheroidal-coordinate system.
Actually, almost the same path integral was investigated in
\cite[p.122]{GROad} in connection with the elliptic-paraboloid
coordinate system on the three-dimensional hyperboloid. Let us set:
$$\lambda^2=\viert-\frac{2m|a|E}{\hbar^2},\quad
   \tilde\kappa^2=-\frac{2mbd^2E}{\hbar^2}\enspace.$$
In \cite{GROad} we have derived the following heuristic path integral identity
\begin{eqnarray}       & &
  \pathint\mu\pathint\nu d^2(\sinh^2\mu+\sin^2\nu)
         \nonumber\\   & &\qquad\times
  \exp\Bigg\{\ih\intt\Bigg[{m\over2}d^2
   (\sinh^2\mu+\sin^2\nu)(\dot\mu^2+\dot\nu^2)
   -\frac{\hbar^2}{2md^2}
    \frac{\lambda^2-\viert}{\sinh^2\mu\sin^2\nu}\Bigg]\dt\Bigg\}
         \nonumber\\   & &
  =d\sqrt{\sin\nu'\sin\nu''\sinh\mu'\sinh\mu''}\sum_{l=0}^\infty
  \frac{2l+1}{\pi}\frac{\Gamma(l-\lambda+1)}{\Gamma(l+\lambda+1)}
  \int_0^\infty p^2dp\,\e^{-\i\hbar p^2T/2m}
         \nonumber\\   & &\qquad\times
 \ps_l^{\lambda\,*}(\cos\nu';p^2d^2)\ps_l^\lambda(\cos\nu'';p^2d^2)
  S_l^{\lambda\,(1)\,*}(\cosh\mu';pd)S_l^{\lambda\,(1)}(\cosh\mu'';pd)
  \enspace.\qquad \vphantom{\bigg)}
\label{path-identity-spheroidal}
\end{eqnarray}
where $S_l^{n\,(1)},\ps_l^n$ are prolate spheroidal wavefunctions
\cite{MESCH}. By considering a proper analytic continuation and observing
\begin{equation}
  \ps^\mu_\nu(x;0)=P_\nu^\mu(x)\enspace,\qquad(|x|\leq1)\enspace,
\label{limit-spheroidal-functions}
\end{equation}
we found the solution ($a>0$, $|\vtheta|<\pi/2,\vrho\in\bbbr$):
\begin{eqnarray}       & &\!\!\!\!\!\!\!\!\!\!
  \pathint{a}\pathint\vtheta
   {\cosh^2a-\cos^2\vtheta\over\cosh^3a\cos^3\vtheta}\pathint\vrho
         \nonumber\\   & &\!\!\!\!\!\!\!\!\!\!\quad\times
   \exp\Bigg[{\i m\over2\hbar}\intt
   {(\cosh^2a-\cos^2\vtheta)(\dot a^2+\dot\vtheta^2)+\dot\vrho^2\over
    \cosh^2a\cos^2\vtheta}\dt -{3\i\hbar T\over 8m}\Bigg]
         \nonumber\\   & &\!\!\!\!\!\!\!\!\!\!
   =\sqrt{\cosh a'\cosh a''}\,\cos\vtheta'\cos\vtheta''
    \int_{\bbbr}{d\kappa\over2\pi}\e^{\i \kappa(\vrho''-\vrho')}
         \nonumber\\   & &\!\!\!\!\!\!\!\!\!\!\quad\times
    \int_0^\infty dp\,\sinh\pi p
    \int_0^\infty{dk\,k\sinh\pi k\over(\cosh^2\pi k+\sinh^2\pi p)^2}
    \,\energyldrei
         \nonumber\\   & &\!\!\!\!\!\!\!\!\!\!\quad\times
   \!\!\sum_{\epsilon,\epsilon'=\pm1}\!\!
   S_{\i p-1/2}^{\i k\,(1)}(\epsilon\tanh a'';\i \kappa)
   S_{\i p-1/2}^{\i k\,(1)\,*}(\epsilon\tanh a';\i \kappa)
   \ps_{\i k-1/2}^{\i p}(\epsilon'\sin\vtheta'';-\kappa^2)
   \ps_{\i k-1/2}^{\i p\,*}(\epsilon'\sin\vtheta';-\kappa^2)\,.
\nonumber\\   &&
\label{path-solution-spheroidal}
\end{eqnarray}    
Let us use (\ref{path-solution-spheroidal}): Of course, the variable
$\vrho$ is omitted, we use only the emerging parameter $\kappa$.
The parameter $\lambda$ in (\ref{path-identity-spheroidal}) and
(\ref{path-elliptic}) are the same up to the factor $|a|$. The main
difference is in the parameters $\kappa$ and $\tilde\kappa$, the
latter being imaginary. Inserting in $\tilde\kappa$ the energy-spectrum
$|a|E=\frac{\hbar^2}{2m}(p^2+\viert)$ gives 
$\tilde\kappa=\i d\sqrt{b(p^2+\viert)/|a|}\equiv\i\tilde p$. Combining
(\ref{path-identity-spheroidal}) and (\ref{path-solution-spheroidal})
yields for the path integral (\ref{path-elliptic}):
\begin{eqnarray}       
&& \!\!\!\!\!\!\!\!\!
G(\omega'',\omega',\vphi'',\vphi';E)
=\sqrt{\cos\vphi'\cos\vphi''}
 \int_0^\infty \frac{dp\,\sinh\pi p}{\frac{\hbar^2}{2m|a|}(p^2+\viert)-E}
    \int_0^\infty{dk\,k\sinh\pi k\over(\cosh^2\pi k+\sinh^2\pi p)^2}
         \nonumber\\   &&\!\!\!\!\!\!\!\!\!\quad\times
   \sum_{\epsilon,\epsilon'=\pm1}\!\!
   S_{\i p-1/2}^{\i k\,(1)}(\epsilon\tanh\omega'';-\tilde p)
   S_{\i p-1/2}^{\i k\,(1)\,*}(\epsilon\tanh\omega';-\tilde p)
   \ps_{\i k-1/2}^{\i p}(\epsilon'\sin\vphi'';\tilde p^2)
   \ps_{\i k-1/2}^{\i p\,*}(\epsilon'\sin\vphi';\tilde p^2)\,.
\nonumber\\   &&
\end{eqnarray}    
The wave-functions have the form:
\begin{equation}
\Psi_{p,k;\epsilon,\epsilon'}(\omega,\vphi)      
=\sqrt{\cos\vphi}\,
\frac{\sqrt{\sinh\pi pk\sinh\pi k}}{\cosh^2\pi k+\sinh^2\pi p}
S_{\i p-1/2}^{\i k\,(1)}(\epsilon\tanh\omega;-\tilde p)\,
\ps_{\i k-1/2}^{\i p}(\epsilon'\sin\vphi;\tilde p^2)\enspace.\qquad
\end{equation}       
Of course, the cases $a=-1, b=1$ simplify the formulas a little bit.
For $b=0$ the spheroidal wave-functions give the limiting case
(\ref{limit-spheroidal-functions}), therefore the solution of the
elliptic-parabolic system on the two-dimensional hyperboloid emerges
\cite{GROad}. This completes the discussion on $\DII$.
    

\setcounter{equation}{0}
\section{Darboux Space $\DIII$}
\message{Darboux Space D_III}
The coordinate systems to be considered in the Darboux space $\DIII$
are as follows:
\begin{eqnarray}
\hbox{($(u,v)$-System)}&& x=v+\i u,\quad y=v-\i u\enspace,
\\
\hbox{(Polar:)}&& \xi=\vrho\cos\vphi,\qquad \eta=\vrho\sin\vphi\enspace,\quad
\qquad\qquad\qquad\,\,\,(\vrho>0,\vphi\in[0,2\pi])\enspace,
\\
\hbox{(Parabolic:)}&& \xi=2\,\e^{-u/2}\cos\frac{v}{2}\qquad
\eta=2\,\e^{-u/2}\sin\frac{v}{2}\enspace,
\nonumber\\
&&u=\ln\frac{4}{\xi^2+\eta^2},\qquad
  v=\arcsin\frac{2\xi\eta}{\xi^2+\eta^2}\enspace,\quad
\quad\,\,\,\,(\xi\in\bbbr,\eta>0)\enspace,
\\
\hbox{(Elliptic:)}&& \xi=d\cosh\omega\cos\vphi,\qquad 
                    \eta=d\sinh\omega\sin\vphi\enspace,\quad
\quad(\omega>0,\vphi\in[-\pi,\pi])\,,\qquad
\\
\hbox{(Hyperbolic:)}&& 
\xi=\frac{\mu-\nu}{2\sqrt{\mu\nu}}+\sqrt{\mu\nu},\qquad
\eta=\i\frac{\mu-\nu}{2\sqrt{\mu\nu}}-\sqrt{\mu\nu}\enspace,\quad\,
(\mu,\nu>0)\enspace.
\end{eqnarray}
For the line element we get (we also display, where the metric is
rescaled in such a way that we set $a=b=1$ \cite{KalninsKMWinter}):
\begin{eqnarray}
\d s^2&=& \big(a\,\e^{-(x+y)/2}+b\,\e^{-(x+y)}\big)\d x\d y
\nonumber\\
\hbox{($(u,v)$-Coordinates:)}     
&=&\e^{-2u}(b+a\,\e^u)(\d u^2+\d v^2)=(\e^{-u}+\e^{-2u})(\d u^2+\d v^2)
\\
\hbox{(Polar:)}&=&(a+\hbox{$\frac{b}{4}$}\vrho^2)(\d\vrho^2+\vrho^2\d\vphi^2)
=(1+\bviert\vrho^2)(\d\vrho^2+\vrho^2\d\vphi^2)\,,
\\
\hbox{(Parabolic:)}
&=&(a+\hbox{$\frac{b}{4}$}(\xi^2+\eta^2))(\d\xi^2+\d\eta^2)
=(1+\bviert(\xi^2+\eta^2))(\d\xi^2+\d\eta^2)\,,
\\
\hbox{(Elliptic:)}
&=&(a+\hbox{$\frac{b}{4}$}d^2(\sinh^2\omega+\cos^2\vphi))d^2
   (\sinh^2\omega+\sin^2\vphi)(\d\omega^2+\d\vphi^2)\,,\qquad
\\
\hbox{(Hyperbolic:)}
&=&(a+\hbox{$\frac{b}{2}$}(\mu-\nu))(\mu+\nu)
\left(\frac{\d\mu^2}{\mu^2}-\frac{\d\nu^2}{\nu^2}\right)\,.
\end{eqnarray}
For the Gaussian curvature we find
\begin{equation}
K=-\frac{ab\,\e^{-3u}}{(b\,\e^{-2u}+a\,\e^{-u})^4}\enspace.
\end{equation}
For e.g. $a=1, b=0$ we recover the two-dimensional flat space with the
corresponding coordinate systems. To assure the positive definiteness
of the metric (1.3), we can require $a,b>0$.

\subsection{The Path Integral in $(u,v)$-Coordinates on $\DIII$}
\message{The Path Integral in (u,v)-Coordinates on D_III}
The classical Lagrangian and Hamiltonian are given by:
\begin{equation}
\CL(u,\dot u,v,\dot v)=\frac{m}{2}\frac{b+a\,\e^u}{\e^{2u}}
(\dot u^2+\dot v^2),\quad
\CH(u,p_u,v,p_v)=\frac{1}{2m}\frac{\e^{2u}}{b+a\,\e^u}(p_u^2+p_v^2)\enspace.
\end{equation}
The canonical momenta are given by
\begin{equation}
p_u=\hi\bigg(\frac{\partial}{\partial u}
-\half\frac{a\,\e^{-u}+2b\,\e^{-2u}}{a\,\e^{-u}+b\,\e^{-2u}}\bigg),\quad
p_v=\hi\frac{\partial}{\partial v}\enspace,
\end{equation}
and for the quantum Hamiltonian we find
\begin{eqnarray}
H&=&-\frac{\hbar^2}{2m}\frac{1}{a\,\e^{-u}+b\,\e^{-2u}}
\bigg(\frac{\partial^2}{\partial u^2}+\frac{\partial^2}{\partial v^2}\bigg)
\enspace,
\\
&=&\frac{1}{2m}
\sqrt{ \frac{1}{a\,\e^{-u}+b\,\e^{-2u}} }
\Big(p_u^2+p_v^2\Big)
\sqrt{ \frac{1}{a\,\e^{-u}+b\,\e^{-2u}} }\enspace.
\end{eqnarray}
Therefore we obtain for the path integral
\begin{eqnarray}
&&K(u'',u',v'',v';T)
\nonumber\\ &&
=\pathint{u}\pathint{v}(a\,\e^{-u}+b\,\e^{-2u})
\exp\left[\frac{\i m}{2\hbar}\int_0^T(a\,\e^{-u}+b\,\e^{-2u})(\dot u^2+\dot v^2)
\dt\right]
\nonumber\\
&&=\int_{-\infty}^\infty\frac{\d E}{2\pi\hbar}\,\e^{-\i ET/\hbar}
\int_0^\infty\d s''K(u'',u',v'',v';s'')\enspace,
\end{eqnarray}
with the time-transformed path integral $K(s'')$ given by 
($f(u)=(a\,\e^{-u}+b\,\e^{-2u})=\sqrt{g}$)
\begin{eqnarray}
&&K(u'',u',v'',v';s'')
\nonumber\\
&&=\pathints{u}\pathints{v}
\exp\left\{\ih\int_0^{s''}\bigg[\frac{m}{2}(\dot u^2+\dot v^2)
+Eb\bigg(\e^{-2u}+\frac{a}{b}\e^{-u}\bigg)\bigg]\d s\right\}\,.\qquad\quad
\label{pathintegral-Morse}
\end{eqnarray}
We observe that the path integral in the variable $u$ is a path
integral for the Morse potential
$$V(x)=\frac{V_0^2\hbar^2}{2m}\Big(\e^{-2x}-2\alpha\,\e^{-x}\Big)$$
with $V_0=\sqrt{-2mbE}/\hbar$ and $\alpha=-a/2b$; the Green's
functions in $u$ and $v$ are given by \cite{GRSh} 
\begin{eqnarray}
&&\!\!\!\!\!\!\!\!
G_u(u'';u';\CE)=
\frac{m\Gamma(\half+\sqrt{-2m\CE}/\hbar+a\sqrt{-2mbE}/2b\hbar)}
{\hbar\sqrt{-2mbE}\,\Gamma(1+2\sqrt{-2m\CE}/\hbar)}\,\e^{(u'+u'')/2}
\nonumber\\ &&\!\!\!\!\!\!\!\!\qquad\times
W_{-a\sqrt{-2mbE}/2b\hbar,\sqrt{-2m\CE}/\hbar}
\bigg(\frac{\sqrt{-8mbE}}{\hbar}\,\e^{-u_<}\!\bigg)
M_{-a\sqrt{-2mbE}/2b\hbar,\sqrt{-2m\CE}/\hbar}
\bigg(\frac{\sqrt{-8mbE}}{\hbar}\,\e^{-u_>}\!\bigg),
\nonumber \\
\\
&&\!\!\!\!\!\!\!\!
G_v(v'';v';\CE)=
\frac{1}{2\pi}\sum_{l=-\infty}^\infty
\frac{\e^{\i l(v''-v')}}{\hbar^2l^2/2m-\CE}\enspace.
\end{eqnarray}
This gives for the Green's  function in $(u,v)$-coordinates the solution
(note $\sqrt{-2m\CE}/\hbar\to+l$)
\begin{eqnarray}
&&G(u'',u',v'',v';E)
=\sum_{l=-\infty}^\infty
\frac{\e^{\i l(v''-v')}}{2\pi}
\frac{m\Gamma(\half+l+a\sqrt{-2mbE}/2b\hbar)}
{\hbar\sqrt{-2mbE}\,\Gamma(1+2l)}\,\e^{(u'+u'')/2}
\nonumber\\ &&\qquad\times
W_{-a\sqrt{-2mbE}/2b\hbar,l}
\bigg(2\frac{\sqrt{-2mbE}}{\hbar}\,\e^{-u_<}\bigg)
M_{-a\sqrt{-2mbE}/2b\hbar,l}
\bigg(2\frac{\sqrt{-2mbE}}{\hbar}\,\e^{-u_>}\bigg)\enspace.\qquad\qquad
\end{eqnarray}
In order to extract the wave-functions we use the following representation
\begin{eqnarray}
&&{1\over\sin\alpha}\exp\big[-(x+y)\cot\alpha\big]
  I_{2\mu}\bigg({2\sqrt{xy}\over\sin\alpha}\bigg)
\label{Identity-Buchholz}
\\  &&=
  {1\over2\pi\sqrt{xy}}\int_{-\infty}^\infty
  {\Gamma(\half+\mu+\i p)\Gamma(\half+\mu-\i p)\over\Gamma^2(1+2\mu)}
  \e^{-2\alpha p+\pi p}M_{+\i p,\mu}(-2\i x)M_{-\i p,\mu}(+2\i y)dp\enspace.
\nonumber
\end{eqnarray}
This relation can be derived by using an integral representation as
given by Buchholz \cite[p.158]{BUCH}. Applying this yields by exploiting
$\CE=\hbar^2l^2/2m$ and evaluating the residuum at $E=\hbar^2p^2/2m$:
($\omega=\sqrt{-2E/m}$, we set $a=b=1$, and utilize the calculation of
\cite{GROb} by inserting the path integral solution of the radial
harmonic oscillator in (\ref{pathintegral-Morse}) together with an
appropriate coordinate transformation) 
\begin{eqnarray}
&&G(u'',u',v'',v';E)
=\sum_{l=-\infty}^\infty
\frac{\e^{\i l(v''-v')}}{2\pi}
\frac{1}{\pi\i}\int\frac{\d\CE}{\frac{\hbar^2l^2}{2m}-\CE}
\sqrt{-\frac{8mE}{\hbar^2}}
\nonumber\\ &&\times
\int_0^\infty\frac{\d\sigma}{\sin\omega\sigma}
\exp\left[\frac{4\i aE}{\hbar}\sigma
-\frac{\sqrt{8mE}}{\hbar}\Big(\e^{-u'}+\e^{-u''}\Big)\cot\omega\sigma\right]
I_{\sqrt{8m\CE}/\hbar}\left(\frac{\sqrt{8mE}\,\e^{-(u'+u'')/2}}
                               {\hbar\sin\omega\sigma}\right)
\nonumber\\ &&
=\sum_{l=-\infty}^\infty\frac{\e^{\i l(v''-v')}}{2\pi}\e^{(u'+u'')/2}
\nonumber\\ &&\qquad\qquad\times
\int_0^\infty\frac{\e^{\pi p/2}\d p}{\frac{\hbar^2p^2}{2m}-E}
\frac{|\Gamma(\half+l+\i p)|^2}{2\pi\Gamma^2(1+2l)}
M_{ \i p/2,l}\Big(-2\i p\,\e^{-u'}\Big)
M_{-\i p/2,l}\Big( 2\i p\,\e^{-u''}\Big)\enspace.
\end{eqnarray}
And we can read off the wave-functions 
\begin{equation}
\Psi_{p,l}(u,v)
=\frac{\e^{\i l v}}{\sqrt{2\pi}}\cdot\frac{\e^{\pi p/4}}{\sqrt{2\pi}}\,
\frac{\Gamma(\half+l+\i p/2)}{\Gamma(1+2l)}
M_{ \i p/2,l}\Big(-2\i\e^{-u}\Big)\enspace,
\end{equation}
respectively with $(a,b)$ re-inserted:
\begin{equation}
\Psi_{p,l}(u,v)=\frac{\e^{\i l v}}{\sqrt{2\pi}}\cdot
 \frac{\e^{\pi p/4}}{\sqrt{2\pi}}\,
\frac{\Gamma(\half+l+\i ap/2\sqrt{b})}{\Gamma(1+2l)}
M_{ \i ap/\sqrt{b}2,l}\Big(-2\i p\sqrt{b}\,\e^{-u}\Big)\enspace.
\end{equation}
Note that this evaluation is almost the same as in the path integral for the
two-dimensional Coulomb potential \cite{GRSh}.
This concludes the discussion of the $(u,v)$-system on $\DIII$.

\subsection{The Path Integral in Polar Coordinates on $\DIII$}
\message{The Path Integral in Polar Coordinates on D_III}
In the coordinates $(\vrho,\vphi)$ the Lagrangian and Hamiltonian take on the
form
\begin{equation}
\CL(\vrho,\dot\vrho,\vphi,\dot\vphi)=\frac{m}{2}(a+\hbox{$\frac{b}{4}$}\vrho^2)
(\dot \vrho^2+\vrho^2\dot\vphi^2),\quad
\CH(\vrho,p_\vrho,\vphi,p_\vphi)=\frac{1}{2m}
\frac{1}{a+\hbox{$\frac{b}{4}$}\vrho^2}
\bigg(p_\vrho^2+\frac{1}{\vrho^2}p_\vphi^2\bigg)\enspace.
\end{equation}
The canonical momenta are given by
\begin{equation}
p_\vrho=\hi\bigg(\frac{\partial}{\partial\vrho}
+\frac{b\vrho}{4a+b\vrho^2}+\frac{1}{2\vrho}\bigg),\quad
p_\vphi=\hi\frac{\partial}{\partial\vphi}\enspace.
\end{equation}
Therefore the quantum Hamiltonian is given by:
\begin{eqnarray}
H&=&-\frac{\hbar^2}{2m}\frac{1}{a+\hbox{$\frac{b}{4}$}\vrho^2}
\bigg(\frac{\partial^2}{\partial \vrho^2}
+\frac{1}{\vrho}\frac{\partial}{\partial\vrho}
+\frac{1}{\vrho^2}\frac{\partial^2}{\partial\vphi^2}\bigg)
\\  &=&
\frac{1}{2m}\sqrt{\frac{1}{a+\hbox{$\frac{b}{4}$}\vrho^2}}\bigg(p_\vrho^2
+\frac{1}{\vrho^2}p_\vphi^2\bigg)  
\sqrt{\frac{1}{a+\hbox{$\frac{b}{4}$}\vrho^2}}
      -(a+\hbox{$\frac{b}{4}$}\vrho^2)\frac{\hbar^2}{8m\vrho^2}\enspace,
\end{eqnarray}
and in this case we have an additional quantum potential $\propto\hbar^2$.
This gives for the path integral 
($f(\vrho)=a+\hbox{$\frac{b}{4}$}\vrho^2=\sqrt{g}$)
\begin{eqnarray}
&&K(\vrho'',\vrho',\vphi'',\vphi';T)
\nonumber\\ &&
=\pathint{\vrho}\pathint{\vphi}(a+\hbox{$\frac{b}{4}$}\vrho^2)\vrho
\nonumber\\ &&\qquad\qquad\qquad\qquad\times
\exp\left\{\ih\int_0^T\left[\frac{m}{2}
(a+\hbox{$\frac{b}{4}$}\vrho^2)(\dot \vrho^2+\vrho^2\dot\vphi^2)
+(a+\hbox{$\frac{b}{4}$}\vrho^2)^{-1}
\frac{\hbar^2}{8m\vrho^2}\right]\dt\right\}
\qquad
\nonumber\\ &&
=\int_{-\infty}^\infty\frac{\d E}{2\pi\hbar}\,\e^{-\i ET/\hbar}
\int_0^\infty\d s''K(\vrho'',\vrho',\vphi'',\vphi';s'')\enspace,
\end{eqnarray}
with the time-transformed path integral $K(s'')$ given by
\begin{eqnarray}
&&K(\vrho'',\vrho',\vphi'',\vphi';s'')
\nonumber\\ &&
=\pathints{\vrho}\vrho\pathints{\vphi}
\exp\left\{\ih\int_0^{s''}\left[\frac{m}{2}(\dot \vrho^2+\vrho^2\dot\vphi^2)
+E(a+\hbox{$\frac{b}{4}$}\vrho^2)
+\frac{\hbar^2}{8m\vrho^2}\right]\d s\right\}
\nonumber\\  &&
=\sum_{l=-\infty}^\infty
\frac{\e^{\i l(\vphi''-\vphi')}}{2\pi\sqrt{\vrho'\vrho''}}
\pathints{\vrho}
\exp\left\{\ih\int_0^{s''}\left[\frac{m}{2}\dot \vrho^2
+E\frac{b}{4}\vrho^2-\hbar^2\frac{l^2-\viert}{2m\vrho^2}\right]\d s''
+\ih aEs''\right\}
\nonumber\\  &&
=\sum_{l=-\infty}^\infty\frac{\e^{\i l(\vphi''-\vphi')}}{2\pi}
\frac{m\omega}{\i\hbar\sin\omega s''}
\exp\bigg[-\frac{m\omega}{2\i\hbar}({\vrho'}^2+{\vrho''}^2)\cot\omega s''
+\ih aEs''\bigg]I_l\bigg(\frac{m\omega \vrho'\vrho''}{\i\hbar\sin\omega
  s''}\bigg) 
\end{eqnarray}
($\omega^2=-Eb/2m$), where I have separated off the $\vphi$-path
integration and in the last step inserted the path integral solution
for the radial harmonic oscillator \cite{GRSh,PI}. We use the
integral representation \cite[p.729]{GRA}:
\begin{eqnarray}
&&\int_0^\infty \bigg(\coth\frac{x}{2}\bigg)^{2\nu}
\exp\bigg(-\frac{a_1+a_2}{2}t\cosh x\bigg)
I_{2\mu}(t\sqrt{a_1a_2}\sinh x)dx
\nonumber\\  &&\qquad\qquad\qquad\qquad
=\frac{\Gamma(\half+\mu-\nu)}{t\sqrt{a_1a_2}\Gamma(1+2\mu)}
M_{\nu,\mu}(a_1t)W_{\nu,\mu}(a_2t)\enspace.
\end{eqnarray}
Therefore we obtain for the entire Green's  function (rescaling $b\to4b^2$)
\begin{eqnarray}
&&G(\vrho'',\vrho',\vphi'',\vphi';E)
\nonumber\\  &&=
\sum_{l=-\infty}^\infty\frac{\e^{\i l(\vphi''-\vphi')}}{2\pi}
\int_0^\infty\frac{m\omega\d s''}{\i\hbar\sin\omega s''}
\exp\bigg[-\frac{m\omega}{2\i\hbar}({\vrho'}^2+{\vrho''}^2)\cot\omega s''
+\ih aEs''\bigg]
I_l\bigg(\frac{m\omega \vrho'\vrho''}{\i\hbar\sin\omega s''}\bigg)\qquad
\nonumber\\  &&
\hbox{(substitution $u=\omega s''=\hbar pbs''/m$, $E=\hbar^2p^2/2m$, 
and Wick-rotation)}
\nonumber\\  &&=
\sum_{l=-\infty}^\infty\frac{\e^{\i l(\vphi''-\vphi')}}{2\pi}
\int_0^\infty\frac{\d u}{\sinh u}
\exp\bigg[-\frac{\i pb}{2}({\vrho'}^2+{\vrho''}^2)\coth u+\i aup\bigg]
I_l\bigg(\frac{\i bp\vrho'\vrho''}{\sinh u}\bigg)
\nonumber\\  &&
\hbox{(substitution $\sinh u=1/\sinh v$)}
\nonumber\\  &&=
\sum_{l=-\infty}^\infty\frac{\e^{\i l(\vphi''-\vphi')}}{2\pi}
\int_0^\infty\d v\bigg(\coth\frac{v}{2}\bigg)^{\i ap}
\exp\bigg[-\frac{\i bp}{2}({\vrho'}^2+{\vrho''}^2)\cosh v\bigg]
I_l(\i bp\vrho'\vrho''\sinh v)
\nonumber\\  &&
\hbox{\ }
\nonumber\\  &&
\hbox{(reinserting $p=\sqrt{2mE}/\hbar$)}
\nonumber\\  &&=
\sum_{l=-\infty}^\infty\frac{\e^{\i l(\vphi''-\vphi')}}
                            {2\pi b\sqrt{\vrho'\vrho''}}
\cdot\sqrt{-\frac{m}{2E}}
\frac{\Gamma[\half(1+l-a\sqrt{-2mE/B}/\hbar)]}{\Gamma(1+l)}
\nonumber\\  &&\qquad\qquad\qquad\times
W_{a\sqrt{-2mE/b}/2\hbar,\frac{l}{2}}\left(b\sqrt{-\frac{2mE}{\hbar^2}}
\,\vrho_>\right)
M_{a\sqrt{-2mE/b}/2\hbar,\frac{l}{2}}\left(b\sqrt{-\frac{2mE}{\hbar^2}}
\,\vrho_<\right)\enspace. 
\end{eqnarray}
In order to extract the wave-functions we use again the the representation
(\ref{Identity-Buchholz}) and obtain
\begin{eqnarray}
&&G(\vrho'',\vrho',\vphi'',\vphi';E)
=\sum_{l=-\infty}^\infty\frac{\e^{\i l(\vphi''-\vphi')}}{2\pi}
\nonumber\\  &&\qquad\times
\frac{1}{2\pi b\sqrt{\vrho'\vrho''}}\int_0^\infty
\frac{\d p\,\e^{\pi p}}{\frac{\hbar^2p^2}{2m}-E}
\frac{|\Gamma[\half(1+l+\i ap)]|^2}{\Gamma^2(1+l)}
M_{\i ap/2,l/2}(-\i bp\vrho')M_{-\i ap/2,l/2}(\i bp\vrho'')\,,\qquad\quad
\end{eqnarray}
and the wave-functions have the form
\begin{equation}
\Psi_{p,l}(\vrho,\vphi)=\frac{\e^{\i l\vphi}}{\sqrt{2\pi}}\cdot
\frac{\e^{\pi p/2}}{\sqrt{2\pi b\vrho}}
\frac{\Gamma[\half(1+l+\i ap)]}{l!}M_{\i ap/2,l/2}(-\i bp\vrho)\enspace.
\end{equation}
Note that this system is very similar to the $(u,v)$-system, the principal
difference being another counting in $l$ and the replacement $\vrho=\e^{-u}$. 

\subsection{The Path Integral in Parabolic Coordinates on $\DIII$}
\message{The Path Integral in Parabolic Coordinates on D_III}
The classical Lagrangian and Hamiltonian are given by 
\begin{equation}
\CL(\xi,\dot\xi,\eta,\dot\eta)
=\frac{m}{2}(a+\hbox{$\frac{b}{4}$}(\xi^2+\eta^2))
(\dot\xi^2+\dot\eta^2),\quad
\CH(\xi,p_\xi,\eta, p_\xi)=
\frac{1}{2m}\frac{1}{a+\hbox{$\frac{b}{4}$}(\xi^2+\eta^2)}
            (p_\xi^2+p_\eta^2)\enspace.
\end{equation}
The canonical momenta are given by
\begin{equation}
p_\xi=\hi\bigg(\frac{\partial}{\partial\xi}
+\frac{b\xi}{a+\hbox{$\frac{b}{4}$}(\xi^2+\eta^2)}\bigg),\quad
p_\eta=\hi\bigg(\frac{\partial}{\partial\eta}
+\frac{b\eta}{a+\hbox{$\frac{b}{4}$}(\xi^2+\eta^2)}\bigg)\enspace.
\end{equation}
and for the quantum Hamiltonian we find
\begin{eqnarray}
H&=&-\frac{\hbar^2}{2m}\frac{1}{a+\hbox{$\frac{b}{4}$}(\xi^2+\eta^2)}
\bigg(\frac{\partial^2}{\partial\xi}+\frac{\partial^2}{\partial\eta^2}\bigg)
\enspace,
\\
&=&\frac{1}{2m}\sqrt{\frac{1}{a+\hbox{$\frac{b}{4}$}(\xi^2+\eta^2)}}
(p_\xi^2+p_\eta^2)\sqrt{\frac{1}{a+\hbox{$\frac{b}{4}$}
(\xi^2+\eta^2)}}\enspace.
\end{eqnarray}
Therefore we obtain for the path integral
\begin{eqnarray}
&&K(\xi'',\xi',\eta'',\eta';T)
\nonumber\\ &&
=\pathint{\xi}\pathint{\eta}(a+\hbox{$\frac{b}{4}$}(\xi^2+\eta^2))
\exp\left[\frac{\i m}{2\hbar}
\int_0^T(a+\hbox{$\frac{b}{4}$}(\xi^2+\eta^2))(\dot\xi^2+\dot\eta^2)\dt\right]
\nonumber\\ &&
=\int_{-\infty}^\infty\frac{\d E}{2\pi\hbar}\,\e^{-\i ET/\hbar}
\int_0^\infty\d s''K(\xi'',\xi',\eta'',\eta';s'')\enspace,
\label{pathintegral-parabolic-1}
\end{eqnarray}
with the time-transformed path integral $K(s'')$ given by (the 
time-transformation function reads $f(\xi,\eta)%
=(a+\hbox{$\frac{b}{4}$}(\xi^2+\eta^2))=\sqrt{g}$) 
\begin{eqnarray}
&&K(\xi'',\xi',\eta'',\eta';s'')
=\pathints{\xi}\pathints{\eta}
\nonumber\\ &&\qquad\qquad\times
\exp\left\{\ih\int_0^{s''}\bigg[\frac{m}{2}(\dot\xi^2+\dot\eta^2)
+E\frac{b}{4}(\xi^2+\eta^2)\bigg]\d s''+a\ih Es''\right\}\enspace.
\label{pathintegral-parabolic-2}
\end{eqnarray}
The path integrals in $\xi$ and $\eta$ are path integrals for the
harmonic oscillator, with $\omega^2=-Eb/2m$. The Green's  function for the
harmonic oscillator in the variable $\xi$ is given by \cite{GRSh}
\begin{equation}
G_\xi(\xi'',\xi';\CE)=\sqrt{\frac{m}{\pi\omega\hbar^2}}\,
\Gamma\bigg(\half-\frac{\CE}{\hbar\omega}\bigg)
D_{-\half+\CE/\hbar\omega}\bigg(\sqrt{\frac{2m\omega}{\hbar}}\,\xi_>\bigg)
D_{-\half+\CE/\hbar\omega}\bigg(-\sqrt{\frac{2m\omega}{\hbar}}\,\xi_<\bigg)
\enspace,
\end{equation}
and similarly for $G_\eta(\eta'',\eta';\CE)$. The $D_\nu(z)$ are
parabolic cylinder-functions \cite[p.1064]{GRA}. This gives ($b\to 4b^2$)
\begin{eqnarray}
&&G(\xi'',\xi',\eta'',\eta';E)
=\int\d\CE\frac{m}{\pi\hbar^2b}\sqrt{-\frac{m}{2E}}\,
\Gamma\bigg(\half+\frac{aE-\CE}{b\hbar}\sqrt{-\frac{m}{2E}}\,\bigg)
\Gamma\bigg(\half+\frac{\CE}{b\hbar}\sqrt{-\frac{m}{2E}}\,\bigg)
\nonumber\\ &&\qquad\qquad\times
D_{-\half+\frac{aE-\CE}{b\hbar}\sqrt{-\frac{m}{2E}}}
\left(\sqrt[4]{-\frac{8mEb^2}{\hbar^2}}\,\xi_>\right)
D_{-\half+\frac{aE-\CE}{b\hbar}\sqrt{-\frac{m}{2E}}}
\left(-\sqrt[4]{-\frac{8mEb^2}{\hbar^2}}\,\xi_<\right)
\nonumber\\ &&\qquad\qquad\times
D_{-\half+\frac{\CE}{b\hbar}\sqrt{-\frac{m}{2E}}}
\left(\sqrt[4]{-\frac{8mEb^2}{\hbar^2}}\,\eta_>\right)
D_{-\half+\frac{\CE}{b\hbar}\sqrt{-\frac{m}{2E}}}
\left(-\sqrt[4]{-\frac{8mEb^2}{\hbar^2}}\,\eta_<\right)\enspace.
\end{eqnarray}
Considering (\ref{pathintegral-parabolic-2}), we observe that it has the
same form as the path integral for the Coulomb potential in two
dimensions in parabolic coordinates which was solved in
\cite{DKb,GROab,GRSh,GROPOa}. In the present case the Coulomb coupling 
$\alpha$ is replaced by $aE/2$ and the energy $E$ by
$bE/4$. Introducing the ``Bohr''-radius $a_B=\hbar^2/m\alpha=2\hbar^2/maE$
find  for the solution of (\ref{pathintegral-parabolic-1}) as follows
\begin{equation}
K(\xi'',\xi',\eta'',\eta';T)
=\sum_{e,o}\int_{\bbbr} d\zeta\int_{\bbbr}dp\,
   \e^{-\i\hbar p^2T/2m}\Psi_{p,\zeta}^{(e,o)\,*}(\xi',\eta')
   \Psi_{p,\zeta}^{(e,o)}(\xi'',\eta'')\enspace,
\end{equation}
and $\sum_{e,o}$ denotes the summation over even and odd states respectively; 
the functions $\Psi_{p,\zeta}^{(e,o)}(\xi,\eta)$ are given by

\newpage\noindent
\begin{eqnarray}
  &&\Psi_{p,\zeta}^{(e,o)}(\xi,\eta)
  ={\e^{\pi/2ap}\over\sqrt{2}4\pi^2}
  \nonumber\\   &&\quad\times
    \left({{\big\vert\Gamma(\viert-\frac{\i}{2p}(1/a_B+\zeta))\big\vert^2
    E^{(0)}_{-\half+\frac{\i}{p}(1/a_B+\zeta)}(\e^{-\i\pi/4}\sqrt{2p}\,\xi)
    E^{(0)}_{-\half-\frac{\i}{p}(1/a_B+\zeta)}(\e^{-\i\pi/4}\sqrt{2p}\,\eta)
    }\atop{
    \big\vert\Gamma({3\over4}-\frac{\i}{2p}(1/a_B+\zeta))\big\vert^2
    E^{(1)}_{-\half+\frac{\i}{p}(1/a_B+\zeta)}(\e^{-\i\pi/4}\sqrt{2p}\,\xi)
    E^{(1)}_{-\half-\frac{\i}{p}(1/a_B+\zeta)}(\e^{-\i\pi/4}\sqrt{2p}\,\eta)
    }}\right)\enspace,
  \nonumber\\   &&
\end{eqnarray}
which are $\delta$-normalized according to \cite{POGOa}
\begin{equation}
  \int_0^\infty d\nu\int_{\bbbr}d\xi(\xi^2+\eta^2)
  \Psi_{p',\zeta'}^{(e,o)\,*}(\xi,\eta)
  \Psi_{p,\zeta}^{(e,o)}(\xi,\eta)
  =\delta(p'-p)\delta(\zeta'-\zeta)\enspace,
\end{equation}
and $\zeta$ is the parabolic separation constant.
The functions $E^{(0)}_\nu(z)$ and $E^{(1)}_\nu(z)$ are even and odd
parabolic cylinder functions in the variable~$z$, respectively \cite{BUCH}:
\begin{equation}
\left.\begin{array}{l}
\displaystyle
E_\nu^{(0)}=\sqrt{2}\,\e^{-z^2/4}
{_1}F_1\bigg(-\frac{\nu}{2},\half;\frac{z^2}{2}\bigg)
=\sqrt{2\pi}\,\bigg(\frac{z^2}{2}\bigg)^{-1/4}
{\cal M}_{\nu/2+1/4,-1/4}\bigg(\frac{z^2}{2}\bigg)
\enspace,\\[3mm]
\displaystyle
E_\nu^{(1)}=2z\,\e^{-z^2/4}
{_1}F_1\bigg(\frac{1-\nu}{2},\frac{3}{2};\frac{z^2}{2}\bigg)
=\sqrt{2\pi}\,\bigg(\frac{z^2}{2}\bigg)^{-1/4}
{\cal M}_{\nu/2+1/4,1/4}\bigg(\frac{z^2}{2}\bigg)\enspace,
\end{array}\qquad\right\}
\end{equation}
${_1}F_1(a;b;z)$ is the confluent hypergeometric function \cite[p.1057]{GRA}, 
and ${\cal M}_{\chi,\mu}(z)=M_{\chi,\mu}(z)/\Gamma(1+2\mu)$. 
Note the relation
\begin{equation}
D_\nu(z)=2^{\nu/2}\sqrt{\frac{\pi}{2}}\,
\left[\frac{E_\nu^{(0)}(z)}{\Gamma(\frac{1-\nu}{2})}-
\frac{E_\nu^{(1)}(z)}{\Gamma(-\frac{\nu}{2})}\right]\enspace.
\end{equation}
This concludes the discussion.

\subsection{The Path Integral in Elliptic Coordinates on $\DIII$}
\message{The Path Integral in Elliptic Coordinates on D_III}
The classical Lagrangian and Hamiltonian are given by
\begin{eqnarray}
\CL(\omega,\dot\omega,\vphi,\dot\vphi)&=&
\frac{m}{2}d^2(a+\hbox{$\frac{b}{4}$}d^2(\sinh^2\omega+\cos^2\vphi))
(\sinh^2\omega+\sin^2\vphi)(\dot\omega^2+\dot\vphi^2)\enspace,
\\
\CH(\omega,p_\omega,\vphi,p_\vphi)&=&
\frac{1}{2m}\frac{p_\omega^2+p_\vphi^2}
{d^2(a+\hbox{$\frac{b}{4}$}d^2(\sinh^2\omega+\cos^2\vphi))
(\sinh^2\omega+\sin^2\vphi)}\enspace.
\end{eqnarray}
The canonical momenta have the form
\begin{eqnarray}
p_\omega&=&\hi\bigg(\frac{\partial}{\partial\omega}
+\frac{bd^2\sinh\omega\cosh\omega}
{4a+bd^2(\sinh^2\omega+\cos^2\vphi)}
+\frac{\sinh\omega\cosh\omega}{\sinh^2\omega+\sin^2\vphi}\bigg)\enspace,
\\
p_\vphi&=&\hi\bigg(\frac{\partial}{\partial\vphi}
-\frac{bd^2\sin\vphi\cos\vphi}
{4a+bd^2(\sinh^2\omega+\cos^2\vphi)} 
+\frac{\sin\vphi\cos\vphi}{\sinh^2\omega+\sin^2\vphi}\bigg)\enspace,
\end{eqnarray}
and for the quantum Hamiltonian we find (we use $\sqrt{g}=
d^2(a+\hbox{$\frac{b}{4}$}d^2(\sinh^2\omega+\cos^2\vphi))
 (\sinh^2\omega+\sin^2\vphi)$)
\begin{eqnarray}
H&=&-\frac{\hbar^2}{2m}\frac{1}{\sqrt{g}}
\bigg(\frac{\partial^2}{\partial\omega^2}
     +\frac{\partial^2}{\partial\vphi^2}\bigg)
=\frac{1}{2m}\frac{1}
{\sqrt[4]{g}}(p_\omega^2+p_\vphi^2)\frac{1}{\sqrt[4]{g}}\enspace.
\end{eqnarray}
This give for the path integral
\begin{eqnarray}
&&K(\omega'',\omega',\vphi'',\vphi';T)
=\pathint{\omega}\pathint{\vphi}\sqrt{g}
\nonumber\\  &&\qquad\qquad\times
\exp\left[\frac{\i md^2}{2\hbar}\int_0^T
(a+\hbox{$\frac{b}{4}$}d^2(\sinh^2\omega+\cos^2\vphi))
(\sinh^2\omega+\sin^2\vphi)(\dot\omega^2+\dot\vphi^2)\dt\right]\qquad
\nonumber\\  &&
=\int_{-\infty}^\infty\frac{\d E}{2\pi\hbar}\,\e^{-\i ET/\hbar}
\int_0^\infty\d s''K(\omega'',\omega',\vphi'',\vphi';s'')\enspace,
\end{eqnarray}
(time-transformation function ($f(\omega,\vphi)=\sqrt{g}$)
with the path integral $K(s'')$ given by
\begin{eqnarray}
&&\!\!\!\!\!\!\!\!\!\!\!\!
K(\omega'',\omega',\vphi'',\vphi';s'')=
\pathints{\omega}\pathints{\vphi}
\nonumber\\  &&\!\!\!\!\!\!\!\!\!\!\!\!\quad\times
\exp\left\{\ih\int_0^{s''}\bigg[\frac{m}{2}(\dot\omega^2+\dot\vphi^2)
+E(a+\hbox{$\frac{b}{4}$}d^2(\sinh^2\omega+\cos^2\vphi))
(\sinh^2\omega+\sin^2\vphi)\bigg]\d s\right\}\,.\qquad
\end{eqnarray}
For this kind of problem we do not have any theory of special
functions to treat with and we leave this intractable path integral as
it stands. $b=0$ gives the elliptic system in $\bbbr^2$ \cite{GROad}.

\subsection{The Path Integral in Hyperbolic Coordinates on $\DIII$}
\message{The Path Integral in Hyperbolic Coordinates on D_III}
The classical Lagrangian and Hamiltonian have the form
\begin{eqnarray}
\CL(\mu,\dot\mu,\nu,\dot\nu)&=&
\frac{m}{2}(a+\hbox{$\frac{b}{2}$}(\mu-\nu))(\mu+\nu)
\bigg(\frac{\dot\mu^2}{\mu^2}-\frac{\dot\nu^2}{\nu^2}\bigg)
\\
\CH(\mu,p_\mu,\nu,p_\nu&=&\frac{1}{2m}
\frac{\mu^2p_\mu^2-\nu^2p_\nu^2}{(a+\hbox{$\frac{b}{2}$}(\mu-\nu))(\mu+\nu)}
\end{eqnarray}
The canonical momentum operators are given by
\begin{eqnarray}
p_\mu&=&\hi\bigg[\frac{\partial}{\partial\mu}+\half\bigg(
+\frac{1}{\mu+\nu}+\frac{b}{a+\hbox{$\frac{b}{2}$}
              (\mu-\nu)}-\frac{1}{\mu}\bigg)\bigg],\\
p_\nu&=&\hi\bigg[\frac{\partial}{\partial\mu}+\half\bigg(
+\frac{1}{\mu+\nu}-\frac{b}{a+\hbox{$\frac{b}{2}$}(\mu-\nu)}-\frac{1}{\nu}\bigg)\bigg]\enspace,
\end{eqnarray}
and the quantum Hamiltonian has the form
\begin{eqnarray}
H&=&-\frac{\hbar^2}{2m}\frac{1}{(a+\hbox{$\frac{b}{2}$}(\mu-\nu))(\mu+\nu)}
\left[\mu^2\bigg(\frac{\partial^2}{\partial\mu^2}-\frac{1}{\mu}
\frac{\partial}{\partial\mu}\bigg)
-\nu^2\bigg(\frac{\partial^2}{\partial\nu^2}-\frac{1}{\nu}
\frac{\partial}{\partial\nu}\bigg)\right]
\\
&=&\frac{1}{2m}\left[
\frac{\mu}{\sqrt{(a+\hbox{$\frac{b}{2}$}(\mu-\nu))(\mu+\nu)}}p_\mu^2
\frac{\mu}{\sqrt{(a+\hbox{$\frac{b}{2}$}(\mu-\nu))(\mu+\nu)}}\right.
\nonumber\\  &&\qquad\qquad\qquad\qquad\left.
-\frac{\nu}{\sqrt{(a+\hbox{$\frac{b}{2}$}(\mu-\nu))(\mu+\nu)}}p_\nu^2
\frac{\nu}{\sqrt{(a+\hbox{$\frac{b}{2}$}(\mu-\nu))(\mu+\nu)}}\right]
\enspace.\qquad
\end{eqnarray}
Note that from each coordinate there comes a quantum potential
$\Delta V=\hbar^2/8m$, however they are canceling each other due 
to the minus-sign in the metric in $\nu$. The path integral has the form
\begin{eqnarray}
K(\mu'',\mu',\nu'',\nu';T)
&=&\pathint{\mu}\pathint{\nu}
\frac{(a+\hbox{$\frac{b}{2}$}(\mu-\nu))(\mu+\nu)}{\mu\nu}
\nonumber\\  &&\qquad\qquad\qquad\qquad\times
\exp\left[\frac{\i m}{2\hbar}\int_0^T(a+\hbox{$\frac{b}{2}$}(\mu-\nu))(\mu+\nu)
\bigg(\frac{\dot\mu^2}{\mu^2}-\frac{\dot\nu^2}{\nu^2}\bigg)\dt\right]
\nonumber\\  
&=&\int_{-\infty}^\infty\frac{\d E}{2\pi\hbar}\,\e^{-\i ET/\hbar}
\int_0^\infty\d s''K(\mu'',\mu',\nu'',\nu';s'')\enspace,
\end{eqnarray}
with the time-transformation function 
$f(\mu,\nu)=(a+\hbox{$\frac{b}{2}$}(\mu-\nu))(\mu+\nu)$,
and the path integral $K(s'')$ is given by
\begin{eqnarray}
&&K(\mu'',\mu',\nu'',\nu';s'')
=\pathints{\mu}\pathints{\nu}\frac{1}{\mu\nu}
\nonumber\\  &&\qquad\qquad\qquad\times
\exp\left\{\ih\int_0^{s''}\left[\frac{m}{2}
\bigg(\frac{\dot\mu^2}{\mu^2}-\frac{\dot\nu^2}{\nu^2}\bigg)
+aE(\mu+\nu)+\half bE(\mu^2-\nu^2)\right]\d s\right\}\,.\qquad
\end{eqnarray}
Each of the last path integrals has a similar form as the one discussed in
\cite{GROb}. One can perform the transformation $\mu=\e^x$,
$\nu=\e^y$. This gives e.g. in the variable $\mu$ in the short-time element
\begin{equation}
\frac{\i m}{2\epsilon\hbar}\frac{(\Delta \mu^{(j)})^2}{\mu^{(j-1)}\mu^{(j)}}
\simeq \frac{\i m}{2\epsilon\hbar}(\Delta y^{(j)})^2
+\frac{\i m}{24\epsilon\hbar}(\Delta y^{(j)})^4
\,\dot=\, \frac{\i m}{2\epsilon\hbar}(\Delta y^{(j)})^2
-\frac{\i\epsilon\hbar}{8m}\enspace,
\end{equation}
where use has been of the identity 
$(\Delta y^{(j)})^4\dot=3(\frac{\i\epsilon\hbar}{m})^2$, which is,
of course, valid only in the sense of fluctuating paths.
Note that a quantum potential $\Delta V=-\hbar^2/8m$ appears.
However, the same potential arises in the transformation $\nu=\e^y$,
but with the opposite sign, and both contributions cancel.
Therefore the path-integration in ($\mu,\nu$) now gives a
path-integration in ($x,y$) of the following form
\begin{eqnarray}
&&K(x'',x',y'',y';s'')=\pathints{x}\pathints{y}
\nonumber\\  &&\qquad\qquad\qquad\times
\exp\left\{\ih\int_0^{s''}\bigg[\frac{m}{2}(\dot x^2-\dot y^2)
+E(\hbox{$\frac{b}{2}$}\e^{2x}+a\,\e^x)
-E(\hbox{$\frac{b}{2}$}\e^{2y}-a\,\e^y)\bigg]\d s\right\}\,,\qquad
\end{eqnarray}
and we find the product of two path integrals for the Morse potential.
Applying the same techniques as in \cite{GROb} we obtain for the 
Green's  function $G(\mu',\mu',\nu'',\nu';E)$
(with the abbreviations $\omega=\sqrt{-bE/m}$ and rescaling $b\to 2b$):
\begin{eqnarray}
&&G(\mu',\mu',\nu'',\nu';E)
=\frac{\hbar}{2\pi\i}\int\d\CE\bigg(\frac{\sqrt{-2mE}}{\hbar^2}\bigg)^2
\nonumber\\  &&\qquad\times
\int_0^\infty\frac{\d\sigma}{\sin\omega\sigma}
\exp\left[\frac{4a\i E}{\hbar}\sigma
+\frac{\sqrt{-2mE}}{\hbar}({\mu'}^2+{\mu''}^2)\cot\omega\sigma\right]
I_{\sqrt{8m\CE}/\hbar}\left(
\frac{2\sqrt{-2mE}\mu'\mu'}{\i\hbar\sin\omega\sigma}\right)
\nonumber\\  &&\qquad\times
\int_0^\infty\frac{\d\tau}{\sin\omega\tau}
\exp\left[-\frac{4a\i E}{\hbar}\tau
+\frac{\sqrt{-2mE}}{\hbar}({\nu'}^2+{\nu''}^2)\cot\omega\tau\right]
I_{\sqrt{8m\CE}/\hbar}\left(
\frac{2\sqrt{-2mE}\nu'\nu'}{\i\hbar\sin\omega\tau}\right)\enspace.
\nonumber\\  &&
\end{eqnarray}
Note that we have due to the minus-sign in $\nu$ in the metric
an additional minus-sign in $\CE$ in the Green's  function in the
variable $\nu$. Using now the same integral formula form
Ref.\cite{BUCH} as before, we get (simplifying $a=b=1$)
\begin{eqnarray}
&&G(\mu',\mu',\nu'',\nu';E)
\nonumber\\  &&
=\frac{1}{2\pi\i}\frac{4^2}{\hbar^2}\int\d\CE\frac{2mE}{\hbar^2}
\int_0^\infty\d\sigma\int_0^\infty\d\tau
\int_{-\infty}^\infty\d p_1\int_{-\infty}^\infty\d p_2\,
\e^{\i(4E/\hbar+2\i\omega p_1)\sigma-\i(4E/\hbar-2\i\omega p_2)\tau}
\nonumber\\  &&\qquad\times
\frac{\e^{\pi(p_1+p_2)}}{\mu'\mu''\nu'\nu''}
\frac{|\Gamma(\half+\tilde\CE+\i p_1)|^2}{\Gamma^2(1+2\tilde\CE)}
\frac{|\Gamma(\half+\tilde\CE+\i p_2)|^2}{\Gamma^2(1+2\tilde\CE)}
\nonumber\\  &&\qquad\times
M_{ \i p_1,\tilde\CE}\left(-2\frac{\sqrt{-2mE}}{\hbar}{\mu'}^2\right)
M_{-\i p_1,\tilde\CE}\left( 2\frac{\sqrt{-2mE}}{\hbar}{\mu''}^2\right)
\nonumber\\  &&\qquad\times
M_{ \i p_2,\tilde\CE}\left(-2\frac{\sqrt{-2mE}}{\hbar}{\nu'}^2\right)
M_{-\i p_2,\tilde\CE}\left( 2\frac{\sqrt{-2mE}}{\hbar}{\nu''}^2\right)
\end{eqnarray}
($\tilde\CE=\sqrt{8m\CE}/\hbar$).
Performing the $\sigma-$ and $\tau$-integrations gives poles for
$p_1$ and $p_2$ yielding $E=\hbar^2p_1^2/2m=\hbar^2p_2^2/2m$, 
as it should be, the e.g.~$p_2$-integration evaluates the residuum and
we obtain:
\begin{eqnarray}
&&G(\mu',\mu',\nu'',\nu';E)
=\int\frac{\lambda\d\lambda}{\mu'\mu''\nu'\nu''}
\int_0^\infty\frac{\e^{2\pi p}\d p}{\frac{\hbar^2p^2}{2m}-E}
\nonumber\\  &&\qquad\times
\frac{|\Gamma(\half+\lambda+\i p)|^4}{4\pi p^2\Gamma^4(1+2\lambda)}
M_{ \i p,\lambda}\Big(-2\i p{\mu'}^2\Big)
M_{-\i p,\lambda}\Big( 2\i p{\mu''}^2\Big)
M_{ \i p,\lambda}\Big(-2\i p{\nu'}^2\Big)
M_{-\i p,\lambda}\Big( 2\i p{\nu''}^2\Big)\enspace,
\nonumber\\  &&
\end{eqnarray}
which gives the normalized wave-functions:
\begin{equation}
\Psi_{p,\lambda}(\mu,\nu)
=\sqrt{\frac{\lambda}{4\pi\mu\nu}}
\frac{|\Gamma(\half+\lambda+\i p)|^2}{\Gamma^2(1+2\lambda)}
\frac{\e^{\pi p}}{p}
M_{ \i p,\lambda}\Big(-2\i p\mu^2\Big)
M_{ \i p,\lambda}\Big(-2\i p\nu^2\Big)\enspace.
\end{equation}
This concludes the discussion on $\DIII$.


\setcounter{equation}{0}
\section{Darboux Space $\DIV$}
\message{Darboux Space D_IV}
Finally, we consider the Darboux space $\DIV$.  We have the coordinate systems:
\begin{eqnarray}
\hbox{($(u,v)$-Coordinates:)}&& x=v+\i u,\quad y=v-\i u\enspace,
\qquad\qquad\qquad\quad\,\,
(u\in(0,\hbox{$\frac{\pi}{2}$}),v\in\bbbr)\enspace,
\\
\hbox{(Equidistant:)}&& 
u=\arctan(\e^\alpha),\quad v=\frac{\beta}{2}\enspace,
\qquad\qquad\qquad\quad\,\,
(\alpha\in\bbbr,\beta\in\bbbr)\enspace,
\\
\hbox{(Horospherical:)}&& 
x=\log\frac{\mu-\i\nu}{2},\quad 
y=\log\frac{\mu+\i\nu}{2}\enspace,\qquad\qquad
(\mu,\nu>0)\enspace,
\\
\hbox{(Elliptic:)}&& 
\mu=d\cosh\omega\cos\vphi,\quad \nu=d\sinh\omega\sin\vphi\enspace,\quad\,\,\,
(\omega>0,\vphi\in(0,\hbox{$\frac{\pi}{2}$}))\enspace.\qquad
\end{eqnarray}
We obtain the following forms of the line-element ($a>2b$,
$a_\pm=(a\pm 2b)/4$):
\begin{eqnarray}
\d s^2&=&-\frac{b[\e^{(x-y)/2}+\e^{(y-x)/2}]+a}
             {\big(\e^{(x-y)/2}-\e^{(y-x)/2}\big)^2}\,\d x\d y
=-\frac{b[\e^{x-y}+\e^{y-x}]+a}{\big(\e^{x-y}-\e^{y-x}\big)^2}\,\d x\d y
\nonumber\\
\hbox{($(u,v)$-Coordinates:)}&=&
\frac{2b\cos u +a}{4\sin^2u}(\d u^2+\d v^2)
\nonumber\\
&=&
\left(\frac{a_+}{\sin^2u}+\frac{a_-}{\cos^2u}\right)(\d u^2+\d v^2)
\quad\hbox{(rescaling ${u\over2}\to u$, ${v\over2}\to v$:)}\enspace,\quad
\\
\hbox{(Equidistant:)}
&=&\frac{a-2b\tanh\alpha}{4}(\d\alpha^2+\cosh^2\alpha\d\beta^2)\enspace,
\\
\hbox{(Horospherical:)}
&=&\left(\frac{a_+}{\nu^2}+\frac{a_-}{\mu^2}\right)
(\d\mu^2+\d\nu^2)\enspace,
\\
\hbox{(Elliptic:)}
&=&\left(\frac{a_-}{\cosh^2\omega\cos^2\vphi}
        +\frac{a_+}{\sinh^2\omega\sin^2\vphi}\right)
(\cosh^2\omega-\cos^2\vphi)(\d\omega^2+\d\vphi^2)\enspace,
\nonumber\\
&=&\left(\frac{a_+}{\sin^2\vphi}+\frac{a_-}{\cos^2\vphi}
+\frac{a_+}{\sinh^2\omega}-\frac{a_-}{\cosh^2\omega}
\right)(\d\omega^2+\d\vphi^2)\enspace.
\end{eqnarray}
We observe that the diagonal term in the metric corresponds to
a P\"oschl--Teller potential, a Rosen--Morse potential, an
inverse-square radial potential, and a P\"oschl--Teller and modified
P\"oschl--Teller, respectively.  In particular, the $(u,v)$ and the
equidistant systems are the same, they just differ in the parameterization.
The limiting cases $a=2b$ and $b=0$ give particular cases for the metric on 
the two-dimensional hyperboloid.

\begin{table}[h]
\caption{\label{cosytab2} Limiting Cases of Coordinate Systems on $\DIV$}
\begin{eqnarray}\begin{array}{l}\vbox{\small\offinterlineskip
\halign{&\vrule#&$\strut\ \hfil\hbox{#}\hfill\ $\cr
\noalign{\hrule}
height2pt&\omit&&\omit&&\omit&\cr
&Metric:&&$\DIV$          &&$\Lambda^{(2)}$\ ($a=2b$)          
                          &&$\Lambda^{(2)}$\ ($b=0$)             &\cr
height2pt&\omit&&\omit&&\omit&&\omit&&\omit&\cr
\noalign{\hrule}\noalign{\hrule}
height2pt&\omit&&\omit&&\omit&&\omit&&\omit&\cr
&$\dfrac{2b\cos u +a}{4\sin^2u}(\d u^2+\d v^2)$
 &&$(u,v)$-Coordinates    &&Equidistant  &&Equidistant           &\cr
height2pt&\omit&&\omit&&\omit&&\omit&&\omit&\cr
\noalign{\hrule}\noalign{\hrule}
height2pt&\omit&&\omit&&\omit&&\omit&&\omit&\cr
&$\dfrac{a-2b\tanh\alpha}{4}(\d\alpha^2+\cosh^2\alpha\d\beta^2)$
 &&Equidistant            &&Equidistant   &&Equidistant            &\cr
height2pt&\omit&&\omit&&\omit&&\omit&&\omit&\cr
\noalign{\hrule}
height2pt&\omit&&\omit&&\omit&&\omit&&\omit&\cr
&$\bigg(\dfrac{a_-}{\mu^2}+\dfrac{a_+}{\nu^2}\bigg)
(\d\mu^2+\d\nu^2)$
  &&Horospherical         &&Horicyclic  &&Semi-circular parabolic&\cr
height2pt&\omit&&\omit&&\omit&&\omit&&\omit&\cr
\noalign{\hrule}
height2pt&\omit&&\omit&&\omit&&\omit&&\omit&\cr
&$\left(\dfrac{a_-}{\cosh^2\omega\cos^2\vphi}
         +\dfrac{a_+}{\sinh^2\omega\sin^2\vphi}\right)$
 &&                       &&            &&                       &\cr
&$\times(\cosh^2\omega-\cos^2\vphi)(\d\omega^2+\d\vphi^2)$
 &&Elliptic        &&Elliptic-Parabolic    &&Hyperbolic-parabolic&\cr
height2pt&\omit&&\omit&&\omit&&\omit&&\omit&\cr
\noalign{\hrule}}}\end{array}\nonumber\end{eqnarray}
\end{table}

\noindent
For the Gaussian curvature we obtain e.g. in the $(u,v)$-system
\begin{equation}
K=-\dfrac{\frac{a_+^2}{\sin^6u}+\frac{a_-^2}{\cos^6u}
     +\frac{a_-a_+}{\sin^4u\cos^4u}}
    {\bigg(\frac{a_+}{\sin^2u}+\frac{a_-}{\cos^2u}\bigg)^3}
\enspace.
\end{equation}
The case $a=2b$ yields $a_-=0$, and
\begin{equation}
K=-\frac{1}{b}\enspace,
\end{equation}
and therefore again a space of constant curvature, the hyperboloid
$\Lambda^{(2)}$ is given for $b>0$. We have set the sign in the 
metric (1.4) in such a way that from $a=2b>0$ the hyperboloid
$\Lambda^{(2)}$ emerges. We could also choose the  metric (1.4) with
the opposite sign, then $a=2b<0$ would 
give the same result. In the following it is understood that we make
this restriction of positive definiteness of the metric and we do not
dwell into the problem of continuation into non-positive definiteness.
Because the $(u,v)$-coordinates and the equidistant system are the
same, we do not  evaluate the path integral in the equidistant system.
In the following we assume $a_+>0$ and $a_+>a_-$.

\subsection{The Path Integral in $(u,v)$-Coordinates Coordinates on $\DIV$}
\message{The Path Integral in (u,v)-Coordinates Coordinates on D_IV}
The classical Lagrangian and Hamiltonian are given by
\begin{eqnarray}
\CL(u,\dot u,v,\dot v)&=&\frac{m}{2}\frac{2b\cos2u+a}{\sin^22u}
(\dot u^2+\dot v^2)\enspace,
\\
\CH(u,p_u,v,p_v)&=&\frac{1}{2m}\frac{\sin^22u}{2b\cos2u+a}(p_u^2+p_v^2)
\enspace.
\end{eqnarray}
The canonical momentum operators are given by
\begin{equation}
p_u=\hi\bigg(\frac{\partial}{\partial u}+2\cot2u
-\frac{2b\sin2u}{2b\cos2u+a}\bigg),\quad 
p_v=\hi\frac{\partial}{\partial v}\enspace,
\end{equation}
and the Hamiltonian operator has the form
\begin{eqnarray}
H&=&-\frac{\hbar^2}{2m}\frac{\sin^22u}{2b\cos2u+a}
\bigg(\frac{\partial^2}{\partial u^2}
+\frac{\partial^2}{\partial v^2}\bigg)
\\
&=&\frac{1}{2m}\frac{\sin2u}{\sqrt{2b\cos2u+a}}
(p_u^2+p_v^2)\frac{\sin2u}{\sqrt{2b\cos2u+a}}\enspace.
\end{eqnarray}
We obtain for the path integral 
\begin{eqnarray}
&&\!\!\!\!\!\!
K(u'',u',v'',v';T)
\nonumber\\  &&\!\!\!\!\!\!
=\pathint{u}\pathint{v}
\left(\frac{a_+}{\sin^2u}+\frac{a_-}{\cos^2u}\right)
\exp\left[\frac{\i m}{2\hbar}\int_0^T
\left(\frac{a_+}{\sin^2u}+\frac{a_-}{\cos^2u}\right)
(\dot u^2+\dot v^2)\dt\right].
\nonumber\\ 
\label{DIV-uv}
\end{eqnarray}
This formulation in $(u,v)$-coordinates is unconvenient. Without the term 
$a_+/\sin^2u$ (\ref{DIV-uv}) would be identical with the path integral
in  the hyperbolic strip \cite{GROf}, which is actually a reformulation of 
equidistant coordinates on the two-dimensional hyperboloid. Following 
\cite{GROf} we perform the coordinate transformation $\cos u=\tanh\tau$. 
Further, we separate off the $v$-path integration, and additionally we 
make a time-transformation with the time-transformation function 
$f={a_+}/{\sin^2u}+{a_-}/{\cos^2u}$. Due to the coordinate transformation 
$\cos u=\tanh\tau$ additional quantum terms appear according to 
\begin{equation}
\exp\left({\i m\over2\epsilon\hbar}
{\big(\Delta u^{(j)}\big)^2\over\cos u^{(j-1)}\cos u^{(j)}}\right)
\dot=
\exp\left[{\i m\over2\epsilon\hbar}\big(\Delta\tau^{(j)}\big)^2
-\i\frac{\hbar}{8m}\left(1+{1\over\cosh^2\tau^{(j)}}\right)\right]\enspace.
\end{equation}
We get for the path integral (\ref{DIV-uv})
\begin{equation}
K(u'',u',v'',v';T)
=\int_{-\infty}^\infty\frac{\d E}{2\pi\hbar}\,\e^{-\i ET/\hbar}
\int_0^\infty\d s''\exp\bigg[\ih\bigg(a_+E-\frac{\hbar^2}{8m}\bigg)s''\bigg]
K(\tau'',\tau',v'',v';s'')\enspace,
\end{equation}
and the time-transformed path integral $K(s'')$ is given by
\begin{eqnarray}
&&K(\tau'',\tau',v'',v';s'')
=\int_{-\infty}^\infty\d k_v\frac{\e^{\i k_v(v''-v')}}{2\pi}
 (\cosh\tau'\cosh\tau'')^{-1/2}
\nonumber\\   &&\qquad\times
\pathints{\tau}\exp\left[\ih\int_0^{s''}\left(\frac{m}{2}\dot\tau^2
+\frac{a_-E}{\sinh^2\tau}-\frac{\hbar^2}{2m}\frac{k_v^2+\viert}{\cosh^2\tau}
\right)\d s\right]\enspace.\qquad\qquad\qquad\qquad
\label{DIV-uv-tau}
\end{eqnarray}
The special case $a_-=0$ gives the wave-functions on the
two-dimensional hyperboloid in equidistant coordinates, respectively
on the hyperbolic strip. Inserting the solution for the modified
P\"oschl--Teller potential and evaluating the Green's function on the cut
yields for the path integral solution on $\DIV$ as follows
($K(u'',u',v'',v';T)=K(\tau'',\tau',v'',v';T)$):
\begin{eqnarray}
K(u'',u',v'',v';T)&=&
\int_{-\infty}^\infty\d k_v\int_0^\infty \d p\,
\e^{-\i TE_p/\hbar} \Psi_{p,k_v}(\tau'',v'')\Psi_{p,k_v}^*(\tau',v')\enspace,
\qquad\qquad\\   
\Psi_{p,k_v}(\tau,v)&=&
\frac{\e^{\i k_vv}}{\sqrt{2\pi a_+\cosh\tau}}\,\Psi_p^{(\eta,\i k)}(\tau)
\enspace,
         \\   
E_p&=&\frac{\hbar^2}{2ma_+}\bigg(p^2+\viert\bigg)\enspace,
\end{eqnarray}
where $\eta^2=\viert-2ma_eE/\hbar^2$ and the wave-functions for the
modified P\"oschl--Teller functions as given in the Appendix B.
Re-inserting $\cos u=\tanh\tau$ gives the solution in terms of the
variable~$u$. 

\subsection{The Path Integral in Horospherical Coordinates on $\DIV$}
\message{The Path Integral in Horospherical  Coordinates on D_IV}
The classical Lagrangian and Hamiltonian are given by
\begin{eqnarray}
\CL(\mu,\dot\mu,\nu,\dot\nu)&=&
\frac{m}{2}\bigg(\frac{a_+}{\nu^2}+\frac{a_-}{\mu^2}\bigg)
(\dot\mu^2+\dot\nu^2)\enspace,
\\
\CH(\mu,p_\mu,\nu,p_\nu)&=&
\frac{1}{2m}\frac{\mu^2\nu^2(p_\mu^2+p_\nu^2)}{a_+\mu^2+a_-\nu^2}
\enspace.
\end{eqnarray}
For the canonical momentum operators we have
\begin{eqnarray}
p_\mu&=&\hi\bigg(\frac{\partial}{\partial\mu}
-\frac{\nu^2a_-/\mu}{a_+\mu^2+a_-\nu^2}\bigg)\enspace,
\\
p_\nu&=&\hi\bigg(\frac{\partial}{\partial\nu}
-\frac{\mu^2a_+/\nu}{a_+\mu^2+a_-\nu^2}\bigg)\enspace,
\end{eqnarray}
and for the quantum Hamiltonian we get
\begin{eqnarray}
H&=&-\frac{\hbar^2}{2m}\frac{\mu^2\nu^2}{a_+\mu^2+a_-\nu^2}
\bigg(\frac{\partial^2}{\partial\mu^2}+
\frac{\partial^2}{\partial\nu^2}\bigg)
\\
&=&\frac{1}{2m}\sqrt{\frac{\mu^2\nu^2}{a_+\mu^2+a_-\nu^2}}\,
(p_\mu^2+p_\nu^2)\,\sqrt{\frac{\mu^2\nu^2}{a_+\mu^2+a_-\nu^2}}\enspace.
\end{eqnarray}
For the path integral we obtain (time-transformation function
$f(\mu,\nu)=a_+/\mu^2+a_-/\nu^2=\sqrt{g}$)
\begin{eqnarray}
K(\mu'',\mu',\nu'',\nu';T)
&=&\pathint{\mu}\pathint{\nu}
\bigg(\frac{a_+}{\nu^2}+\frac{a_-}{\mu^2}\bigg)
\nonumber\\  &&\qquad\times
\exp\left[\frac{\i m}{2\hbar}\int_0^T
\bigg(\frac{a_+}{\nu^2}+\frac{a_-}{\mu^2}\bigg)
(\dot\mu^2+\dot\nu^2)\dt\right]\qquad\qquad
\nonumber\\  &=&
\int_{-\infty}^\infty\frac{\d E}{2\pi\hbar}\,\e^{-\i ET/\hbar}
\int_0^\infty\d s''K(\mu'',\mu',\nu'',\nu';s'')\enspace,
\end{eqnarray}
and the time-transformed path integral $K(s'')$ is given by
\begin{eqnarray}
&&K(\mu'',\mu',\nu'',\nu';s'')
\nonumber\\   &&
=\pathints{\mu}\pathints{\nu}
\exp\left\{\ih\int_0^{s''}\bigg[\frac{m}{2}(\dot\mu^2+\dot\nu^2)
+E\bigg(\frac{a_+}{\nu^2}+\frac{a_-}{\mu^2}\bigg)\bigg]
\d s\right\}\enspace.
\nonumber\\   &&
\end{eqnarray}
These two path integrals can be solved by means of the solution for the
inverse-square radial potential, by following the approach of
\cite{GROad} for the semi-circular-parabolic system on the
two-dimensional hyperboloid. We insert the path integral solution for
the inverse-square radial potential (kernel and Green's function) and obtain 
\begin{eqnarray}
&&G(\mu'',\mu',\nu'',\nu';E)=\int_0^\infty\d s''K(\mu'',\mu',\nu'',\nu';s'')
\nonumber\\   &&
=\frac{4m^2}{\hbar^3}\sqrt{\mu'\mu''\nu'\nu''}\int\frac{\d\CE}{2\pi\i}
I_{\lambda}\bigg(\sqrt{2m\CE}\frac{\nu_<}{\hbar}\bigg)
K_{\lambda}\bigg(\sqrt{2m\CE}\frac{\nu_>}{\hbar}\bigg)
I_{\tilde\lambda}\bigg(\sqrt{-2m\CE}\frac{\mu_<}{\hbar}\bigg)
K_{\tilde\lambda}\bigg(\sqrt{-2m\CE}\frac{\mu_>}{\hbar}\bigg)
\nonumber\\   &&
=\frac{m^2}{\hbar^3}\sqrt{\mu'\mu''\nu'\nu''}
\int_0^\infty\frac{\d s''}{s''}\int\frac{\d\CE}{2\pi\i}\,\e^{\i\CE s''/\hbar}
\nonumber\\   &&\qquad\times
\exp\bigg[\frac{m}{2\i\hbar s''}({\nu'}^2+{\nu''}^2)\bigg]
I_\lambda\bigg(\frac{\i m\nu'\nu''}{\hbar s''}\bigg)
I_{\tilde\lambda}\bigg(\sqrt{-2m\CE}\frac{\mu_<}{\hbar}\bigg)
K_{\tilde\lambda}\bigg(\sqrt{-2m\CE}\frac{\mu_>}{\hbar}\bigg)\,.
\end{eqnarray}
I have used the abbreviations (assuming $a_+>a_-$)
\begin{equation}
 \lambda^2=\viert-\frac{2mE}{\hbar^2}a_+,\quad
\tilde\lambda^2=\viert-\frac{2mE}{\hbar^2}a_-\enspace,\quad
\tilde p^2=\frac{a_-}{a_+}\bigg(p^2+\viert\bigg)-\viert\enspace.
\end{equation}
For the $\mu$-dependent part one uses the dispersion relation 
(\ref{DispersionIlambda}) together with the integral representation
\cite[p.725]{GRA} 
\begin{equation}
  \int_0^\infty\e^{-x/2-(z^2+w^2)/2x}
  K_\nu\bigg({zw\over x}\bigg){dx\over x}=2K_\nu(z)K_\nu(w)\enspace.
\end{equation}
In order to analyze the $\nu$-dependent part, we first rewrite the
$\nu$-dependent part of the Green's  function according to
\begin{equation}
   I_\lambda(-\i k\mu_<)K_\lambda(-\i k\mu_>)
   ={\i\pi\over2}J_\lambda(k\mu_<)H_\lambda^{(1)}(k\mu_>)\enspace.
\end{equation}
and then the wavefunctions on the cut are then obtained by using
\begin{eqnarray}
   \Psi_{k,p}(\mu'')\Psi_{k,p}^*(\mu')&\propto&\Big[
   J_{-\i p}(k\mu'')H_{-\i p}^{(1)}(k\mu')
   -J_{\i p}(k\mu'')H_{\i p}^{(1)}(k\mu')\Big]
        \nonumber\\
  &=&\sinh\pi p H_{\i p}^{(1)}(k\mu'')H_{-\i p}^{(1)}(k\mu')\enspace,
\end{eqnarray}
($\lambda=-\i p$) and the relation of the Hankel-function, i.e., $H_
\nu^{(1)}(z)=\i\big[ \e^{-\i\nu\pi}J_\nu(z)-J_{-\nu}(z)\big]/$ $\sin\pi
\nu$.  Therefore with $\CE=\hbar^2\kappa/2m$:
\begin{eqnarray}
&&G(\mu'',\mu',\nu'',\nu';E)
=\frac{\sqrt{\mu'\mu''\nu'\nu''}}{2\pi^2}
\nonumber\\   &&\qquad\times\left\{
\int\frac{\d\kappa}{2\pi\i}
I_{\tilde\lambda}\big(-\i\sqrt{\kappa}\,\mu_<\big)
K_{\tilde\lambda}\big(-\i\sqrt{\kappa}\,\mu_>\big)
\int_0^\infty\frac{\d p\,p\sinh\pi p}{\frac{\hbar^2}{2ma_+}(p^2+\viert)-E}
K_{\i p}\big(\sqrt{\kappa}\,\nu'\big)
K_{\i p}\big(\sqrt{\kappa}\,\nu'\big)\right.
\nonumber\\   &&\qquad\qquad\qquad\qquad
+(\mu\leftrightarrow\nu)\Bigg\}
\nonumber\\   &&
=\frac{1}{8\pi^2}\sqrt{\mu'\mu''\nu'\nu''}
\int\d\kappa\int_0^\infty\frac{\d p\,p\sinh\pi p\sinh\pi\tilde p}
{\frac{\hbar^2}{2ma_+}(p^2+\viert)-E}
\nonumber\\   &&\qquad\times
\bigg[K_{\i p}\big(\sqrt{\kappa}\,\nu'\big)
K_{\i p}\big(\sqrt{\kappa}\,\nu'\big)
H_{-\i\tilde p}^{(1)}\big(\sqrt{\kappa}\,\mu'\big)
H_{-\i\tilde p}^{(1)}\big(\sqrt{\kappa}\,\mu''\big)
+(\mu\leftrightarrow\nu)\bigg]\enspace.
\end{eqnarray}
I have taken into account that the final result must be
symmetrical in $\mu$ and $\nu$ which also accounts for the additional
factor $\bhalf$. The wave-functions thus have the form
\begin{equation}
\Psi_{p,\kappa}(\mu,\nu)=
\frac{\sqrt{\mu\nu}}{2\sqrt{2}\,\pi}\sqrt{p\sinh\pi p\sinh\pi\tilde p}
\Big[K_{\i p}\big(\sqrt{\kappa}\,\nu\big)
H_{-\i\tilde p}^{(1)}\big(\sqrt{\kappa}\,\mu\big)
+(\mu\leftrightarrow\nu)\Big]\enspace.
\end{equation}
Note that $a_-=0$ gives the horicyclic path integral on the two-dimensional 
hyperboloid. Then $\tilde\lambda=\pm\half$, and the corresponding 
Bessel-functions give exponentials, therefore we obtain the wave-functions 
of the horicyclic system. This result completes the calculation.

\subsection{The Path Integral in Elliptic Coordinates on $\DIV$}
\message{The Path Integral in Elliptic Coordinates on D_IV}
The classical Lagrangian and Hamiltonian are given by
\begin{eqnarray}
\CL(\omega,\dot\omega,\vphi,\dot\vphi)&=&
\frac{m}{2}\left(\frac{a_-}{\cosh^2\omega\cos^2\vphi}
         +\frac{a_+}{\sinh^2\omega\sin^2\vphi}\right)
(\cosh^2\omega-\cos^2\vphi)(\dot\omega^2+\dot\vphi^2)\enspace,
\nonumber\\
\\
\CH(\omega,p_\omega,\vphi,p_\vphi)&=&
\frac{1}{2m}\left(\frac{a_-}{\cosh^2\omega\cos^2\vphi}
         +\frac{a_+}{\sinh^2\omega\sin^2\vphi}\right)^{-1}
\frac{p_\omega^2+p_\vphi^2}{\cosh^2\omega-\cos^2\vphi}\enspace.
\end{eqnarray}
For the canonical momentum operators we get
\begin{eqnarray} 
p_\omega&=&\hi\bigg[\frac{\partial}{\partial\omega}
+g^{-1/2}
\bigg(\frac{a_-\tanh\omega}{\sinh^2\omega}-
      \frac{a_+\coth\omega}{\cosh^2\omega}\bigg)\bigg]
\\
p_\vphi&=&\hi\bigg[\frac{\partial}{\partial\vphi}
+g^{-1/2}\bigg(\frac{a_-\tan\vphi}{\sin^2\vphi}
              -\frac{a_+\cot\vphi}{\cos^2\vphi}\bigg)\bigg]\enspace.
\end{eqnarray}
the Hamiltonian operator is given by
\begin{eqnarray} 
H&=&
-\frac{\hbar^2}{2m}\left(\frac{a_-}{\cosh^2\omega\cos^2\vphi}
         +\frac{a_+}{\sinh^2\omega\sin^2\vphi}\right)^{-1}
\frac{1}{\cosh^2\omega-\cos^2\vphi}
\bigg(\frac{\partial^2}{\partial\omega^2}
     +\frac{\partial^2}{\partial\vphi^2}\bigg)
\\ &=&
\frac{1}{2m}\frac{1}{\sqrt[4]{g}}(p_\omega^2+p_\vphi^2)
\frac{1}{\sqrt[4]{g}}\enspace.
\end{eqnarray} 
For the path integral in elliptic coordinates we obtain
(note that for $a_-=0$ we get the path integral in
elliptic-parabolic coordinates on the two-dimensional hyperboloid)
\begin{eqnarray} 
&&K(\omega'',\omega',\vphi'',\vphi';T)
=\pathint{\omega}\pathint{\vphi}\sqrt{g}
\nonumber\\   &&\qquad\times
\exp\left[\frac{\i m}{2\hbar}\int_0^T
\left(\frac{a_+}{\sin^2\vphi}+\frac{a_-}{\cos^2\vphi}
+\frac{a_+}{\sinh^2\omega}-\frac{a_-}{\cosh^2\omega}
\right)(\dot\omega^2+\dot\vphi^2)\dt\right]\enspace.
\label{pathintegralDIV-elliptic}
\end{eqnarray}
In order to obtain a convenient form to evaluate 
(\ref{pathintegralDIV-elliptic}) we perform the coordinate transformation
$\cos\vphi=\tanh\tau$ in the same way as in (\ref{DIV-uv}).
Performing the time-transformation with $f(\omega,\vphi)=\sqrt{g}$,
the time-transformed path integral $K(s'')$ is given by
\begin{equation} 
K(\omega'',\omega',\tau'',\tau';T)
=\int_{-\infty}^\infty\frac{\d E}{2\pi\hbar}\,\e^{-\i ET/\hbar}
\int_0^\infty\d s''
\exp\bigg[\ih\bigg(a_+E-\frac{\hbar^2}{8m}\bigg)\bigg]
K(\omega'',\omega',\tau'',\tau';s'')\enspace,
\end{equation}
and the time-transformed path integral $K(s'')$ is given by
\begin{eqnarray}
&&\!\!\!\!\!\!\!\!
K(\omega'',\omega',\tau'',\tau';s'')
=\pathints{\omega}\pathints{\tau}\cosh\tau
\nonumber\\   &&\!\!\!\!\!\!\!\!\quad\times
\exp\left\{\ih\int_0^{s''}\left[\frac{m}{2}(\dot\tau^2+\cosh^2\tau\dot\omega^2)
+\frac{a_+E}{\sinh^2\tau}+\frac{1}{\cosh^2\tau}\bigg(
 \frac{a_+E}{\sinh^2\omega}-\frac{a_-E}{\cosh^2\omega}-\frac{\hbar^2}{8m}\bigg)
\right]\d s\right\}.\qquad
\end{eqnarray} 
The $\omega$-path integration is separated by means of a modified
P\"oschl--Teller potential with $\eta^2=\viert+2mEa_+/\hbar^2$,
$\nu^2=\viert+2mEa_-/\hbar^2$ and the $\tau$-path integration is the
same as in (\ref{DIV-uv-tau}). This gives the solution:
\begin{eqnarray}
K(\omega'',\omega',\tau'',\tau';T)
&=&\int_{-\infty}^\infty\d k_v\int_0^\infty \d p\,\e^{-\i TE_p/\hbar} 
\Psi_{p,k}^*(\omega',\tau')\Psi_{p,k}(\omega'',\tau'')\enspace,\qquad
         \\   
\Psi_{p,k}(\omega,\tau)&=&(\cosh\tau)^{-1/2}\,
\Psi_k^{(\eta,\nu)}(\omega)\Psi_p^{(\eta,\i k)}(\tau)\enspace,
         \\   
E_p&=&\frac{\hbar^2}{2ma_+}\bigg(p^2+\viert\bigg)\enspace,
\end{eqnarray}
and we can re-insert $\cos\vphi=\tanh\tau$. 
This concludes the discussion on the Darboux space $\DIV$. 

\newpage\noindent%
\setcounter{equation}{0}%
\section{Summary and Discussion}
\message{Summary and Discussion}
In this paper I have discussed path integration on Darboux Spaces, labeled
by $\DI$ to $\DIV$. We set up the metrics following Kalnins et al.
\cite{KalninsKMWinter,KalninsKWinter}. In each of these spaces the 
Schr\"odinger equation, respectively the path integral were separable in
several coordinate systems. Our results are  summarized in Table
\ref{solutions}. 

In the Darboux space $\DI$ we found the solutions in the $(u,v)$- and rotated
$(u,v)$-coordinates. A closed expression for the Green's functions
could be found, however, the wave-function are only implicitly known
because of the boundary conditions which must be imposed on the
system. A solution in displaced parabolic coordinates was not possible 
due to its quartic anharmonic structure of the transformed dynamics. 

In the Darboux space $\DII$ I succeeded in writing down the Green
functions and the corresponding expansions into the wave-functions. I
found the expressions in the four coordinate systems, i.e. 
$(u,v)$-coordinates, polar, parabolic and elliptic coordinates. 
Several path integral techniques from former studies were
indispensable tools in the considerations. We stressed the limiting
case of the hyperbolic plane, i.e. the two-dimensional hyperboloid. 
The Green's function and the wave-functions were determined in 
the soluble systems with the general feature that the energy-spectrum 
has the form: $E=\frac{\hbar^2p^2}{2m|a|}(p^2+\viert)$. The additional 
zero-point energy $E_0=\frac{\hbar^2p^2}{8m|a|}$ is a characteristic 
feature the quantum motion on spaces with negative curvature~\cite{GRSc}. 

In the Darboux space $\DIII$ I found the solutions in the $(u,v)$-system, 
the closely related polar system, the $(u,v)$-system, and the hyperbolic
coordinate system. In elliptic coordinates no solution could be found.
The Green's function and the wave-functions were determined in the 
soluble systems with the general feature that the energy-spectrum has
the form: $E=\frac{\hbar^2p^2}{2m}$, which is different in its zero-point
valued form $\DII$.

In the Darboux space $\DIV$ we found solutions in $(u,v)$-coordinates, 
horospherical, and elliptic coordinates. Here also the
limiting case to the two-dimensional hyperboloid was shortly mentioned. 
The Green's functions  and the wave-functions were calculated in the
separable coordinate system. The energy-spectrum has
the form: $E=\frac{\hbar^2p^2}{2ma_+}(p^2+\viert)$, similarly as on $\DII$.

\begin{table}[h]
\caption{\label{solutions} Solutions of the path integration in
  Darboux spaces}
\begin{eqnarray}\begin{array}{l}\vbox{\small\offinterlineskip
\halign{&\vrule#&$\strut\ \hfil\hbox{#}\hfill\ $\cr
\noalign{\hrule}
height2pt&\omit&&\omit&\cr
&Space and            &&Solution in terms of the wave-functions  &\cr
&Coordinate System    &&                                         &\cr
height2pt&\omit&&\omit&\cr
\noalign{\hrule}\noalign{\hrule}
height2pt&\omit&&\omit&\cr
&$\DI$                &&          &\cr
height2pt&\omit&&\omit&\cr
\noalign{\hrule}
height2pt&\omit&&\omit&\cr
&$(u,v)$-Coordinates         &&Product of Airy functions      &\cr
&Rotated $(u,v)$-Coordinates &&Product of Airy functions      &\cr
&Displaced parabolic  &&no solution                           &\cr
height2pt&\omit&&\omit&\cr
\noalign{\hrule}\noalign{\hrule}
height2pt&\omit&&\omit&\cr
&$\DII$               &&          &\cr
height2pt&\omit&&\omit&\cr
\noalign{\hrule}
height2pt&\omit&&\omit&\cr
&$(u,v)$-Coordinates  &&Exponential times K-Bessel function   &\cr
&Polar                &&Legendre times K-Bessel function      &\cr
&Parabolic            &&Product of W-Whittaker functions      &\cr
&Elliptic             &&Spheroidal wave-functions             &\cr
height2pt&\omit&&\omit&\cr
\noalign{\hrule}\noalign{\hrule}
height2pt&\omit&&\omit&\cr
&$\DIII$              &&          &\cr
height2pt&\omit&&\omit&\cr
\noalign{\hrule}
height2pt&\omit&&\omit&\cr
&$(u,v)$-Coordinates  &&Exponential times M-Whittaker functions    &\cr
&Polar                &&Exponential times M-Whittaker functions    &\cr
&Parabolic            &&Product of parabolic cylinder functions    &\cr
&Elliptic             &&No solution                                &\cr
&Hyperbolic           &&Product of M-Whittaker functions           &\cr
height2pt&\omit&&\omit&\cr
\noalign{\hrule}\noalign{\hrule}
height2pt&\omit&&\omit&\cr
&$\DIV$               &&          &\cr
height2pt&\omit&&\omit&\cr
\noalign{\hrule}
height2pt&\omit&&\omit&\cr
&$(u,v)$-Coordinates  &&Exponential times Legendre function          &\cr
&Equidistant          &&Exponential times Legendre function          &\cr
&Horospherical        &&Product of K- and $H^{(1)}$-Bessel functions &\cr
&Elliptic             &&Product of Legendre functions                &\cr
height2pt&\omit&&\omit&\cr
\noalign{\hrule}}}\end{array}\nonumber\end{eqnarray}
\end{table}

We were able to solve the various path integral representations, because we
have now to our disposal not only the basic path integrals for the harmonic 
oscillator, the linear oscillator, the radial harmonic oscillator, and
the modified P\"oschl--Teller Potential, but also path integral
identities derived from path integration on harmonic spaces like the
elliptic and spheroidal path integral representations with its more 
complicated special functions~\cite{GROad,GKPSa,GRSh}. 
This includes also numerous transformation
techniques to find a particular solution based on one of the basic solutions.
Various Green's  function analysis techniques can be applied to find not
only an expression for the Green's  function but also for the
wave-functions and the energy spectrum. 

The present study continues the analysis of path integrals on curved space
\cite{GROad} with the simple case of the two-dimensional Euclidean space with
its four separating coordinate systems (Cartesian, polar, elliptic and
parabolic) up to the complicated case of the three-dimensional hyperboloid
with its 34 separating coordinate systems.

In our papers \cite{GROPOa}--\cite{GROPOd} we have studied super-integrable
potentials on spaces of constant curvature, i.e. flat space \cite{GROPOa},
spheres \cite{GROPOb} and two and three-dimensional hyperboloids 
\cite{GROPOa,GROPOb}. In the Euclidean flat spaces $\bbbr^2$ and $\bbbr^3$ a
complete list was given, and were called Smorodinski--Winternitz
potentials \cite{WSUF}. We have extended this study by introducing
corresponding potentials on spaces with (non-zero) constant curvature,
i.e. on spheres and hyperboloids. Further studies along these lines for
superintegrability on spaces with constant curvature were given by
Kalnins et al. \cite{MKP1} on the complex $2$-sphere (five coordinate
systems which separate the Laplace-Beltrami equation), on the complex
Euclidean space $E_{2,C}$ \cite{MKP2} (six coordinate
systems which separate the Laplace-Beltrami equation), 
in $\bbbr^2$ and on the two-dimensional sphere (with emphasis on the
polynomial solutions of the superintegrable potentials)
\cite{MKP3}, and on the two-dimensional hyperboloid \cite{MKP4,MKP5}.
In the latter also two potentials were studied which have until-then not
been considered. The focus in those studies were of course on the
harmonic oscillator (with its deformations and generalizations) and the Coulomb
potential. As it turns, out in all those space an harmonic oscillator and a
Coulomb potential could be defined and solved in various coordinate
representations. In particular, the three-dimensional Coulomb potential
problem separates in spherical, conical, parabolic, and prolate spheroidal
coordinates. These feature translated also into the three-dimensional sphere
and the three-dimensional hyperboloid. This particular feature of the
Coulomb system has its origin in its super-integrability, i.e., beside the
energy and the angular momentum conservation we have an additional conserved
quantity, the Lenz--Runge vector.  

\newpage
Therefore we have explicitly shown that path integral calculations are
not only possible in flat space (with several potential problems), or
in spaces with non-vanishing constant curvature, but is also applicable
in spaces of non-constant curvature.

The most serious drawback of the path integral method in comparison to
the operator method is that in the operator method we can investigate
the more complicated parametric coordinate systems in terms of
Lam\'e-polynomials. For this kind of coordinate systems a path
integral approach exists for the spheroidal wave-functions (elliptic
wave-functions in $\bbbr^2$ and spheroidal wave-functions in $\bbbr^3$) 
based on the theory of Meixner and Sch\"afke \cite{MESCH}, and for the
three-dimensional sphere \cite{GKPSa}. 

It is therefore quite naturally the raise the question of super-integrable
system in Darboux spaces, in fact one of the intentions of  
\cite{KalninsKMWinter,KalninsKWinter}. And indeed, analogies of an oscillator
and a Coulomb potential can be found.  
However, the freedom of choice of free parameters seems somewhat limited in
comparison with the spaces of constant curvature. This can be understood by
the feature of the Darboux spaces that the corresponding metric includes
already a complicated ``potential''-term, i.e. the metric almost
equals a superintegrable potential in $\bbbr^2$. The time-transformation 
function is in almost all cases equals to $\sqrt{g}$ which is most obvious in 
the case of $\DIV$. These issues will
be discussed in more detail in a future publication \cite{GROPOe}.

\subsection*{Acknowledgments}
I would like to thank George Pogosyan (JINR Dubna) for helpful discussions on 
the properties of coordinate systems and superintegrability, and a
critical reading of the manuscript.


\begin{appendix}
\section{Formulation of the Path Integral in Curved Spaces}
\message{Formulation of the Path Integral in Curved Spaces}
In order to set up our notation for path integrals on curved manifolds
we proceed in a canonical way. To
avoid unnecessary overlap with our Table of Path Integrals \cite{GRSh}
I give in the following only the essential information required for the
path integral representation on curved spaces. For more details
concerning ordering prescriptions, transformation techniques, pertur%
bation expansions, point interactions, and boundary conditions I refer
to \cite{GRSh}, where also listings of the application of Basic Path
Integrals will be presented. In the following $\vec q$ denote some 
D-dimensional coordinates. We start
by considering the classical Lagrangian corresponding to the line
element $\d s^2=g_{ab}\d q^a\d q^b$ of the classical motion in some
$D$-dimensional Riemannian space
\begin{equation}
   \CL_{Cl}(\vec q,\dot{\vec q})
   ={m\over2}\bigg({\d s\over\dt}\bigg)^2-V(\vec q)
   ={m\over2}g_{ab}(\vec q)\dot q^a\dot q^b-V(\vec q)\enspace.
\end{equation}
The quantum Hamiltonian is {\em constructed} by means of the
Laplace-Beltrami operator
\begin{equation}
   H=-\hbarm\Delta_{LB}+V(\vec q)
   =-\hbarm{1\over\sqrt{g}}{\partial\over\partial q^a}g^{ab}\sqrt{g}
     {\partial\over\partial q^b}+V(\vec q)
\label{NUMAk}
\end{equation}
as a {\em definition} of the quantum theory on a curved space. 
Here are $g=\det{(g_{ab})}$ and $(g^{ab})=(g_{ab})^{-1}$.
The scalar product for wavefunctions on the manifold reads $(f,g)=\int
\d\vec q\sqrt{g}f^*(\vec q)g(\vec q)$, and the momentum operators which
are hermitian with respect to this scalar product are given by
\begin{equation}
p_a=\hi\bigg({\partial\over\partial q^a}+{\Gamma_a\over2}\bigg)\enspace,
  \qquad\Gamma_a={\partial\ln\sqrt{g}\over\partial q^a}\enspace.
\label{NUMAb}
\end{equation}
In terms of the momentum operators (\ref{NUMAb}) we can rewrite $\b H$
by using a product according to $g_{ab}=h_{ac}h_{cb}$ \cite{GRSh}. Then
we obtain for the Hamiltonian (\ref{NUMAk}) (PF - {\em P}roduct-{\em
F}orm)
\begin{equation}
  \b H=-\hbarm\Delta_{LB}+V(\vec q)
  ={1\over2m}h^{ac}p_ap_bh^{cb}+\Delta V_{PF}(\vec q)+V(\vec q)\enspace,
\end{equation}
and for the path integral
\begin{eqnarray}       & &
  K(\vec q'',\vec q';T)
         \nonumber\\   & &
   =\pathintG{\vec q}{PF}\sqrt{g(\vec q)}\exp\bigg\{\ih\intt
     \bigg[{m\over2}h_{ac}(\vec q)h_{cb}(\vec q)\dot q^a\dot q^b
   -V(\vec q)-\Delta V_{PF}(\vec q)\bigg]\dt\bigg\}
         \nonumber\\   & &
  =\limN\Norm^{ND/2}\prod_{k=1}^{N-1}\int\d\vec q_k\sqrt{g(\vec q_k)}
         \nonumber\\   & &\qquad \times
  \exp\Bigg\{\ih\sum_{j=1}^N\bigg[{m\over2\epsilon}
  h_{bc}(\vec q_j)h_{ac}(\vec q_{j-1})\Delta q_j^a\Delta q_j^b
  -\epsilon V(\vec q_j)-\epsilon\Delta V_{PF}(\vec q_j)\bigg]\Bigg\}
  \enspace.
\label{NUMAa}
\end{eqnarray}
$\Delta V_{PF}$ denotes the well-defined quantum potential
\begin{equation}
  \Delta V_{PF}(\vec q)=\hbaram
  \Big[g^{ab}\Gamma_a\Gamma_b+2(g^{ab}\Gamma_b)_{,b}+{g^{ab}}_{,ab}\Big]
  +\hbaram\Big(2h^{ac}{h^{bc}}_{,ab}-{h^{ac}}_{,a}{h^{bc}}_{,b}
                -{h^{ac}}_{,b}{h^{bc}}_{,a}\Big)
\end{equation}
arising from the specific lattice formulation (\ref{NUMAa}) of the path
integral or the ordering prescription for position and momentum
operators in the quantum Hamiltonian, respectively. We have used
the abbreviations $\epsilon=(t''-t')/N\equiv T/N$, $\Delta\vec q_j=\vec
q_j-\vec q_{j-1}$, $\vec q_j=\vec q(t'+j\epsilon)$ $(t_j=t'+\epsilon j,
j=0,\dots,N)$ and we interpret the limit $N\to \infty$ as equivalent to
$\epsilon\to0$, $T$ fixed. The lattice representation can be obtained
by exploiting the composition law of the time-evolution operator $U=\exp
(-\i HT/\hbar)$, respectively its semi-group property. 

Note that the first summand on $\Delta V$ corresponds to the quantum potential
of the Weyl-ordered Hamiltonian, respectively a mid-point prescription of the
path integral. Note also that in the case that the metric tensor is diagonal
to the unit tensor, i.e. $(g_{ab}=f^2\delta_{ab})$ we obtain
\begin{equation}
\Delta V((\vec q)=\hbar^2\frac{D-2}{8m}\sum_a\frac{(4-D)f_{,a}^2+2f\cdot
  f_{,aa}}{f^4}\enspace.
\label{DeltaVPF}
\end{equation}
This gives the important special case that for $D=2$: $\Delta V=0$, a property
which is quite useful for the considered two-dimensional Darboux spaces.

The path integral representation (\ref{NUMAa}) is not explicitly 
evaluable in many cases, in particular if explicitly coordinate-dependent
metric terms are present, or potentials like the Coulomb
potential.  Here the so-called ``time transformation'' comes into
play which leads in combination with  ``coordinate transformation''  to
general ``space-time transformations'' (also ``Duru--Kleinert transformation'' 
\cite{DKa,DKb,KLEo} in path integrals. The time
transformation is implemented \cite{KLEo} 
by introducing a new ``pseudo-time'' $s''$. In order to do
this, one first makes use  of the operator identity
(one-dimensional case)
\begin{equation}
 {1\over H-E}=f_r(x,t){1\over f_l(x,t)(H-E)f_r(x,t)}f_l(x,t)\enspace,
\end{equation}
where $H$ is the Hamiltonian corresponding to the path integral $K(t'',
t')$, and $f_{l,r}(x,t)$ are functions in $q$ and $t$, multiplying from
the left or from the right, respectively, onto the operator $(H-E)^{-1}
$. Secondly, one introduces a new pseudo-time $s''$ and assumes that the
constraint
\begin{equation}
\ints ds f_l\big(F(q(s),s)\big)\cdot f_r\big(F(q(s),s)\big)=T=t''-t'
\end{equation}
has for all admissible paths a unique solution $s''>0$ given by
\begin{equation}
  s''=\intt{dt\over f_l(x,t)f_r(x,t)}=\intt{ds\over{F'}^2(q(s),s)}
  \enspace.
\end{equation}
Here one has made the choice $f_l\big(F(q(s),s)\big)=f_r\big(F(q(s),s)
\big)=F'\big(q(s),s)\big)$ in order that in the final result the
metric coefficient in the kinetic energy term is equal to one. A
convenient way to derive the corresponding transformation formul\ae\
uses the energy dependent Green's  function $G(E)$ of the kernel $K(T)$
defined by
\begin{equation}
  G(q'',q';E)=\bigg<q''\bigg|{1\over H-E-\i\epsilon}\bigg|q'\bigg>
  =\ih\int_0^\infty dT \e^{\i(E+\i\epsilon)T/\hbar}K(q'',q';T)\enspace.
\end{equation}
For the one-dimensional path integral one obtains the following
transformation formula
\begin{eqnarray}
  K(x'',x';T)&=&\int_{\bbbr}{dE\over2\pi\i}\e^{-\i ET/\hbar}
  G(q'',q';E)\enspace, \\
  G(q'',q';E)&=&\ih\Big[F'(q'')F'(q')\Big]^{1/2}
  \int_0^\infty ds''\hat K(q'',q';s'')\enspace,
\end{eqnarray}
with the transformed path integral $\hat K$ having the form
\begin{eqnarray}       & &\!\!\!\!\!\!\!
  \hat K(q'',q';s'')=\limN\Norm^{1/2}\prod_{k=1}^{N-1}\int dq_k
         \nonumber\\   & &\!\!\!\!\!\!\!\qquad\times
  \exp\Bigg\{\ih\sum_{j=1}^N\Bigg[{m\over2\epsilon}(\Delta q_j)^2
       -\epsilon {F'}^2(\bar q_j)\Big(V(F(\bar q_j))-E\Big)
       -\epsilon\Delta V(\bar q_j)\Bigg]\Bigg\}
                  \\   & &\!\!\!\!\!\!\!
  \equiv\pathints{q}\exp\Bigg\{\ih\ints\Bigg[{m\over2}\dot q^2
     -{F'}^2(q)\Big(V(F(q)-E\Big)-\Delta V(q)\Bigg]\d s\Bigg\}
  \enspace,\qquad\qquad
\end{eqnarray}
and with the quantum potential $\Delta V$ given by
\begin{equation}
  \Delta V(q)=\hbaram\Bigg(3{{F''}^2\over{F'}^2}
     -2{F'''\over F'}\Bigg)\enspace.
\end{equation}
Note that $\Delta V$ has the form of a Schwarz derivative of $F$. A
rigorous lattice derivation is far from being trivial and has been
discussed elsewhere \cite{FLM,GRSh,KLEo}. 
 
Let us consider a pure time transformation in a path integral. Let
($\vec q$ a $D$-dimensional coordinate)
\begin{equation}
  G(\vec q'',\vec q';E)=\sqrt{f(\vec q')f(\vec q'')}
  \ih\int_0^\infty ds''\Big<\vec q''\Big|
  \exp\Big(-\i s''\sqrt{f}\,(H-E)\sqrt{f}/\hbar\Big)\Big|{\vec q}'\Big>
  \enspace,
\end{equation}
which corresponds to the introduction of the ``pseudo-time'' $s''=\intt
ds/f(\vec q(s))$ and we assume that the Hamiltonian $H$ is product
ordered. Then
\begin{equation}
  G(\vec q'',\vec q';E)=\ih(f'f'')^{\half(1-D/2)}
  \int_0^\infty\tilde K(\vec q'',\vec q';s'')\d s''\enspace,
\end{equation}
with the path integral
\begin{eqnarray}       & &
  \tilde K(\vec q'',\vec q';s'')
         \nonumber\\   & &
=\pathints{{\vec q}}\sqrt{\tilde g}  
\exp\left\{\ih\ints\bigg[{m\over2}\tilde h_{ac}\tilde h_{cb}\dot
  q^a\dot q^b-f\Big(V(\vec q)+\Delta V_{PF}(\vec q)-E\Big)
  \bigg]\d s\right\}\enspace.\qquad
\label{NUMAe}
\end{eqnarray}
Here $\tilde h_{ac}=h_{ac}/\sqrt{f}$, $\sqrt{\tilde g}=\det
(\tilde h_{ac})$ and (\ref{NUMAe}) is of the canonical product form.
Note that for $D=2$ the prefactor gives unity.

This latter path integral technique of ''time-transformation'' is used in this
paper in almost all cases in order to solve the corresponding path integrals
in the various coordinate systems on Darboux spaces. Of course, the
time-transformation is used in such a way that the metric term $(g_{ab})$ is
transformed to unity. 

In our calculations we have in all cases a metric which is diagonal,
and in almost all cases is of the form
$g_{ab}=f^2\delta_{ab}$. This has the consequence that the quantum
potential $\Delta V=0$ and the term $\tilde h_{ac}$ can be transformed
to unity. This simplifies the calculations significantly.

\section{Some Important Path Integral Solutions and Identities}
\message{Some Important Path Integral Solutions and Identities}
In this Appendix we cite some important path integral solutions,
in particular for the (radial) harmonic oscillator, the linear potential,
and for the modified P\"oschl--Teller potential. 

\subsection{The Path Integral for the Radial Harmonic Oscillator}
\message{The Path Integral for the Radial Harmonic Oscillator}
 The calculation of the path integral for the radial harmonic oscillator
has first been performed by Peak and Inomata \cite{PI}. 
For a comprehensive bibliography, see \cite{GRSh}. We have the path
integral representation ($r>0$)
\begin{eqnarray}
  &&\pathint{r}
  \exp\left[\ih\int_{t'}^{t''}\left({m\over2}
  \big(\dot r^2-\omega^2r^2)
      -\hbar^2{\lambda^2-\viert\over2mr^2}\right)\dt\right]
  \nonumber\\   &&=
  {m\omega\sqrt{r'r''}\over\i\hbar\sin\omega T}
  \exp\bigg[-{m\omega\over2\i\hbar}({r'}^2+{r''}^2)\cot\omega T\bigg]
  I_\lambda\bigg({m\omega r'r''\over\i\hbar\sin\omega T}\bigg)
\end{eqnarray}
$I_\lambda(z)$ is a modified Bessel function.
The energy dependent kernel (Green's function $G(E)$) is given by
\begin{eqnarray}
  &&\ih\int_0^\infty\d E\,\e^{\i ET/\hbar}\pathint{r}
 \exp\left[\ih\int_{t'}^{t''}\left({m\over2}
  \big(\dot r^2-\omega^2r^2)
      -\hbar^2{\lambda^2-\viert\over2mr^2}\right)\dt\right]
  \nonumber\\   &&=
  {\Gamma[\half(1+\lambda-E/\hbar\omega)]\over\hbar\omega\sqrt{r'r''}\,
   \Gamma(1+\lambda)}
  W_{E/2\hbar\omega,\lambda/2}\bigg({m\omega\over\hbar}r^2_>\bigg)
  M_{E/2\hbar\omega,\lambda/2}\bigg({m\omega\over\hbar}r^2_<\bigg)
  \enspace.
\end{eqnarray}
Here $M_{\mu,\nu}(z)$ and $W_{\mu,\nu}(z)$ are Whittaker functions.

\subsection{Green's Function the Linear Potential}
\message{Green's Function for the Linear Potential}
The energy dependent kernel for the linear potential is given by 
\begin{eqnarray}
   &&\ih\int_0^\infty\d E\,\e^{\i ET/\hbar}\pathint{x}
   \exp\left[\ih\int_0^T\bigg({m\over2}\dot x^2-kx\bigg)\dt\right]
\nonumber\\   &&
   ={4\over3}{m\over\hbar^2}\bigg[
   \bigg(x'-{E\over k}\bigg)\bigg(x''-{E\over k}\bigg)\bigg]^{1/2}
  K_{1/3}\left[{\sqrt{8mk}\over3\hbar}
   \bigg(x_>-{E\over k}\bigg)^{3/2}\right]
    I_{1/3}\left[{\sqrt{8mk}\over3\hbar}
   \bigg(x_<-{E\over k}\bigg)^{3/2}\right]\enspace.
\nonumber\\ 
\end{eqnarray}

\subsection{Green's Function for the Harmonic Oscillator}
\message{Green's Function for the Harmonic Oscillator}
The energy dependent kernel for the harmonic oscillator is given by
\begin{eqnarray}
&&\ih\int_0^\infty\d E\,\e^{\i ET/\hbar}
\pathint{x}\exp\left[\frac{\i m}{2\hbar}
\int_0^T(\dot x^2-\omega^2x^2)\dt\right]
\nonumber\\   &&
  =\sqrt{m\over\pi\hbar^3\omega}\,
   \Gamma\bigg(\half-{E\over\hbar\omega}\bigg)
  D_{-\half+{E\over\hbar\omega}}\left(\sqrt{2m\omega\over\hbar}\,
   x_>\right)
  D_{-\half+{E\over\hbar\omega}}\left(-\sqrt{2m\omega\over\hbar}\,
   x_<\right)\,.
\end{eqnarray}
The $D_\nu(z)$ are parabolic cylinder functions. 

\subsection{The Modified P\"oschl--Teller Potential}
\message{The Modified Poschl-Teller Potential}
The path integral solution for the modified P\"oschl--Teller
potential can be achieved by means of the $\SU(1,1)$-path
integral. For a comprehensive bibliography, see \cite{GRSh}. 
We have \cite{BJb,DURb,FLM,KLEMUS}
\begin{eqnarray}
  &&\pathint{r}
  \exp\left\{\ih\int_{t'}^{t''}\left[{m\over2}\dot r^2
   -\hbarm \bigg({\eta^2-{1\over4}\over\sinh^2r}
   -{\nu^2-{1\over4}\over\cosh^2r}\bigg)\right]\dt\right\}
  \nonumber\\   &&
  =\sum_{n=0}^{N_M}\Psi_n^{(\eta,\nu)\,*}(r')
  \Psi_n^{(\eta,\nu)}(r'')
  \exp\bigg\{{\i\hbar T\over2m}\Big[2(k_1-k_2-n)-1\Big]^2\bigg\}
   \nonumber\\   &&\qquad\qquad\times
  +\int_0^\infty dp\,\Psi_p^{(\eta,\nu)\,*}(r')
  \Psi_p^{(\eta,\nu)}(r'') \exp\bigg(-{\i\hbar T\over2m}p^2\bigg)
  \enspace.
\end{eqnarray}
Let us introduce the numbers $k_1,k_2$ defined by:
$k_1=\half(1\pm\nu)$, $k_2=\half(1\pm\eta)$, where the correct sign
depends on the boundary-conditions for $r\to0$ and $r\to\infty$,
respectively. In particular for $\eta^2={1\over4}$, i.e.\
$k_2={1\over4},{3\over4}$, we obtain wavefunctions with even and odd
parity, respectively. The number $N_M$ denotes the maximal number of
states with $0,1,\dots,N_M<k_1-k_2-\half$. The bound state
wavefunctions read as ($\kappa=k_1-k_2-n$)
\begin{eqnarray}
  \Psi_n^{(\eta,\nu)}(r)
  &=&N_n^{(\eta,\nu)}(\sinh r)^{2k_2-\half}
                    (\cosh r)^{-2k_1+{3\over2 }}
  \nonumber\\   &&\qquad\times
  {_2}F_1(-k_1+k_2+\kappa,-k_1+k_2-\kappa+1;2k_2;-\sinh^2r)
  \\
  N_n^{(\eta,\nu)}
  &=&{1\over\Gamma(2k_2)}
  \bigg[{2(2\kappa-1)\Gamma(k_1+k_2-\kappa)
                     \Gamma(k_1+k_2+\kappa-1)\over
    \Gamma(k_1-k_2+\kappa)\Gamma(k_1-k_2-\kappa+1)}\bigg]^{1/2}\enspace.
\end{eqnarray}
The scattering states are given by:
 \begin{eqnarray}
 \Psi_p^{(\eta,\nu)}(r)
  &=&N_p^{(\eta,\nu)}(\cosh r)^{2k_1-\half}(\sinh r)^{2k_2-\half}
  \nonumber\\   &&\qquad\qquad\times
  {_2}F_1(k_1+k_2-\kappa,k_1+k_2+\kappa-1;2k_2;-\sinh^2r)
  \\
  N_p^{(\eta,\nu)}
  &=&{1\over\Gamma(2k_2)}\sqrt{p\sinh\pi p\over2\pi^2}
  \Big[\Gamma(k_1+k_2-\kappa)\Gamma(-k_1+k_2+\kappa)
  \nonumber\\   &&\qquad\qquad\times
  \Gamma(k_1+k_2+\kappa-1)\Gamma(-k_1+k_2-\kappa+1)\Big]^{1/2}\enspace,
\end{eqnarray}
[$\kappa=\half(1+\i p)$]. ${_2}F_1(a,b;c;z)$ is the hypergeometric function.
The Green's function has the form 
\begin{eqnarray}
 &&\ih\int_0^\infty\d E\,\e^{\i ET/\hbar}\pathint{r}
  \exp\left\{\ih\intt\left[{m\over2}\dot r^2
   -\hbarm\bigg({\eta^2-\viert\over\sinh^2r}
   -{\nu^2-\viert\over\cosh^2r}\bigg)\right]\dt\right\}
  \nonumber\\   &&
  ={m\over\hbar^2}{\Gamma(k_1-L_\nu)\Gamma(L_\nu+k_1+1)\over
   \Gamma(k_1+k_2+1)\Gamma(k_1-k_2+1)}
  (\cosh r'\cosh r'')^{-(k_1-k_2)}(\tanh r'\tanh r'')^{k_1+k_2+1/2}
  \nonumber\\   &&\qquad\times
  {_2}F_1\bigg(-L_\nu+k_1,L_\nu+k_1+1;k_1-k_2+1;
                          {1\over\cosh^2r_<}\bigg)
  \nonumber\\   &&\qquad\times
  {_2}F_1\bigg(-L_\nu+k_1,L_\nu+k_1+1;k_1+k_2+1;\tanh^2r_>\bigg)\enspace.
\end{eqnarray}

\end{appendix}


\input cyracc.def
\font\tencyr=wncyr10
\font\tenitcyr=wncyi10
\font\tencpcyr=wncysc10
\def\cyrrm{\tencyr\cyracc}
\def\cyrit{\tenitcyr\cyracc}
\def\cyrcp{\tencpcyr\cyracc}
\addcontentsline{toc}{section}{References}%
\renewcommand{\baselinestretch}{0.95}%
\noindent%

\end{document}